\documentclass[12pt, draftclsnofoot, onecolumn]{IEEEtran}
\usepackage{cite}
\usepackage{graphicx}
\usepackage{amsmath}
\usepackage{array}
\usepackage{mdwmath}
\usepackage{amssymb}
\usepackage{mdwtab}
\usepackage{stfloats}
\usepackage[tight,footnotesize]{subfigure}
\usepackage{amsmath,amsthm}
\usepackage{threeparttable}
\usepackage{color}
\usepackage{url}
\usepackage{algpseudocode}
\usepackage{algorithm}
\usepackage{multicol}
\algrenewcommand{\algorithmicrequire}{\textbf{Input:}}
\algrenewcommand{\algorithmicensure}{\textbf{Output:}}

\newtheorem{theorem}{Theorem}
\newtheorem{remark}{Remark}
\newtheorem{lemma}{Lemma}

\begin{document}

\title{Wireless Performance Evaluation of Building Layouts: Closed-Form Computation\\ of Figures of Merit}
\author{Jiliang~Zhang,~\textit{Senior Member~IEEE},~Andr\'es~Alay\'on~Glazunov,~\textit{Senior Member IEEE},
 and Jie~Zhang,~\textit{Senior Member IEEE}\
\thanks{
Jiliang Zhang is with the Department of Electronic and Electrical Engineering, The University of Sheffield, Sheffield, UK. 
Andr\'es Alay\'on Glazunov is with Department of Electrical Engineering, University of Twente, Enschede, Overijssel, Netherlands, and also with Department of Electrical Engineering, Chalmers University of Technology, Gothenburg, Sweden.
Jie Zhang is with the Department of Electronic and Electrical Engineering, the University of Sheffield, Sheffield, UK, and also with Ranplan Wireless Network Design Ltd., Cambridge, UK.
}
\thanks{The research is funded by EUROSTARS Project BuildWise (11088).}
}
\markboth{IEEE}%
{Shell \MakeLowercase{\textit{et al.}}: Bare Demo of IEEEtran.cls for Journals}
\maketitle
\begin{abstract}
This paper presents a part of our ground-breaking work on evaluation of buildings in terms of wireless friendliness in the building-design stage. The main goal is to devise construction practices that provide for a good performance of wireless networks deployed in buildings.
In this paper, the interference gain (IG) and power gain (PG) are defined as two figures of merit (FoM) of the wireless performance of buildings. 
The FoMs bridge the gap between building design and wireless communications industries.
An approach to derive exact closed-form equations for these FoMs is proposed for the first time. The derived analytic expressions facilitate straightforward and more computationally efficient numerical evaluation of the proposed FoMs as compared to Monte Carlo simulations for well-known indoor propagation models. It is shown that the derived closed-form expression can be readily employed to evaluate the impact of building properties, such as the sizes and the aspect ratios (ARs) of rooms, on the wireless performance.
The proposed approach sheds light to architects on evaluation and design of wireless-friendly building layouts. 

\end{abstract}

\IEEEpeerreviewmaketitle

\section{Introduction}

It has been predicted that mobile traffic will increase by up to $1000$ times in the next decade, with over $80\%$ of mobile traffic taking place indoors \cite{indoors80}. 
6G systems will meet the required mobile traffic through the combined effects of densification of the wireless network through small cells and the large-scale deployment of massive multiple-input multiple-output (MIMO) array antennas \cite{6Gdefine}. The large number of small cells are anticipated to enhance coverage and spectral efficiency \cite{sc2,sc3,sc4,sc5,jz1,jz2,jz3}, whereas in massive MIMO, array antennas are scaled up to increase reliability and spectral efficiency \cite{massive1, massive2, massive3,xg0}. On the other hand, wireless communication plays an important role in the realization of the ``smart building" paradigm, or more broadly, in the ``smart city" vision \cite{smartcity}. Different types of buildings have an intrinsic wireless performance which is independent of how densely small cells are deployed, or the massive MIMO array antenna deployed \cite{patent,tvtbwp}. 
The present work is the first step towards providing guidance, to both civil engineers and architects, on incorporating the wireless performance of buildings at the design stage. 

Building design and construction has mainly focused on improving safety, visual and thermal comfort, and indoor air quality \cite{tallbuilding}. In the last decade, energy efficiency became an important metric in building design tools \cite{buildingenergy}. The operational energy required by a building can be evaluated by taking into account the lighting, the heating, the ventilation and the cooling as well as the provision of hot water \cite{cibse}. 
At present, mobile operators are already among the top energy consumers \cite{r1}, and an increasing trend of power consumption of wireless networks can be observed in different generations of wireless systems \cite{r2,r3}.  
Therefore, the energy saving is worthwhile to be considered in the design of green buildings. However, the indoor power consumption of wireless networks has not been considered.

One important factor impacting the power consumption in small-cell networks is the building-blockage.
In \cite{LOS1,LOS2,LOS3}, the LOS probability is analytically and empirically modelled.
In \cite{TB}, an analytic expression for the coverage rate is provided by approximating the line-of-sight (LOS) region as a fixed LOS ball.
In \cite{wallgeneration}, four random wall generation methods to place blockage objects in an indoor scenario were proposed and analyzed, where walls are either distributed randomly, semi-deterministically or heuristically. 
However, previous evaluations of the blockage effect proposed that buildings form a random process of shapes.  The LOS probability was there assumed to be irrelevant for different directions of the link. This model seems to be promising in providing insights on the configuration of a wireless network in a homogeneous indoor scenario, where walls are assumed to be uniformly distributed in a building. However, in specific building designs, whether a wireless link is blocked by structures not only depends on the distance between the transmitter and the receiver, but also on their relative position and direction in the floor plan of the building. 
That is, the blockage condition of the wireless signal depends on the considered direction. Therefore, the homogeneous assumption is not suitable for evaluating specific building designs.

Rather than random blockages, ray-tracing/ray-launching techniques are employed to evaluate the indoor performance of wireless networks (i.e., for users located indoors) for specific building designs in \cite{raytracing1,raytracing2,raytracing3,raytracing4,raytracing5,raytracing6,ZHL}. 
However, results attained through this approach are only valid for a specific network deployment, and are not general enough to evaluate the wireless performance of buildings, where all possible positions of transmitters and receivers in the building under design are needed to be evaluated. 
In this case, ray-tracing/ray-launching techniques become computationally expensive, and therefore it is not practical to evaluate the wireless performance of a building layout in a reasonable time.

Moreover, wireless networks are traditionally designed and deployed when the buildings have been built. Before the networks are deployed, wireless communications engineers have to carry out ray-tracing/ray-launching based simulations for specific deployments of wireless networks. In other words, wireless communications engineers adapt a network to suit its deployment scenarios without modifying any property of the building layout. 
However, for an existing building with a specific intrinsic wireless performance determined by the building layout, the indoor wireless networks may not achieve an acceptable performance regardless how densely small cells or MIMO array antenna elements are deployed. 
The wireless performance of building layout has to be taken into account carefully in the building design stage, which means that building designers, both civil engineers and architects, requires guidance on how to predict and evaluate the wireless performance of a building under design.

Taking wireless performance of buildings into account will become indispensable for the design of future smart buildings in which appliances and many other devices and ``things" will all be connected wirelessly. To address the challenges posed by the building evaluation in terms of wireless performance, ground-breaking works are needed. We are engaged in the collaborative project Build-Wise \cite{eurostars}. Supported by this project, figures of merits (FoMs)
reflecting the intrinsic wireless performance of buildings have been defined considering the impact of blockage, bounce reflections, and waveguide effects introduced by indoor building structures employing the open space scenario as a benchmark for wireless network performance evaluation \cite{patent,tvtbwp}. 
The proposed FoMs bridge the gap between building design and wireless communications, laying the foundation to an improved wireless service indoors by optimising building layouts.
However, the computation of the FoMs is based on Monte Carlo simulations, which is computationally expensive since an infinite number of transmit elements is assumed in the FoM definition. Therefore, an analytic framework is critically needed for fast computation of the FoMs.


The main contribution of this paper is a mathematical framework to evaluate the wireless performance of buildings.
The proposed approach facilitates straightforward and more computationally efficient numerical evaluation of the proposed FoMs.
Firstly, two FoMs, i.e., the interference gain (IG) and power gain (PG) are defined to capture the impact of the building layout on the power of interference signals and intended signals, respectively.
Secondly, an approach to derive exact FoMs in closed form is proposed for polygonal-shaped rooms. By using this approach, the FoMs can be quickly and exactly computed given building layout tiled by polygonal rooms.
Thirdly, the analytic approach is validated through Monte Carlo simulations for typical indoor path gain models.
This proposed analytic framework provides a direct and practical quantification of the wireless performance of a building.
Finally, the impact of the building layout, e.g., their sizes and the aspect ratios (AR) of rectangular rooms, on wireless performance FoMs is analyzed based on the proposed analytic approach.

The remainder of this paper is organized as follows. Section II summarizes the notation and main assumptions made in this paper. In Section III, two FoMs, i.e.,  the IG, and the PG are introduced to measure the wireless performance of a building. In Section IV, an approach to compute the closed-form expressions of the proposed FoMs is provided. In Section V, the approach is validated through Monte Carlo simulations. In Section VI, the impact of building layout on FoMs is analyzed. Conclusions are provided in Section VII.

\section{Notation and assumption}
\label{Sec:II}
\subsection{Notation}
We use the following notations throughout this paper.
1) Polylogarithm $\mathrm{Li}_s(z) \triangleq \sum\limits_{k=1}^\infty {\frac{z^k}{k^s}}$.
2) Beta function $B(x,y)\triangleq\int_0^1t^{x-1}(1-t)^{y-1}\mathrm{d}t=\frac{\Gamma(x)\Gamma(y)}{\Gamma(x+y)}$.
3) Gamma function $\Gamma(z) \triangleq \int_0^\infty x^{z-1} e^{-x} \mathrm{d}x=\frac{1}{z} \prod\limits_{n=1}^{\infty} \frac{\left(1 + \frac{1}{n}\right)^z}{1+\frac{z}{n}}$.
4) Hypergeometric function ${}_2F_1(a,b;c;z) \triangleq\frac{1}{B(b,c-b)} \int_0^1 x^{b-1} (1-x)^{c-b-1}(1-zx)^{-a}\mathrm{d}x= \sum\limits_{n=0}^\infty \frac{(a)_n (b)_n}{(c)_n} \frac{z^n}{n!}$, where $(x)_{n}\triangleq \prod\limits_{k=0}^{n-1}(x-k)$.
5) ``$\land$'', ``$\lor$'', and ``$\lnot$'' denote the logical conjunction, 
the logical disjunction, 
and the logical complement, respectively. 
``$\top$'' and ``$\bot$'' denote the logical tautology and logical contradiction, respectively.
6) $\mathrm{Im}[z]$ denotes the imaginary part of a complex number $z$.
7) In the open space scenario, $I_{\mathrm{O}}$ and $P_{\mathrm{O}}$ denote the power of interference signals and the power of intended signals, respectively. In the LOS scenario, $I_{\mathrm{L}}$ and $P_{\mathrm{L}}$ denote the power of interference signals and the power of intended signals, respectively. In the non-LOS (NLOS) scenario, $I_{\mathrm{N}}$ and $P_{\mathrm{N}}$ denote the power of interference signals and the power of intended signals, respectively.
8) $\gamma_{\mathrm{O}}$ and $\gamma_{\mathrm{B}}$ denote the signal-to-interference-plus-noise-ratio (SINR) in the open space and the indoor scenario, respectively.
9) $G_{\mathrm{O}}$, $G_{\mathrm{L}}$, and $G_{\mathrm{N}}$ denote path gain models in the open space, the LOS, and the NLOS scenarios, respectively.
10) $P_{\mathrm{T}}$ $[\mathrm{W}\mathrm{m}^{-2}] $ denotes the transmit power density, i.e., transmit power per unit area.
11) $P_{\mathrm{th}}$ $[\mathrm{W}\mathrm{m}^{-2}] $ denotes the threshold power level determined by the sensitivity of the receiver also defined per unit area.
12) $R_{\mathrm{O}}$, $R_{\mathrm{L}}$ and $R_{\mathrm{N}}$ denote the coverage distances under open space, LOS and NLOS scenarios, respectively.
13) $g_{\mathrm{I}}$ and $g_{\mathrm{P}}$ denote the IG and the PG of buildings, respectively.
14) $\delta (\xi)$ is defined as a function of inequality $\xi$. If $\xi$ holds, $\delta (\xi)=1$. Otherwise,   $\delta (\xi)=0$. 

\subsection{Assumptions}
We use the following assumptions throughout this paper.
\begin{enumerate}
\item 
The FoMs proposed in this paper is designed to evaluate a building layout in terms of wireless performance.
Instead of any specifically defined radio network, the idea of the approach is to evaluate the maximum achievable performance gain of indoor wireless networks in a building. 
In this paper, $P_{\mathrm{T}}$ $[\mathrm{W}/\mathrm{m}^{2}] $ denotes the transmit power density, i.e., transmit power per unit area. For a small cell network with an intensity of $\Lambda\ \mathrm{[BSs/m^2]}$, the power of each small cell base station is $P_{\mathrm{T}}/\Lambda\ \mathrm{[W/BS]}$.
To figure out the maximum achievable performance gain, we need to use the best case of indoor wireless network deployment even though it might be not realistic in existing networks. 
It is predictable that the average SINR for a downlink reference user equipment (UE)  with a fixed sensitivity increases with the intensity of transmit elements with a fixed total transmit power. Therefore, an infinite number of transmit elements is assumed in our derivations, i.e., small cells \cite{6Gdefine} or equivalent antenna elements \cite{dmmimo} are assumed to be uniformly distributed in both the indoor and the outdoor environment.

\item Multi-slope path gain models \cite{twolobe} for the open space, the LOS and the NLOS scenarios are employed in this paper. Path gain is defined in the general form as
\begin{eqnarray}
G(R)\triangleq \frac{P_{\mathrm{R}}}{P_{\mathrm{T}}},
\end{eqnarray}
where 
$P_{\mathrm{R}}$ and $P_{\mathrm{T}}$ denote the received and the transmit power densities, respectively,
and $R$ denotes the horizontal distance from the transmitter to the receiver. 

The path gain in the open space scenario is given by \cite[Eqs. (4,5)]{tworay}
\begin{eqnarray}
\label{PL_os}
G_{\mathrm{O}}(R)=
\min\left\{ 1, \left(\frac{\lambda}{4\pi} \right)^2R^{-2}, \left(h_{\mathrm{T}}h_{\mathrm{R}}\right)^2{R^{-4}} \right\},
\end{eqnarray}
where $h_\mathrm{T}$ and $h_\mathrm{R}$ are the heights of the transmit and the receive antenna, respectively; $f_{\mathrm{c}}$ is the center frequency,
$\lambda\triangleq \frac{c}{f_{\mathrm{c}}}$ denotes the corresponding wavelength, $c=3\times 10^8 \mathrm{[m/s]}$ is the speed of light.

Path gain models for the LOS and NLOS indoor scenarios can then be computed as
\begin{eqnarray}
&&\label{PLLOS}
G_{\mathrm{L}}(R)=
\left\{
\begin{array}{l l}
G_{\mathrm{O}}(R),& R\leq1~\mathrm{m},\\
 \left(\frac{\lambda}{4\pi} \right)^2R^{-n_{\mathrm{L}}},&R>1~\mathrm{m}.
\end{array}
\right.
\\&&
\label{GN}
G_{\mathrm{N}}(R)=
\left\{
\begin{array}{l l}
G_{\mathrm{O}}(R),& R\leq1~\mathrm{m},\\
 \left(\frac{\lambda}{4\pi} \right)^2R^{-n_{\mathrm{N}}},&R>1~\mathrm{m},
\end{array}
\right.
\end{eqnarray}
where the numerical values of $n_{\mathrm{L}}$ and $n_{\mathrm{N}}$ are determined by the type of environment
including building materials, shape and size of rooms, etc. Following the energy conservation law, path gains $G_{\mathrm{O/L/N}}$ have to be less than one. 
\item 
In this paper, we specialize our results to $n_{\mathrm{L}}=1.73$, and $n_{\mathrm{N}}=3.19$ following the 3GPP indoor channel model working at 0.5-100 GHz \cite[Table 7.4.1-1]{3GPP}.  
Since the 3GPP model is defined for $R>1~\mathrm{m}$, we assume that $G_{\mathrm{L}}(R)=G_{\mathrm{N}}(R)=G_{\mathrm{O}}(R)$, at $R\leq1~\mathrm{m}$.
\item
We constrain ourselves to $R_{s}>1$ $\rm m$, $s\in\left\{\mathrm{O},\mathrm{L},\mathrm{N}\right\}$, and therefore $P_{\mathrm{th}}<P_{\mathrm{T}}\left(\frac{\lambda}{4\pi}\right)^2$.\item In this paper, the detectable power is defined as the receive power from transmit elements satisfying the condition
\begin{eqnarray}
P_{\mathrm{T}}G_{s}(R)>P_{\mathrm{th}},  s\in\{\mathrm{O, L, N}\},
\end{eqnarray}
where $P_{\mathrm{th}}$ $[\mathrm{W}\mathrm{m}^{-2}] $ is a threshold power level determined by the sensitivity of the UE receiver.
\item $R_{\mathrm{O}}$, $R_{\mathrm{L}}$ and $R_{\mathrm{N}}$ denote the coverage distances under open space, LOS and NLOS scenarios, respectively and satisfy the identities $P_{\mathrm{T}}G_{\mathrm{O}}(R_{\mathrm{O}})=P_{\mathrm{th}}$, $P_{\mathrm{T}}G_{\mathrm{L}}(R_{\mathrm{L}})=P_{\mathrm{th}}$, and $P_{\mathrm{T}}G_{\mathrm{N}}(R_{\mathrm{N}})=P_{\mathrm{th}}$. Hence, it can be shown after some straightforward algebraic manipulations that
\begin{eqnarray}
&&\label{RO}
R_{\mathrm{O}}=
\left\{
\begin{array}{l l}
\sqrt{\frac{P_{\mathrm{T}}}{P_{\mathrm{th}}}} \frac{\lambda}{4\pi}, & \frac{P_{\mathrm{T}}}{P_{\mathrm{th}}}<\frac{(4\pi)^4\left({h_{\mathrm{T}}h_{\mathrm{R}}}\right)^2}{\lambda^4},\\
\left(\frac{P_{\mathrm{T}}}{P_{\mathrm{th}}}\right)^{\frac{1}{4}} \sqrt{h_{\mathrm{T}}h_{\mathrm{R}}}, &
\frac{P_{\mathrm{T}}}{P_{\mathrm{th}}}\geq
\frac{(4\pi)^4\left({h_{\mathrm{T}}h_{\mathrm{R}}}\right)^2}{\lambda^4},
\end{array}
\right.
\\&&\label{RL}
R_{\mathrm{L}}=\left(\frac{P_{\mathrm{T}}}{P_{\mathrm{th}}}\right)^{\frac{1}{n_{\mathrm{L}}}} \left(\frac{\lambda}{4\pi} \right)^{\frac{2}{n_{\mathrm{L}}}},
\\&&\label{RN}
R_{\mathrm{N}}=\left(\frac{P_{\mathrm{T}}}{P_{\mathrm{th}}}\right)^{\frac{1}{n_{\mathrm{N}}}} \left(\frac{\lambda}{4\pi} \right)^{\frac{2}{n_{\mathrm{N}}}}.
\end{eqnarray}
\item 
In the best case, the reference UE can make use of all the detectable power, which is constrained by the sensitivity of its receiver. The receive power is in turn assumed to be the result of maximum ratio transmissions (MRT) in distributed massive MIMO networks or coordinated multipoint transmission (CoMP) in ultra-dense small-cell networks \cite{comp1,comp2}, where all network elements transmit the same signal for the reference UE continuously at full power. To simplify the analysis, we further assume that all transmit elements use the same frequency band.
\item In cooperative ultra-dense small-cell/massive MIMO systems, channel responses are smoothed by the extremely large spatial diversity as a result of the favorable action of the law of large numbers. In essence, small-scale fading is negligible \cite{5Gdefine}, therefore, only large-scale fading is considered in this paper.
\end{enumerate}

\begin{figure}[!t]
  \centering
  \includegraphics [width=2.5in]{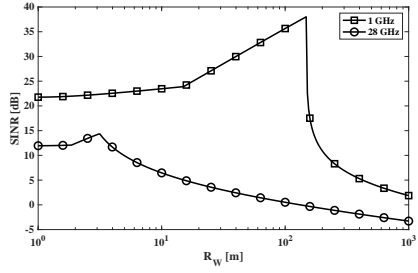}
\caption{Impact of the radii $R_{\mathrm{W}}$ of a circular room on the SINR of the reference UE at the center of it.
$\sigma=-93$ dB, 
$P_\mathrm T=-30$ dBW/m$^2$,
$P_\mathrm {th}=-100$ dBW/m$^2$,
and
$h_\mathrm T=h_\mathrm R=1.2$ m.
} \label{circ}
\end{figure}

With all the assumptions, 
we first observe the impact of the room structure on a reference UE located in the center of a circular room with a radius of $R_\mathrm{W}$. As shown in Fig. \ref{circ}, the SINR of the reference UE varies significantly for different values of $R_\mathrm{W}$. Thus, it is necessary to consider this impact on the building design stage. 
In practice, rooms in a building are not circular in general, and the evaluation of the building is therefore much more complicated. In the following Section III, we will design the FoMs to evaluate the wireless friendliness of building layouts.
\section{FoM definition}

In order to meaningfully evaluate the wireless performance of buildings, the impact of buildings on wireless networks has to be well understood. To help in the analysis we employ the open space scenario with path gain model (\ref{PL_os}) as a benchmark or reference. Then, the impact of buildings on the wireless networks can be evaluated by comparing the SINR of the signals received indoors with that in the open space scenario.

The expression of the SINR in the open space is given as follows.
Using a polar-coordinate system $(R,\theta)$ by choosing the UE as the reference point and an arbitrary angle as the reference 0-angle in the open space, the power of the intended signal is given by the integral of the receive power from all transmit elements with a distance that is less than $R_{\mathrm{O}}$, i.e.,
\begin{eqnarray}
\label{POdefeq}
P_{\mathrm{O}}=\int_0^{2\pi} \int_0^{R_{\mathrm{O}}} P_{\mathrm{T}}G_{\mathrm{O}}(R) R\mathrm{d}R\mathrm{d}\theta.
\end{eqnarray}
Whereas, the power of the interference signal is given by
\begin{eqnarray}
\label{IOdefeq}
I_{\mathrm{O}}=\int_0^{2\pi} \int_{R_{\mathrm{O}}}^{+\infty} P_{\mathrm{T}}G_{\mathrm{O}}(R) R\mathrm{d}R\mathrm{d}\theta.
\end{eqnarray}

Thus, the SINR in an open space scenario is given by
\begin{eqnarray}
\label{snros}
\gamma_{\mathrm{O}}=\frac{P_{\mathrm{O}}}{I_{\mathrm{O}}+\sigma^2}.
\end{eqnarray}

Then, the SINR of a UE in the considered building is given as follows.
For an arbitrary transmit element located at $(R, \theta)$, the transmission from the element to the UE could be either LOS or NLOS. If the transmission is LOS, the received signal is considered as an intended signal if $R<R_{\mathrm{L}}$. Otherwise, if $R\geq R_{\mathrm{L}}$, the received signal is considered as an interference signal. Similarly, if the transmission is NLOS, the intended signal satisfies $R<R_{\mathrm{N}}$, and the interference signal satisfies $R\geq R_{\mathrm{N}}$. For a given $(R,\theta)$, we use a Boolean variable $\mathcal{V}(R,\theta)$ to denote that ``the transmission from the transmit element at $(R,\theta)$ to the UE is LOS''. Accordingly, $\lnot\mathcal{V}(R,\theta)$ denotes that ``the transmission from the transmit element at $(R,\theta)$ to the UE is NLOS''. If a signal transmitted from $(R,\theta)$ is an intended signal, we have $[\mathcal{V}(R,\theta)\land (R<R_{L})]\lor[\lnot\mathcal{V}(R,\theta)\land (R<R_{N})]$. Then, the power of intended signals in the considered building is given by
\begin{eqnarray}
P_{\mathrm{B}}= \underbrace{\int_{\mathcal{V}(R,\theta)\land (R<R_{L})} P_{\mathrm{T}}G_{\mathrm{L}}(R) R\mathrm{d}R\mathrm{d}\theta}_{P_{\mathrm{L}}}
+\underbrace{\int_{\lnot\mathcal{V}(R,\theta)\land (R<R_{N})} P_{\mathrm{T}}G_{\mathrm{N}}(R) R\mathrm{d}R\mathrm{d}\theta}_{P_{\mathrm{N}}},
\end{eqnarray}
where $P_{\mathrm{L}}$ and $P_{\mathrm{N}}$, respectively, denote receive power of intended signals from LOS and NLOS small cells/antenna elements, and can be computed separately since $\mathcal{V}(R,\theta)\land (R<R_{L})$ and $\lnot\mathcal{V}(R,\theta)\land (R<R_{N})$ are exclusive.
Similarly, the power of the interference signal is given by
\begin{eqnarray}
I_{\mathrm{B}}=\underbrace{\int_{\mathcal{V}(R,\theta)\land (R\geq R_{L})} P_{\mathrm{T}}G_{\mathrm{L}}(R) R\mathrm{d}R\mathrm{d}\theta}_{I_{\mathrm{L}}}
+\underbrace{\int_{\lnot\mathcal{V}(R,\theta)\land (R\geq R_{N})} P_{\mathrm{T}}G_{\mathrm{N}}(R) R\mathrm{d}R\mathrm{d}\theta}_{I_{\mathrm{N}}},
\end{eqnarray}
where $I_{\mathrm{L}}$ and $I_{\mathrm{N}}$, respectively, denote the LOS and the NLOS received powers of interference signals.
Therefore, we have the expression of the indoor SINR as
\begin{eqnarray}
\label{snrb}
\gamma_{\mathrm{B}}=\frac{P_{\mathrm{B}}}{I_{\mathrm{B}}+\sigma^2}=\frac{P_{\mathrm{L}}+P_{\mathrm{N}}}{I_{\mathrm{L}}+I_{\mathrm{N}}+\sigma^2}.
\end{eqnarray}

Combining (\ref{snros}) and (\ref{snrb}), we obtain
\begin{eqnarray}
\label{metricdef}
\gamma_{\mathrm{B}}
=\frac{{I_{\mathrm{O}}+\sigma^2}}{I_{\mathrm{L}}+I_{\mathrm{N}}+\sigma^2}
\frac{P_{\mathrm{L}}+P_{\mathrm{N}}}{P_{\mathrm{O}}}\gamma_{\mathrm{O}}
=g_{\mathrm{I}}g_{\mathrm{P}}\gamma_{\mathrm{O}},
\end{eqnarray}
and we define two FoMs of a building's wireless performance as \cite{patent,tvtbwp}
\begin{eqnarray}
&&
\label{GI}
g_{\mathrm{I}}\triangleq\frac{{I_{\mathrm{O}}+\sigma^2}}{I_{\mathrm{L}}+I_{\mathrm{N}}+\sigma^2},
\\
&&
\label{GP}
g_{\mathrm{P}}\triangleq \frac{P_{\mathrm{L}}+P_{\mathrm{N}}}{P_{\mathrm{O}}},
\end{eqnarray}
where $g_{\mathrm{I}}$ and $g_{\mathrm{P}}$ denote the IG and the PG, respectively. The variables $I_{\mathrm{O}}$ and $P_{\mathrm{O}}$ denote the power of interference signals and the power of intended signals in the open space scenario, respectively. In the LOS scenario, $I_{\mathrm{L}}$ and $P_{\mathrm{L}}$ denote the power of interference signals and the power of intended signals, respectively. In the NLOS scenario, $I_{\mathrm{N}}$ and $P_{\mathrm{N}}$ denote the power of interference signals and the power of intended signals from the transmit elements in the floor where the UE located, respectively. 
$\sigma^2$ denotes the power of thermal noise. 

\begin{remark}
The FoMs $g_{\mathrm{I}}$ and $g_{\mathrm{P}}$ are intrinsic wireless performance characteristics of buildings. They represent the effective change in the SINR when coverage is provided within the building relative to the SINR in the open space scenario. On one hand, $g_{\mathrm{I}}$ captures the impact of the building on the power of interference signals due to blockage, while, on the other hand, $g_{\mathrm{P}}$ captures the impact of the building on the power of intended signals.
\end{remark}

The proposed FoMs and their analytic computation approach, will provide civil engineers and architects  guidance to evaluate the wireless friendliness of a building layout under their design. If FoMs are lower than expected, an acceptable wireless performance will not be achievable.
For communication engineers, the FoM is a performance gain upper bound provided by the buildings under design. To achieve the upper bound, Assumptions 1) and 7) stated in Section II need to be approached.

In our model we consider an infinite number of small cells/antenna elements. Therefore, analytic expressions are comparatively more computationally efficient than performing Monte Carlo simulations to approximately compute the FoMs. Also, analytic results can provide accurate results and lead to a more straightforward comparison and optimization of the building layout. Therefore, we propose next an approach to derive $g_{\mathrm{I}}$ and $g_{\mathrm{P}}$ in closed form for evaluating the wireless performance of buildings.

%
%
%

\section{Analytic computation of FoM}
Next we provide analytic equations for intended and interference signals starting with the open space scenario ($I_{\mathrm{O}}$ and $P_{\mathrm{O}}$) and later for the LOS and NLOS scenarios too (i.e., $I_{\mathrm{L}}$, $P_{\mathrm{L}}$, $I_{\mathrm{N}}$, and $P_{\mathrm{N}}$, respectively).  

\begin{theorem}
\label{osPItheorem}
In the open space scenario, the power of interference signals $I_{\mathrm{O}}$ and the power of intended signals $P_{\mathrm{O}}$ are computed by (\ref{Iosfinal}) and (\ref{PRosfinal}), respectively.

\begin{eqnarray}
&&\hspace{-0.2in}
\label{Iosfinal}
I_{\mathrm{O}}=
\left\{
\begin{array}{l l}
\frac{P_{\mathrm{T}}\lambda^2}{8\pi}\left\{
\frac{1}{2}
+  \ln\left(\frac{16\sqrt{P_{\mathrm{th}}}\pi^2 h_{\mathrm{T}}h_{\mathrm{R}}}{\sqrt{P_{\mathrm{T}}}\lambda^2}\right)
\right\}, 
& \frac{P_{\mathrm{T}}}{P_{\mathrm{th}}}<\frac{(4\pi)^4\left({h_{\mathrm{T}}h_{\mathrm{R}}}\right)^2}{\lambda^4},  \\
\pi\sqrt{P_{\mathrm{th}}P_{\mathrm{T}}} h_{\mathrm{T}}h_{\mathrm{R}}, 
& \frac{P_{\mathrm{T}}}{P_{\mathrm{th}}}\geq\frac{(4\pi)^4\left({h_{\mathrm{T}}h_{\mathrm{R}}}\right)^2}{\lambda^4}.
\end{array}
\right.
\\&&\hspace{-0.2in}
\label{PRosfinal}
P_{\mathrm{O}}=
\left\{
\begin{array}{l l}
\frac{P_{\mathrm{T}}\lambda^2}{16\pi}
\left[1+ \ln \left(\frac{P_{\mathrm{T}}}{P_{\mathrm{th}}}\right)\right], 
& \frac{P_{\mathrm{T}}}{P_{\mathrm{th}}}<\frac{(4\pi)^4\left({h_{\mathrm{T}}h_{\mathrm{R}}}\right)^2}{\lambda^4}, \\
\frac{P_{\mathrm{T}}\lambda^2}{16\pi}+
\frac{P_{\mathrm{T}}\lambda^2}{8\pi}\ln\left(\frac{16\pi^2 h_{\mathrm{T}}h_{\mathrm{R}}}{\lambda^2}\right)
+ \frac{\lambda^4{P_{\mathrm{T}}}}{2^9\pi^3(h_{\mathrm{T}}h_{\mathrm{R}})^{2}}-\frac{\pi P_{\mathrm{th}}}{2}, 
& \frac{P_{\mathrm{T}}}{P_{\mathrm{th}}}\geq\frac{(4\pi)^4\left({h_{\mathrm{T}}h_{\mathrm{R}}}\right)^2}{\lambda^4}.
\end{array}
\right.
\end{eqnarray}
\end{theorem}
\begin{IEEEproof}
See Appendix A.
\end{IEEEproof}
The parameters of the equations have been defined above in Section \ref{Sec:II} and are omitted here for the sake of compactness.

\begin{figure}[!t]
\begin{minipage}[t]{0.5\textwidth}
  \centering
  \includegraphics [width=2.5in]{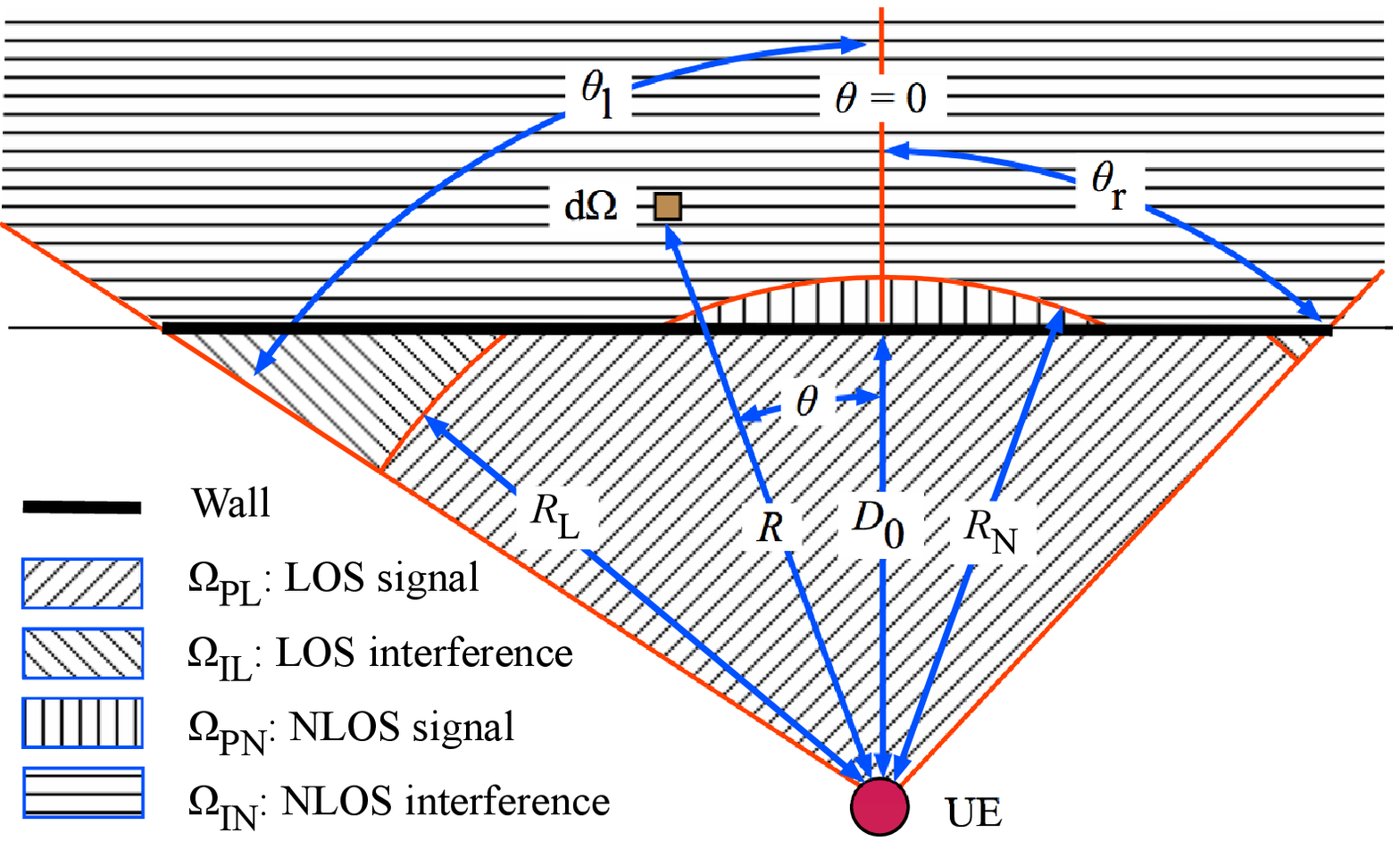}
\caption{An example of the toy model and its $\Omega_{\mathrm{PL}}$, $\Omega_{\mathrm{PN}}$, $\Omega_{\mathrm{IL}}$, and $\Omega_{\mathrm{IN}}$.} \label{toymodel}
\end{minipage}
\begin{minipage}[t]{0.44\textwidth}
  \centering
  \includegraphics [width=2.5in]{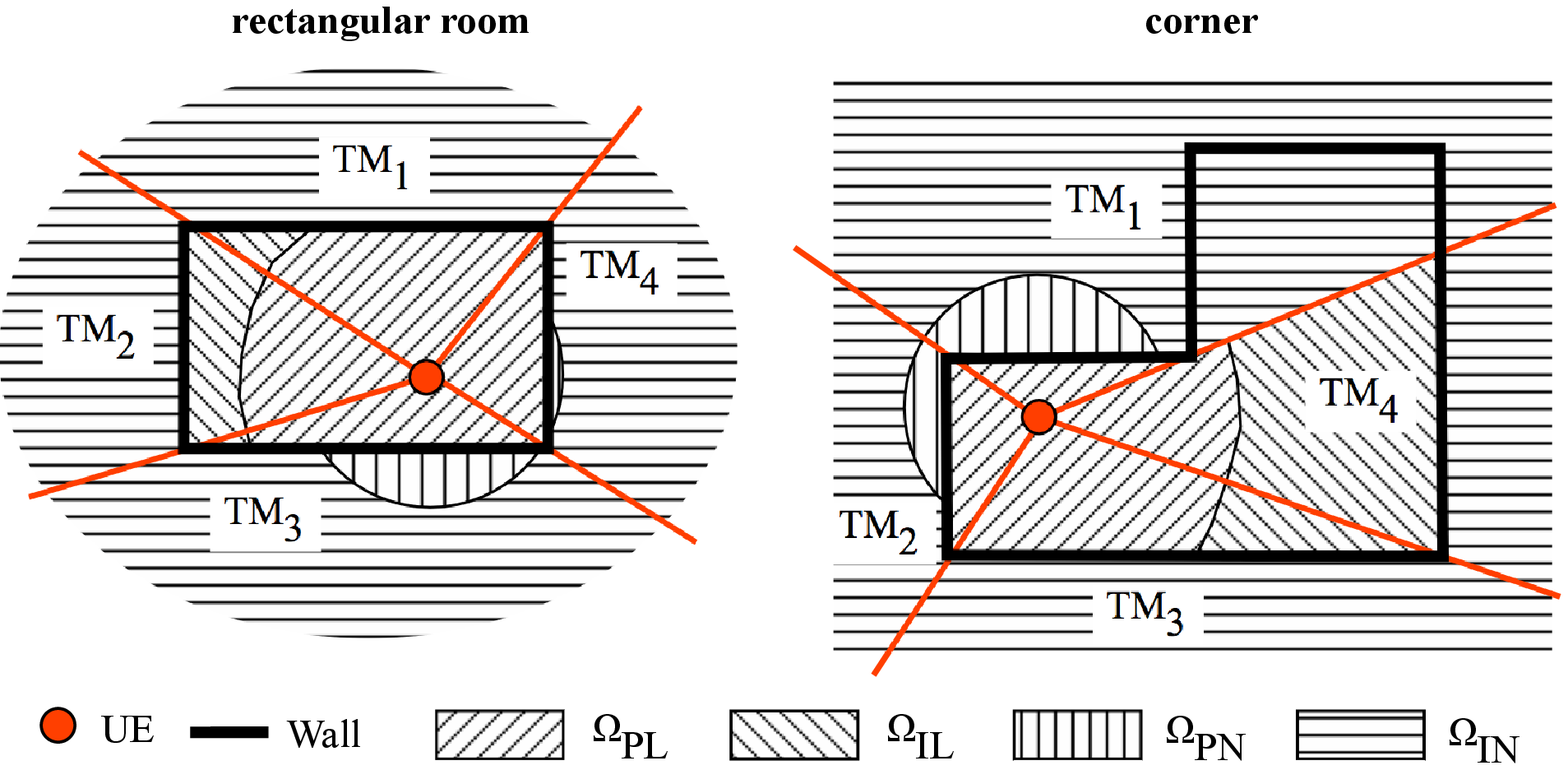}
\caption{Examples of dividing a room into toy models.} \label{examples}
\end{minipage}
\end{figure}

The computation of $I_{\mathrm{L}}$, $P_{\mathrm{L}}$, $I_{\mathrm{N}}$, and $P_{\mathrm{N}}$ is not straightforward since the indoor scenarios are not homogeneous.
Therefore, in order to facilitate their derivation, we propose a toy model (TM) as shown in Fig. \ref{toymodel}. In the TM, we consider a circular sector with an infinite radius. A reference (or probing) UE is located at the vertex of the sector. A wall of finite length is modeled as a straight line segment across the sector with endpoints lying on the radii of the circular sector. The distance from the UE to the wall is denoted as $D_0$.
A polar-coordinate system $(R, \theta)$ is introduced by choosing the UE as the reference point. The perpendicular direction to the wall is assumed to be the reference $0$-angle. $\theta_{\mathrm{l}}$ and $\theta_{\mathrm{r}}$ denote angles of the left and right end points of the wall, respectively. With these definitions,  $-\frac{\pi}{2}<\theta_{\mathrm{l}}<\theta_{\mathrm{r}}<\frac{\pi}{2}$.

It is worthwhile to note that the proposed TM of a wall can be used to represent any polygonal-shaped room. Indeed, any room can be modelled by adding several TMs, i.e., one for each wall that can be seen by the reference UE in the room. Some examples are given in Fig.~\ref{examples}. For a specific direction, whether an area is LOS or NLOS is determined by the closest wall.

For a room consisting of $N_{\mathrm{TM}}$ TMs, we then have
\begin{eqnarray}
&&\label{tmp1}
I_{\mathrm{L}}=\sum\limits_{n_{\mathrm{TM}}=1}^{N_{\mathrm{TM}}}I_{\mathrm{L},n_{\mathrm{TM}}},
\\
&&\label{tmp2}
P_{\mathrm{L}}=\sum\limits_{n_{\mathrm{TM}}=1}^{N_{\mathrm{TM}}}P_{\mathrm{L},n_{\mathrm{TM}}},
\\
&&\label{tmp3}
I_{\mathrm{N}}=\sum\limits_{n_{\mathrm{TM}}=1}^{N_{\mathrm{TM}}}I_{\mathrm{N},n_{\mathrm{TM}}},
\\
&&\label{tmp4}
P_{\mathrm{N}}=\sum\limits_{n_{\mathrm{TM}}=1}^{N_{\mathrm{TM}}}P_{\mathrm{N},n_{\mathrm{TM}}}.
\end{eqnarray}

For known closed-form expressions of $I_{\mathrm{L},n_{\mathrm{TM}}}$, $P_{\mathrm{L},n_{\mathrm{TM}}}$, $I_{\mathrm{N},n_{\mathrm{TM}}}$, and $P_{\mathrm{N},n_{\mathrm{TM}}}$, it is straightforward to compute $I_{\mathrm{L}}$, $P_{\mathrm{L}}$, $I_{\mathrm{N}}$, and $P_{\mathrm{N}}$ by (\ref{tmp1}-\ref{tmp4}).
For an arbitrary $n_{\mathrm{TM}}$ (i.e., TM of a wall contribution to the computed signals) we have
\begin{eqnarray}
&&\label{PLdefin}
P_{\mathrm{L},n_{\mathrm{TM}}}
=\int_{\Omega_{\mathrm{PL}}}
{P_{\mathrm{T}}}{G_{\mathrm{L}}(R)}\mathrm{d}\Omega,
\\
&&\label{ILdefin}
I_{\mathrm{L},n_{\mathrm{TM}}}
=\int_{\Omega_{\mathrm{IL}}}
{P_{\mathrm{T}}}{G_{\mathrm{L}}(R)}\mathrm{d}\Omega,
\\
&&\label{PNdefin}
P_{\mathrm{N},n_{\mathrm{TM}}}
=\int_{\Omega_{\mathrm{PN}}}
{P_{\mathrm{T}}}{G_{\mathrm{N}}(R)}\mathrm{d}\Omega,
\\
&&\label{INdefin}
I_{\mathrm{N},n_{\mathrm{TM}}}
=\int_{\Omega_{\mathrm{IN}}}
{P_{\mathrm{T}}}{G_{\mathrm{N}}(R)}\mathrm{d}\Omega,
\end{eqnarray}
where
$\Omega_{\mathrm{PL}}$, $\Omega_{\mathrm{IL}}$, $\Omega_{\mathrm{PN}}$, and $\Omega_{\mathrm{IN}}$ denote the areas contributing to the LOS intended signals, the  LOS interference signals, the NLOS intended signals, and the NLOS interference signals at the reference UE, respectively.
Expressed in polar-coordinates, the straight line equation of the wall is given by
$R=\frac{D_0}{\cos{\theta}}$.
For any infinitesimal area at position  $(R, \theta)$, the area is a LOS area if
$R<\frac{D_0}{\cos{\theta}}.$
Else, the area is NLOS if
$R\geq\frac{D_0}{\cos{\theta}}.
$
Therefore, $\Omega_{\mathrm{IL}}$, $\Omega_{\mathrm{PL}}$, $\Omega_{\mathrm{IN}}$, and $\Omega_{\mathrm{PN}}$ are, respectively, defined as
\begin{eqnarray}
&&
\label{omegaPL}
\Omega_{\mathrm{PL}}
\triangleq\left\{\left(R,\theta\right)\left|
\left(R<\min\left\{R_{\mathrm{L}},\frac{D_0}{\cos{\theta}}\right\}\right)\land\left(-\frac{\pi}{2}<\theta_\mathrm{l}<\theta<\theta_\mathrm{r}<\frac{\pi}{2}\right)
\right.\right\},
\\
&&
\label{omegaIL}
\Omega_{\mathrm{IL}}
\triangleq\left\{\left(R,\theta\right)\left|
\left(R_{\mathrm{L}}\leq R< \frac{D_0}{\cos{\theta}}\right)\land\left(-\frac{\pi}{2}<\theta_\mathrm{l}<\theta<\theta_\mathrm{r}<\frac{\pi}{2}\right)
\right.\right\},
\\
&&
\label{omegaPN}
\Omega_{\mathrm{PN}}
\triangleq\left\{\left(R,\theta\right)\left|
\left(\frac{D_0}{\cos{\theta}}\leq R < R_{\mathrm{N}}\right)\land\left(-\frac{\pi}{2}<\theta_\mathrm{l}<\theta<\theta_\mathrm{r}<\frac{\pi}{2}\right)
\right.\right\},
\\
&&
\label{omegaIN}
\Omega_{\mathrm{IN}}
\triangleq\left\{\left(R,\theta\right)\left|
\left(R\geq\max\left\{R_{\mathrm{N}},\frac{D_0}{\cos{\theta}}\right\}\right)\land\left(-\frac{\pi}{2}<\theta_\mathrm{l}<\theta<\theta_\mathrm{r}<\frac{\pi}{2}\right)
\right.\right\}.
\end{eqnarray}

In a TM with a given set of ($P_{\mathrm{T}}$,  $P_{\mathrm{th}}$, $\theta_{\mathrm{r}}$, $\theta_{\mathrm{l}}$, $D_0$, $\lambda$), the integrals $P_{\mathrm{L},n_{\mathrm{TM}}}$, $I_{\mathrm{L},n_{\mathrm{TM}}}$, $P_{\mathrm{N},n_{\mathrm{TM}}}$ and $I_{\mathrm{N},n_{\mathrm{TM}}}$ are computed by following Theorems  \ref{PRLF}-\ref{INF}.

\begin{theorem}
 \label{PRLF}
For an arbitrary $n_{\mathrm{TM}}$, $P_{\mathrm{L},n_{\mathrm{TM}}}$ is computed by {\rm TABLE \ref{TABLEPRL}} in closed form, where
$P_{\mathrm{L},1}$, $P_{\mathrm{L},2}$, and $P_{\mathrm{L},3}$ are given by the following expressions
\begin{eqnarray}
&&
\label{PRLC1}
P_{\mathrm{L},1}=P_T\left(\frac{\lambda}{4\pi}\right)^2\left(\theta_{\mathrm{r}}-\theta_{\mathrm{l}}\right)\left[\frac{1}{2}-\ln\left(\frac{\lambda}{4\pi}\right)\right]+P_T\left(\frac{\lambda}{4\pi}\right)^2Z_0(\theta_{\mathrm{l}},\theta_{\mathrm{r}},1,R_{\mathrm{L}},n_{\mathrm{L}}-1),
\\&&
\label{PRLC2}
P_{\mathrm{L},2}=
P_T\left(\frac{\lambda}{4\pi}\right)^2Z_1\left(\max\{\theta_{\mathrm{l}},-\theta_{\mathrm{L1}}\},\min\{\theta_{\mathrm{r}},\theta_{\mathrm{L1}}\},R_{\mathrm{L}},D_0,n_{\mathrm{L}}-1\right),
\\
&&
\label{PRLC3}
P_{\mathrm{L},3}=
P_T\left(\frac{\lambda}{4\pi}\right)^2Z_1\left(\max\{\theta_{\mathrm{l}},-\theta_{\mathrm{L2}}\},\min\{\theta_{\mathrm{r}},\theta_{\mathrm{L2}}\},1,D_0,1\right)\nonumber\\
&&\hspace{0.45in}-P_T\left(\frac{\lambda}{4\pi}\right)^2Z_1\left(\max\{\theta_{\mathrm{l}},-\theta_{\mathrm{L2}}\},\min\{\theta_{\mathrm{r}},\theta_{\mathrm{L2}}\},1,D_0,n_{\mathrm{L}}-1\right),
\end{eqnarray}
\begin{eqnarray}
&&
\label{thata_L1}
\theta_{\mathrm{L1}}=\arccos\left(\frac{D_0}{R_{\mathrm{L}}}\right),
\\&&
\label{thata_L2}
\theta_{\mathrm{L2}}=\arccos\left(D_0\right),
\end{eqnarray}
\begin{eqnarray}
&&
\label{z0final}
Z_0(z_1,z_2,z_3,z_4,z_5)=\left\{
\begin{array}{ll}
(z_2-z_1) \ln\left(\frac{z_4}{z_3}\right),&z_5=1,\\
\frac{(z_2-z_1)(z_4^{1-z_5}-z_3^{1-z_5})}{1-z_5},&z_5\neq 1,
\end{array}
\right.
\\&&
\label{z1final}
Z_1(z_1,z_2,z_3,z_4,z_5)=\left\{
\begin{array}{ll}
\frac{z_4^2[\tan(z_2)-\tan(z_1)]}{2}+\frac{(z_1-z_2)z_3^2}{2}, &z_5=-1,\\ 
z_4\ln\left(\frac{\tan(z_2) + \sec(z_2) }{\tan(z_1) + \sec(z_1) }\right)+(z_1-z_2)z_3, &z_5=0,\\ 
(z_2-z_1)\ln \left(\frac{2z_4}{z_3}\right)+\frac{\mathrm{Im}\left[\mathrm{Li}_2(-e^{2jz_2})-\mathrm{Li}_2(-e^{2jz_1})\right]}{2}, &z_5=1,\\ 
\left\{
\begin{array}{l}
\frac{B\left(\frac{1}{2},\frac{z_5}{2}\right) \left[\mathrm{sgn}(z_2)-\mathrm{sgn}(z_1)\right] z_4^{1-z_5}}{2(1-z_5)}
\\+\frac{(z_1-z_2)z_3^{1-z_5}
}{1-z_5}  +
\frac{z_4^{1-z_5}}{(1-z_5)z_5}\times
\\
\left[
 \mathrm{sgn}(z_1)\cos^{z_5}(z_1)\ _2F_1\left(\frac{z_5}{2},\frac{1}{2}; \frac{z_5+2}{2}, \cos^2(z_1)\right)
\right.
 \\
 \left.-\mathrm{sgn}(z_2)\cos^{z_5}(z_2)\ _2F_1\left(\frac{z_5}{2},\frac{1}{2}; \frac{z_5+2}{2}, \cos^2(z_2)\right)
\right],
\end{array}
\right. &\mathrm{else},
\end{array}
\right.
\end{eqnarray}
where $-\frac{\pi}{2}<z_1<z_2<\frac{\pi}{2}$, $z_3>0$, $z_4>0$, and $z_5>-1$. 
\end{theorem}
\begin{IEEEproof}
See Appendix B.
\end{IEEEproof}

\begin{table}[!t]
\renewcommand{\arraystretch}{1.3}
\caption{Computation of $P_{\mathrm{L}, n_{\mathrm{TM}}}$, where $P_{\mathrm{L},1}$, $P_{\mathrm{L},2}$, and $P_{\mathrm{L},3}$ are respectively computed by (\ref{PRLC1}), (\ref{PRLC2}), and (\ref{PRLC3}).}
\label{TABLEPRL}
\centering
\begin{tabular}{c|c|c|c}
\hline
Case&$D_0$&$\theta$& $P_{\mathrm{L}, n_{\mathrm{TM}}}$\\
\hline
1&$1\leq D_0<R_{\mathrm{L}}$
&$(\theta_{\mathrm{l}}<\theta_{\mathrm{L1}})\land(\theta_{\mathrm{r}}>-\theta_{\mathrm{L1}})$
&$P_{\mathrm{L},1}+P_{\mathrm{L},2}$\\
\hline
2&$\frac{\lambda}{4\pi}<D_0<1$
&$(\theta_{\mathrm{l}}<\theta_{\mathrm{L2}})\land(\theta_{\mathrm{r}}>-\theta_{\mathrm{L2}})$
&$P_{\mathrm{L},1}+P_{\mathrm{L},2}+P_{\mathrm{L},3}$\\
\hline
3&$\frac{\lambda}{4\pi}<D_0<1$
&$(\theta_{\mathrm{l}}>\theta_{\mathrm{L2}})\lor(\theta_{\mathrm{r}}<-\theta_{\mathrm{L2}})\land\left[(\theta_{\mathrm{l}}<\theta_{\mathrm{L1}})\land(\theta_{\mathrm{r}}>-\theta_{\mathrm{L1}})\right]$
&$P_{\mathrm{L},1}+P_{\mathrm{L},2}$\\
\hline
4& \multicolumn{2}{c|}{else}& $P_{\mathrm{L},1}$\\
\hline
\end{tabular}
\end{table}

\begin{theorem}
 \label{ILF}
$I_{\mathrm{L},n_{\mathrm{TM}}}$ is computed by {\rm TABLE \ref{TABLEIL}} in closed form, where
\begin{eqnarray}
&&
\label{ILC1}
I_{\mathrm{L},1}=P_T\left(\frac{\lambda}{4\pi}\right)^2Z_1\left(\theta_{\mathrm{l}},\theta_{\mathrm{r}},R_{\mathrm{L}},D_0,n_{\mathrm{L}}-1\right),
\\&&
\label{ILC2}
I_{\mathrm{L},2}=
P_T\left(\frac{\lambda}{4\pi}\right)^2Z_1\left(\theta_{\mathrm{l}},-\theta_{\mathrm{L1}},R_{\mathrm{L}},D_0,n_{\mathrm{L}}-1\right),
\end{eqnarray}
\begin{eqnarray}
\label{ILC3}
I_{\mathrm{L},3}=
P_T\left(\frac{\lambda}{4\pi}\right)^2Z_1\left(\theta_{\mathrm{L1}},\theta_{\mathrm{r}},R_{\mathrm{L}},D_0,n_{\mathrm{L}}-1\right),
\end{eqnarray}
where $\theta_{\mathrm{L1}}$ and $Z_1$ are computed by (\ref{thata_L1}) and (\ref{z1final}),  respectively.
\end{theorem}
\begin{IEEEproof}
See Appendix C.
\end{IEEEproof}

\begin{table}[!t]
\renewcommand{\arraystretch}{1.3}
\caption{Computation of $I_{\mathrm{L}, n_{\mathrm{TM}}}$, where $I_{\mathrm{L},1}$, $I_{\mathrm{L},2}$, and $I_{\mathrm{L},3}$ are respectively computed by (\ref{ILC1}), (\ref{ILC2}), and (\ref{ILC3}).}
\label{TABLEIL}
\centering
\begin{tabular}{c|c|c|c}
\hline
Case&$D_0$&$\theta$& $I_{\mathrm{L}, n_{\mathrm{TM}}}$\\
\hline
1&$R_{\mathrm{L}}\leq D_0$
&$\theta_{\mathrm{l}}<\theta_{\mathrm{r}}$
&$I_{\mathrm{L},1}$
\\
\hline
2&$R_{\mathrm{L}}> D_0$
&
$\left(\theta_{\mathrm{r}}<-\theta_{\mathrm{L1}}\right)\lor\left(\theta_{\mathrm{l}}>\theta_{\mathrm{L1}}\right)$
&$I_{\mathrm{L},1}$\\
\hline
3&$R_{\mathrm{L}}> D_0$
&$\left(\theta_{\mathrm{l}}<\theta_{\mathrm{L1}}\right)\land\left(-\theta_{\mathrm{L1}}<\theta_{\mathrm{r}}\right)$
&$
\begin{array}{l}
I_{\mathrm{L},2}\delta(\theta_{\mathrm{l}}<-\theta_{\mathrm{L1}})
\\+I_{\mathrm{L},3}\delta(\theta_{\mathrm{l}}>\theta_{\mathrm{L1}})
\end{array}
$\\
\hline
4& \multicolumn{2}{c|}{else}&0\\
\hline
\end{tabular}
\end{table}

\begin{theorem}
 \label{PRNF}
$P_{\mathrm{N},n_{\mathrm{TM}}}$ is computed by {\rm TABLE \ref{TABLEPRN}} in closed form, where $P_{\mathrm{N},1}$ and $P_{\mathrm{N},2}$ are respectively computed by 
\begin{eqnarray}
&&
\label{PRNC1}
P_{\mathrm{N},1}=-
P_\mathrm{T}\left(\frac{\lambda}{4\pi}\right)^2Z_1\left(\max\{\theta_{\mathrm{l}},-\theta_{\mathrm{N1}}\},\min\{\theta_{\mathrm{r}},\theta_{\mathrm{N1}}\},R_{\mathrm{N}},D_0,n_{\mathrm{N}}-1\right),
\\&&
\label{PRNC2}
P_{\mathrm{N},2}=-
P_\mathrm{T}\left(\frac{\lambda}{4\pi}\right)^2Z_1\left(\max\{\theta_{\mathrm{l}},-\theta_{\mathrm{N2}}\},\min\{\theta_{\mathrm{r}},\theta_{\mathrm{N2}}\},1,D_0,1\right)\nonumber\\
&&\hspace{0.45in}+P_\mathrm{T}\left(\frac{\lambda}{4\pi}\right)^2Z_1\left(\max\{\theta_{\mathrm{l}},-\theta_{\mathrm{N2}}\},\min\{\theta_{\mathrm{r}},\theta_{\mathrm{N2}}\},1,D_0,n_{\mathrm{N}}-1\right),
\\
&& 
\label{thata_N1}
\theta_{\mathrm{N1}}=\arccos\left(\frac{D_0}{R_{\mathrm{N}}}\right),
\\&&
\label{thata_N2}
\theta_{\mathrm{N2}}=\arccos\left(D_0\right),
\end{eqnarray}
where $Z_0$ and $Z_1$ are computed by (\ref{z0final}) and (\ref{z1final}), respectively.
\end{theorem}
\begin{IEEEproof}
See Appendix D.
\end{IEEEproof}

\begin{table}[!t]
\renewcommand{\arraystretch}{1.3}
\caption{Computation of $P_{\mathrm{N}, n_{\mathrm{TM}}}$, where $P_{\mathrm{N},1}$ and $P_{\mathrm{N},2}$ are computed by (\ref{PRNC1}) and (\ref{PRNC2}), respectively.}
\label{TABLEPRN}
\centering
\begin{tabular}{c|c|c|c}
\hline
Case&$D_0$&$\theta$& $P_{\mathrm{N}, n_{\mathrm{TM}}}$\\
\hline
1&$1\leq D_0<R_{\mathrm{N}}$
&$(\theta_{\mathrm{l}}<\theta_{\mathrm{N1}})\land(\theta_{\mathrm{r}}>-\theta_{\mathrm{N1}})$
&$P_{\mathrm{N},1}$\\
\hline
2&$\frac{\lambda}{4\pi}<D_0<1$
&$(\theta_{\mathrm{l}}<\theta_{\mathrm{N2}})\land(\theta_{\mathrm{r}}>-\theta_{\mathrm{N2}})$
&$P_{\mathrm{N},1}+P_{\mathrm{N},2}$\\
\hline
3&$\frac{\lambda}{4\pi}<D_0<1$
&$(\theta_{\mathrm{l}}>\theta_{\mathrm{N2}})\lor(\theta_{\mathrm{r}}<-\theta_{\mathrm{N2}})\land\left[(\theta_{\mathrm{l}}<\theta_{\mathrm{N1}})\land(\theta_{\mathrm{r}}>-\theta_{\mathrm{N1}})\right]$
&$P_{\mathrm{N},1}$\\
\hline
4& \multicolumn{2}{c|}{else}& 0\\
\hline
\end{tabular}
\end{table}

\begin{theorem}
\label{INF}
 $I_{\mathrm{N},n_{\mathrm{TM}}}$ is computed by {\rm TABLE \ref{TABLEIN}} in closed form, where
\begin{eqnarray}
&&\hspace{-0.65in}
\label{INC1}
I_{\mathrm{N},1}=-P_{\mathrm{T}}\left(\frac{\lambda}{4\pi}\right)^2Z_1\left(\theta_{\mathrm{l}},\theta_{\mathrm{r}},+\infty,D_0,n_{\mathrm{N}}-1\right),
\\&&\hspace{-0.65in}
\label{INC2}
I_{\mathrm{N},2}=
P_{\mathrm{T}}\left(\frac{\lambda}{4\pi}\right)^2
 Z_1\left(\max\{\theta_{\mathrm{l}},-\theta_{\mathrm{N1}}\},\min\{\theta_{\mathrm{r}},\theta_{\mathrm{N1}}\},R_{\mathrm{N}},D_0,n_{\mathrm{N}}-1\right),
\end{eqnarray}
where $\theta_{\mathrm{N1}}$, and $Z_1$ are computed by (\ref{thata_N1}) and (\ref{z1final}), respectively.
\end{theorem}
\begin{IEEEproof}
See Appendix  E.
\end{IEEEproof}

\begin{table}[!t]
\renewcommand{\arraystretch}{1.3}
\caption{Computation of $I_{\mathrm{N}, n_{\mathrm{TM}}}$, where $I_{\mathrm{N},1}$ and $I_{\mathrm{N},2}$ are computed by (\ref{INC1}), and (\ref{INC2}), respectively.}
\label{TABLEIN}
\centering
\begin{tabular}{c|c|c|c}
\hline
Case&$D_0$&$\theta$& $I_{\mathrm{N}, n_{\mathrm{TM}}}$\\
\hline
1&$R_{\mathrm{N}}\leq D_0$
&$\theta_{\mathrm{l}}<\theta_{\mathrm{r}}$
&$I_{\mathrm{N},1}$
\\
\hline
2&$R_{\mathrm{N}}> D_0$
&
$\left(\theta_{\mathrm{r}}<-\theta_{\mathrm{N1}}\right)\lor\left(\theta_{\mathrm{l}}>\theta_{\mathrm{N1}}\right)$
&$I_{\mathrm{N},1}$\\
\hline
3&$R_{\mathrm{N}}> D_0$
&$\left(\theta_{\mathrm{l}}<\theta_{\mathrm{N1}}\right)\land\left(-\theta_{\mathrm{N1}}<\theta_{\mathrm{r}}\right)$
&$I_{\mathrm{N},1}+I_{\mathrm{N},2}$\\
\hline
\end{tabular}
\end{table}

\begin{figure} [!t]
\centering
 \includegraphics [width=5in]{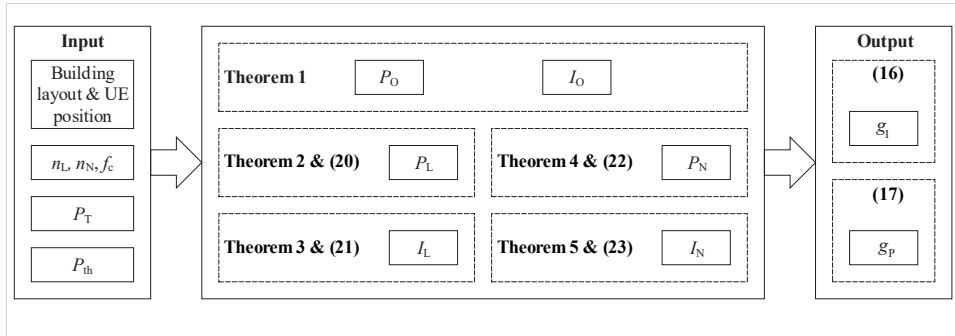}
\caption{The procedure of computing  $g_{\mathrm{I}}$ and $g_{\mathrm{P}}$.
}\label{computingproc}
\end{figure}

\begin{remark}
\label{finalremark}
The computation of the $g_\mathrm{I}$ and $g_\mathrm{P}$ follows the flow chart shown in Fig. \ref{computingproc}. The input parameters are the building layout, the reference UE position, $n_{\mathrm{L}}$, $n_\mathrm{N}$, $f_c$, $P_\mathrm{T}$ and $P_\mathrm{th}$. First, Theorems  \ref{PRLF}, \ref{ILF}, \ref{PRNF}, and \ref{INF}  are used to compute signals corresponding to the open space, LOS and NLOS scenarios. Second, the obtained (\ref{tmp1}-\ref{tmp4}) and (\ref{Iosfinal}-\ref{PRosfinal}) are substituted into (\ref{GI}-\ref{GP}) to obtain the FoM.
\end{remark}


\section{Comparison theoretical results with Monte-Carlo simulations}

In this section, we compare the results obtained by the derived closed-form expressions and the Monte Carlo simulations. 
The analytic results used in the following Figs. \ref{validationPRosIos}-\ref{cornormmwave} are summarized in TABLE \ref{anar}.
In Sections V-VI, we use $h_\mathrm T=h_\mathrm R=1.2$ m without further specification.

\begin{table}[!t]
\renewcommand{\arraystretch}{1.3}
\caption{Analytic results used in Figs. \ref{validationPRosIos}-\ref{cornormmwave}}
\label{anar}
\centering
\begin{tabular}{c|c}
\hline
{\rm Figures} & {\rm associating analytical results.}\\
\hline
{Fig. \ref{validationPRosIos}}                                             & { (\ref{Iosfinal}) and (\ref{PRosfinal})}\\
{Figs. \ref{validationtoymodel}-\ref{validationtoymodelmmwave}} & {Theorems 2-5}\\
{Figs. \ref{rect}-\ref{cornormmwave}} & {Remark 2}\\
\hline
\end{tabular}
\end{table}

In Fig. \ref{validationPRosIos}, we compare the results obtained by the derived closed-form expressions of $P_{\mathrm{O}}$ and $I_{\mathrm{O}}$, i.e., (\ref{Iosfinal}) and (\ref{PRosfinal}), with Monte Carlo simulations. 
The numerical results show an excellent agreement between the analytic, i.e., closed-form, results and the Monte Carlo simulations.

\begin{figure}[!t]
\centering
\subfigure[$f_c=1$ GHz]{\includegraphics[width=1.5in]{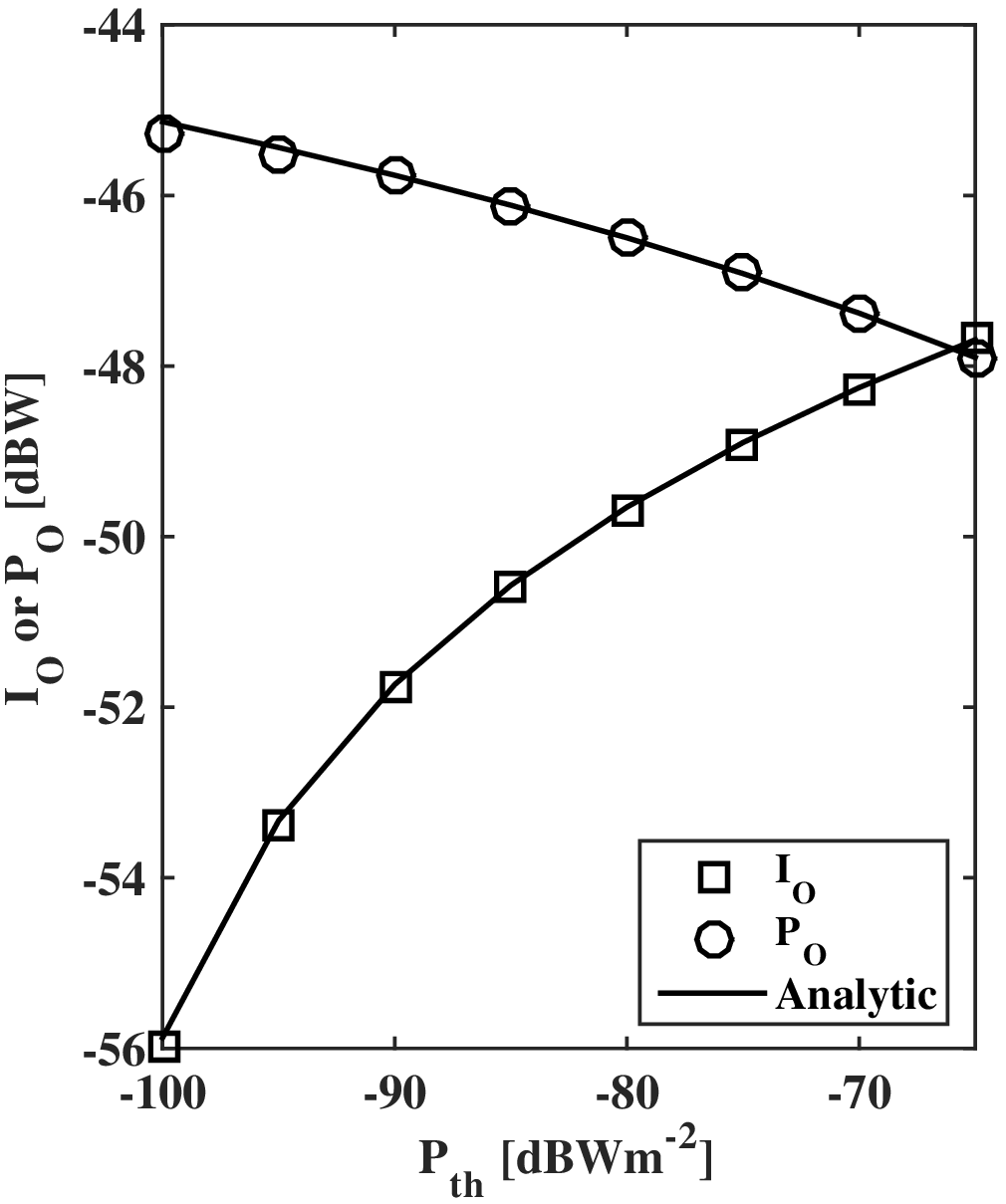}}
\subfigure[$f_c=28$ GHz]{\includegraphics[width=1.5in]{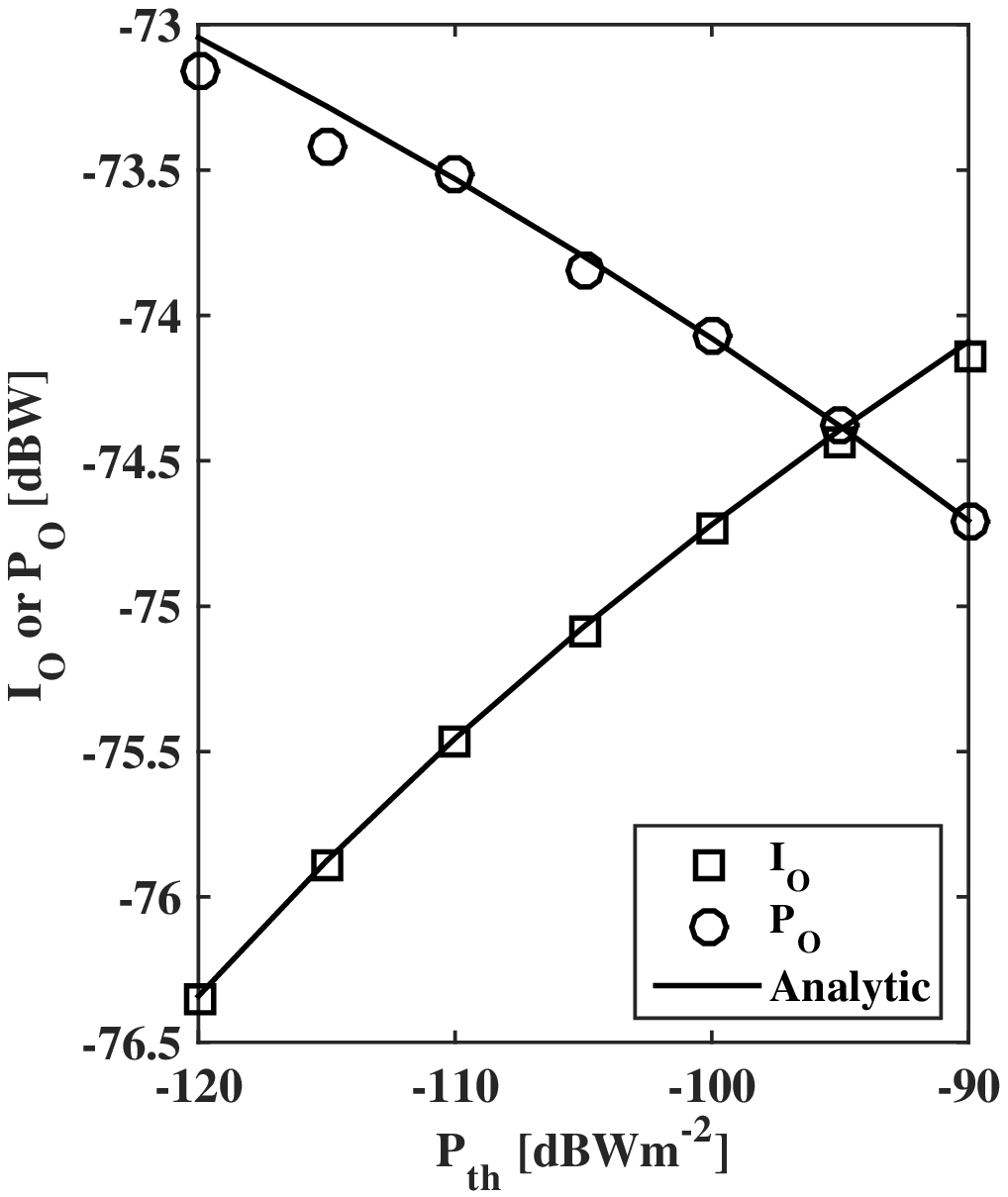}}
\caption{Comparison of analytic $P_{\mathrm{O}}$ and $I_{\mathrm{O}}$, computed by Theorem 1, with Monte-Carlo simulations. $P_{\mathrm{T}}=-30\ \mathrm{dBWm^{-2}}$. }\label{validationPRosIos}
\end{figure}
\begin{figure} [!t]
\centering
\subfigure[$\theta_{\mathrm{r}}=-0.4$]{\includegraphics[width=1.5in]{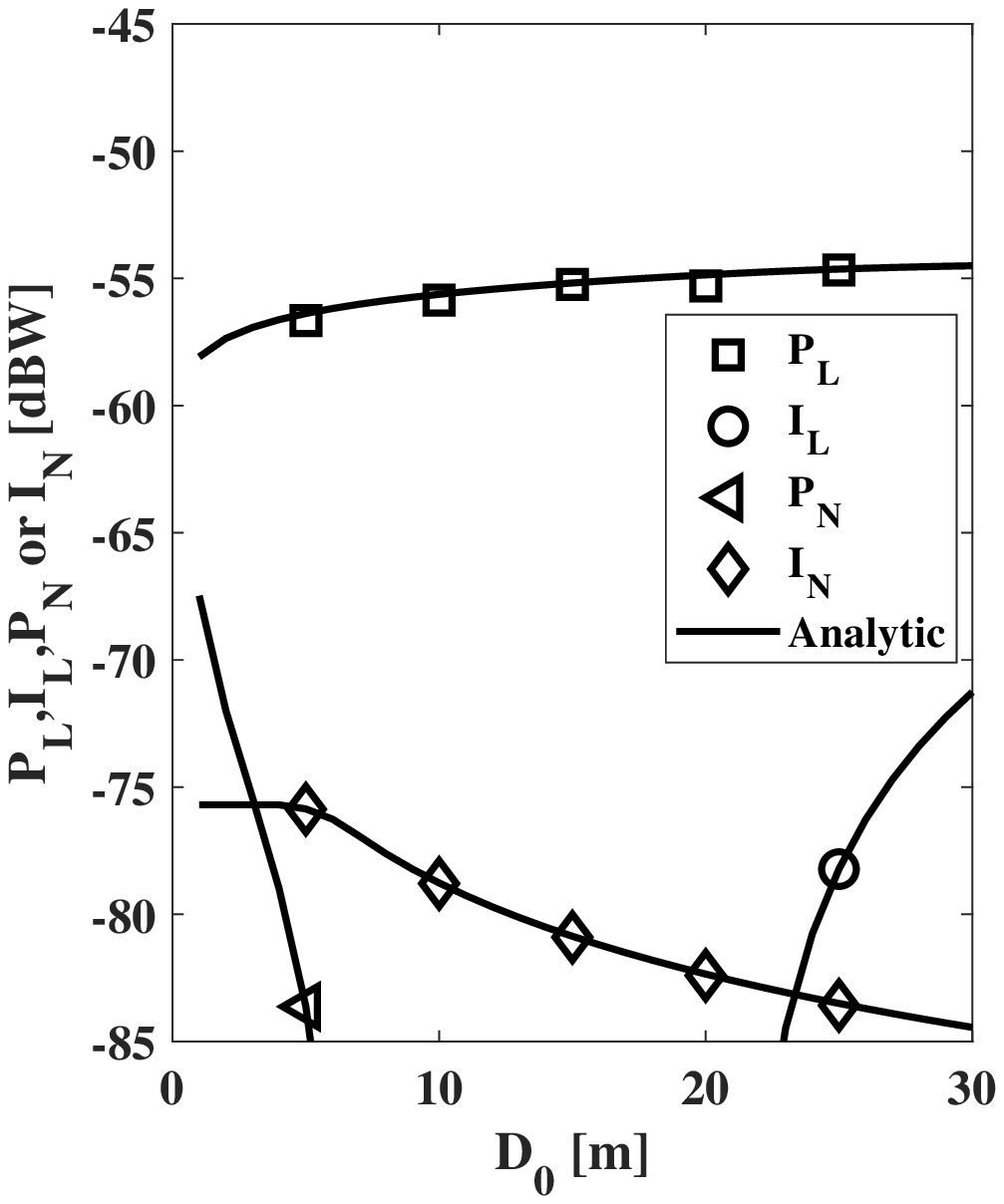}}
\subfigure[$\theta_{\mathrm{r}}=0$]{\includegraphics[width=1.5in]{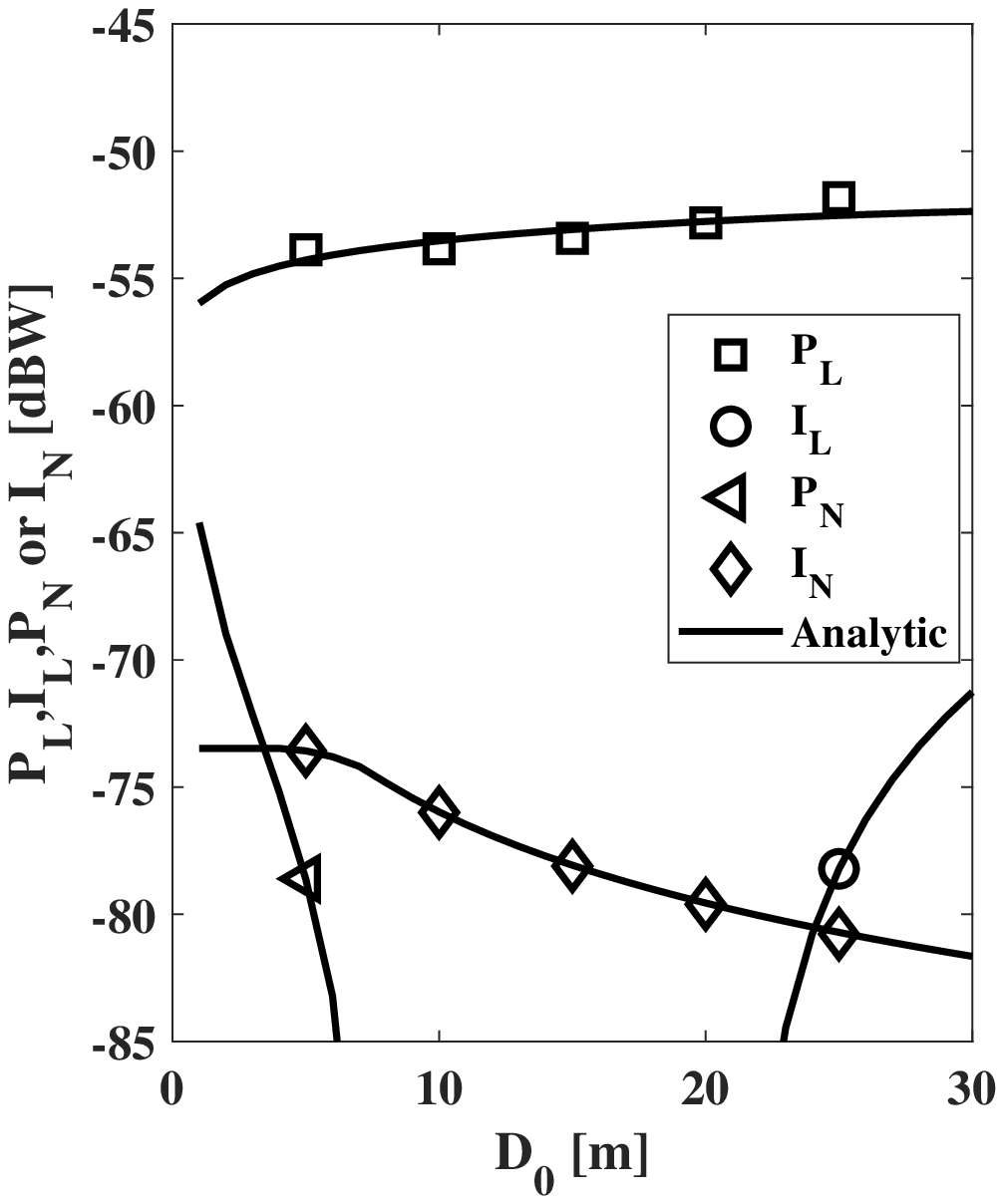}}
\subfigure[$\theta_{\mathrm{r}}=0.4$]{\includegraphics[width=1.5in]{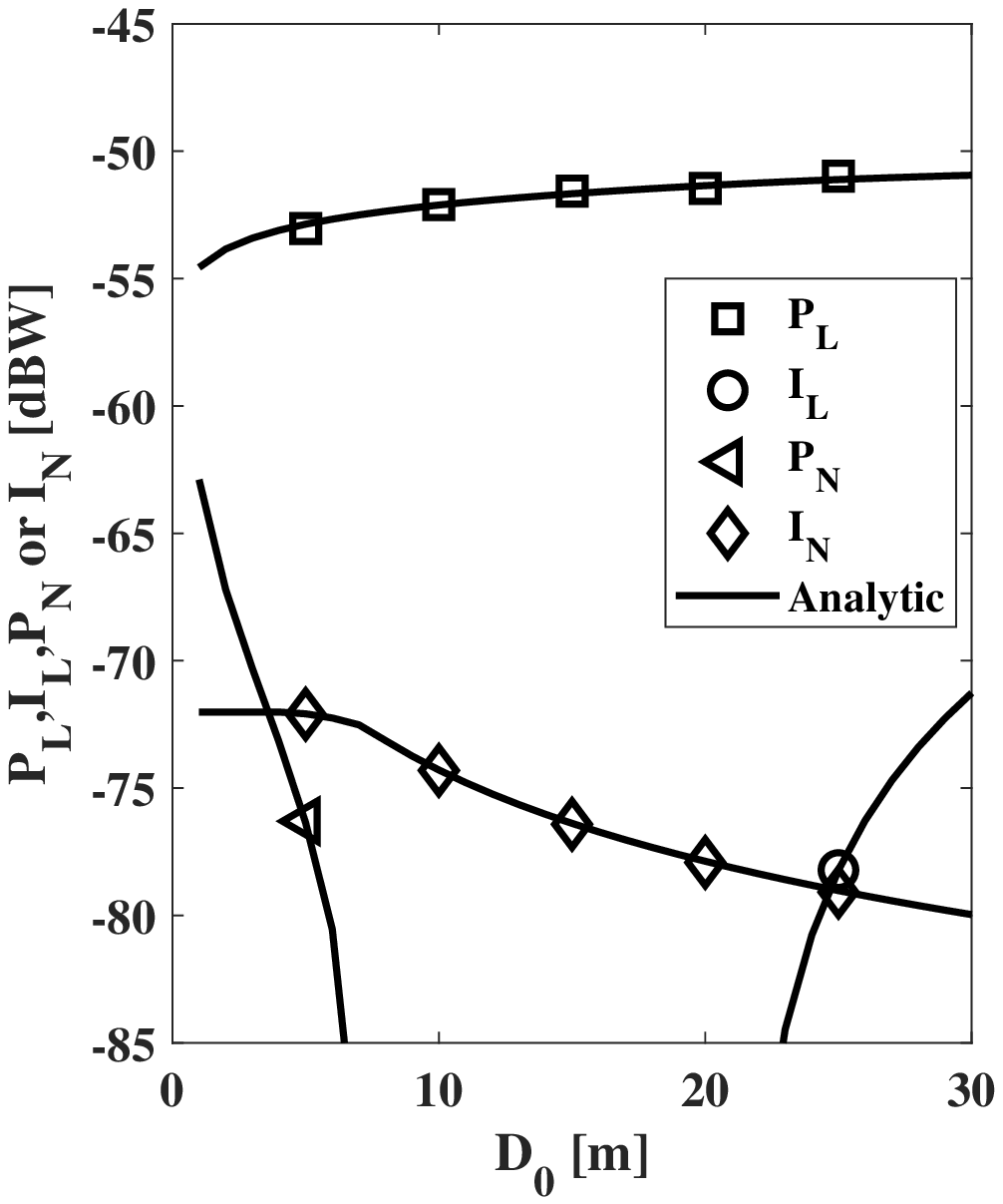}}
\subfigure[$\theta_{\mathrm{r}}=1$]{\includegraphics[width=1.5in]{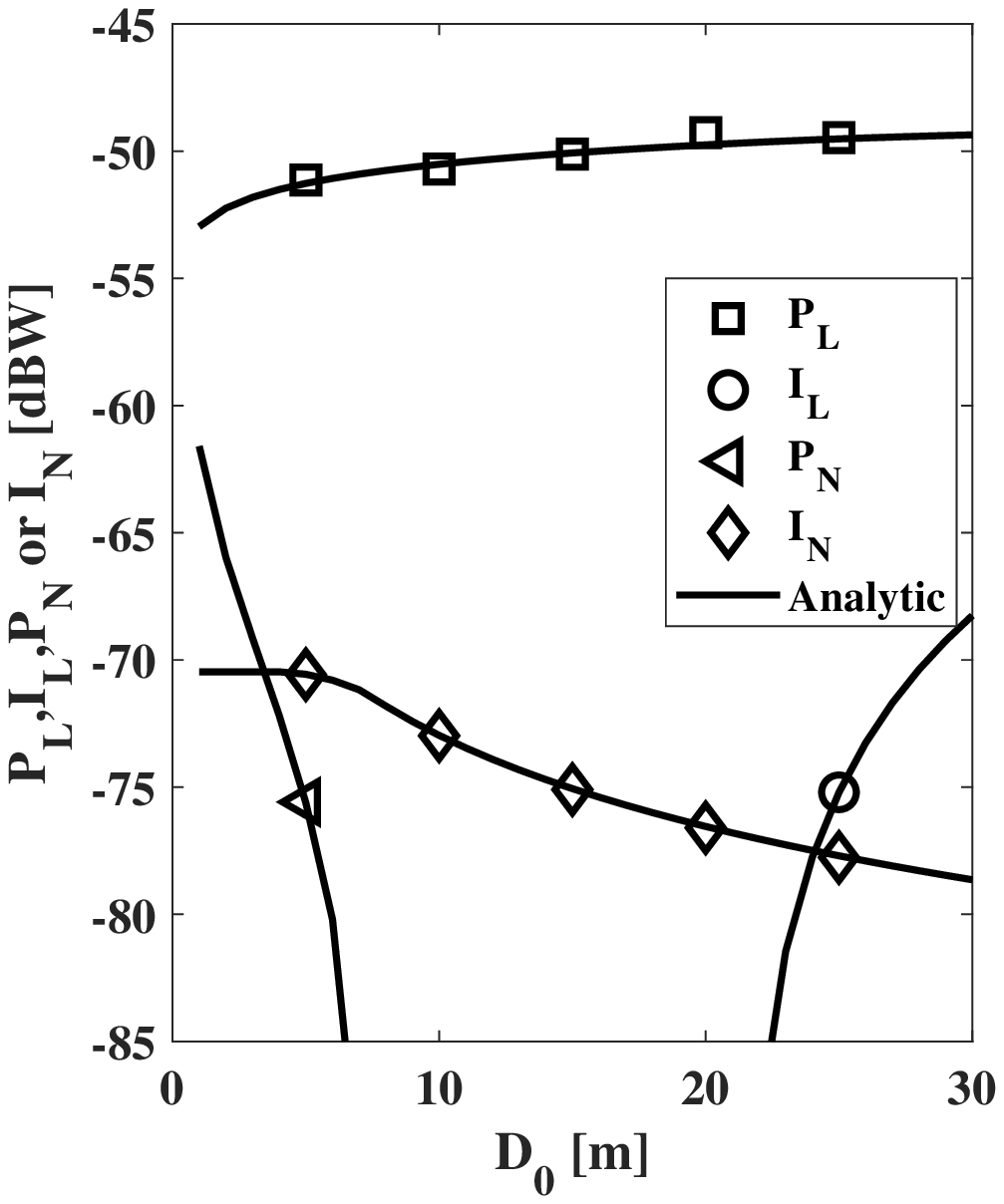}}
\caption{Comparison of analytic $P_{\mathrm{L}}$, $P_{\mathrm{N}}$, $I_{\mathrm{L}}$ and $I_{\mathrm{N}}$, computed by Theorems 2-5, with Monte-Carlo simulations. $f_c=1 \mathrm{GHz}$ $P_{\mathrm{T}}=-30$ $\mathrm{dBWm^{-2}}$, $P_{\mathrm{th}}=-75$ $\mathrm{dBWm^{-2}}$, and $\theta_{\mathrm{l}}=-1$. Markers are generated from Monte Carlo simulations. Solid lines are analytic results.
}\label{validationtoymodel}
\centering
\subfigure[$\theta_{\mathrm{r}}=-0.4$]{\includegraphics[width=1.5in]{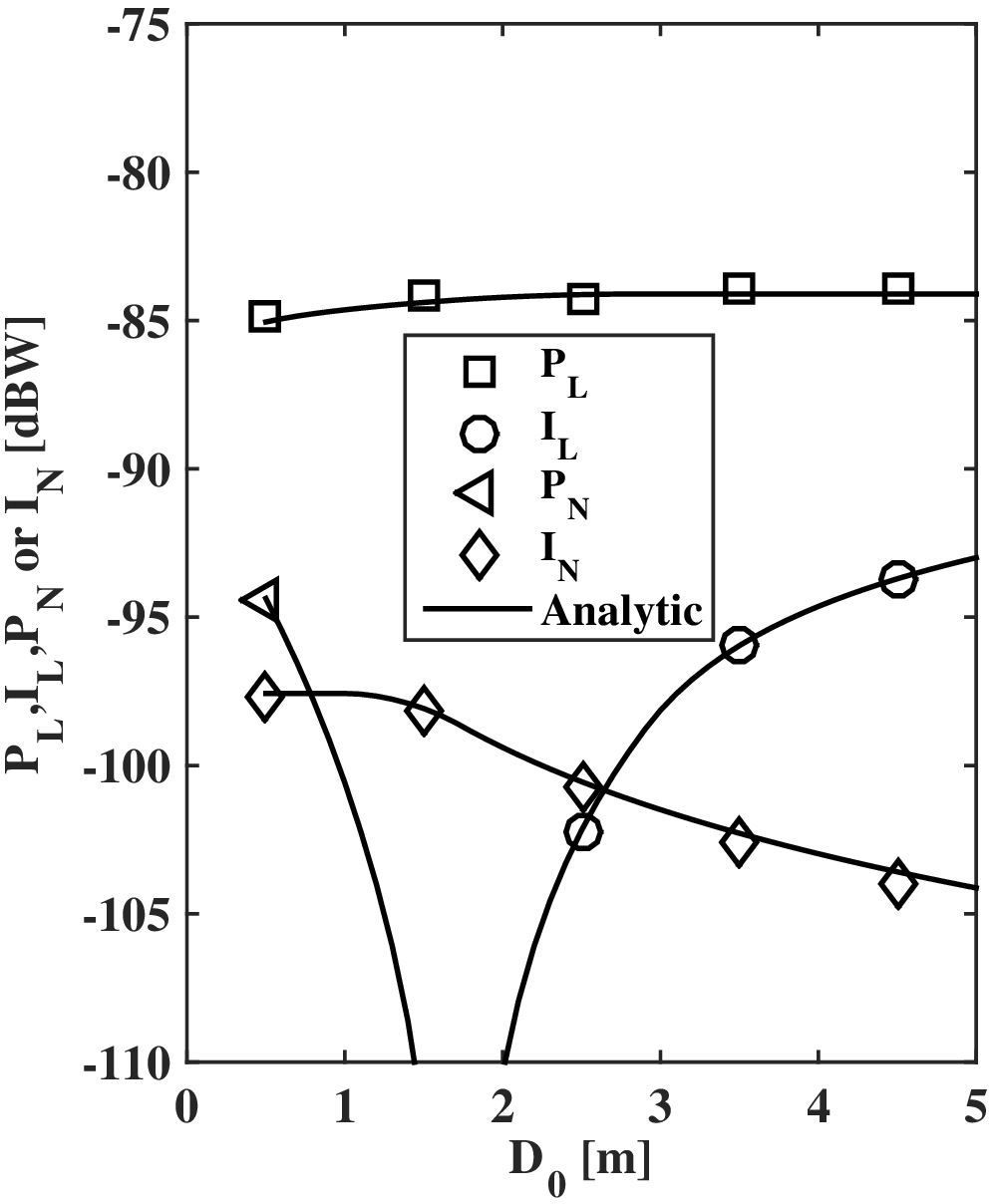}}
\subfigure[$\theta_{\mathrm{r}}=0$]{\includegraphics[width=1.5in]{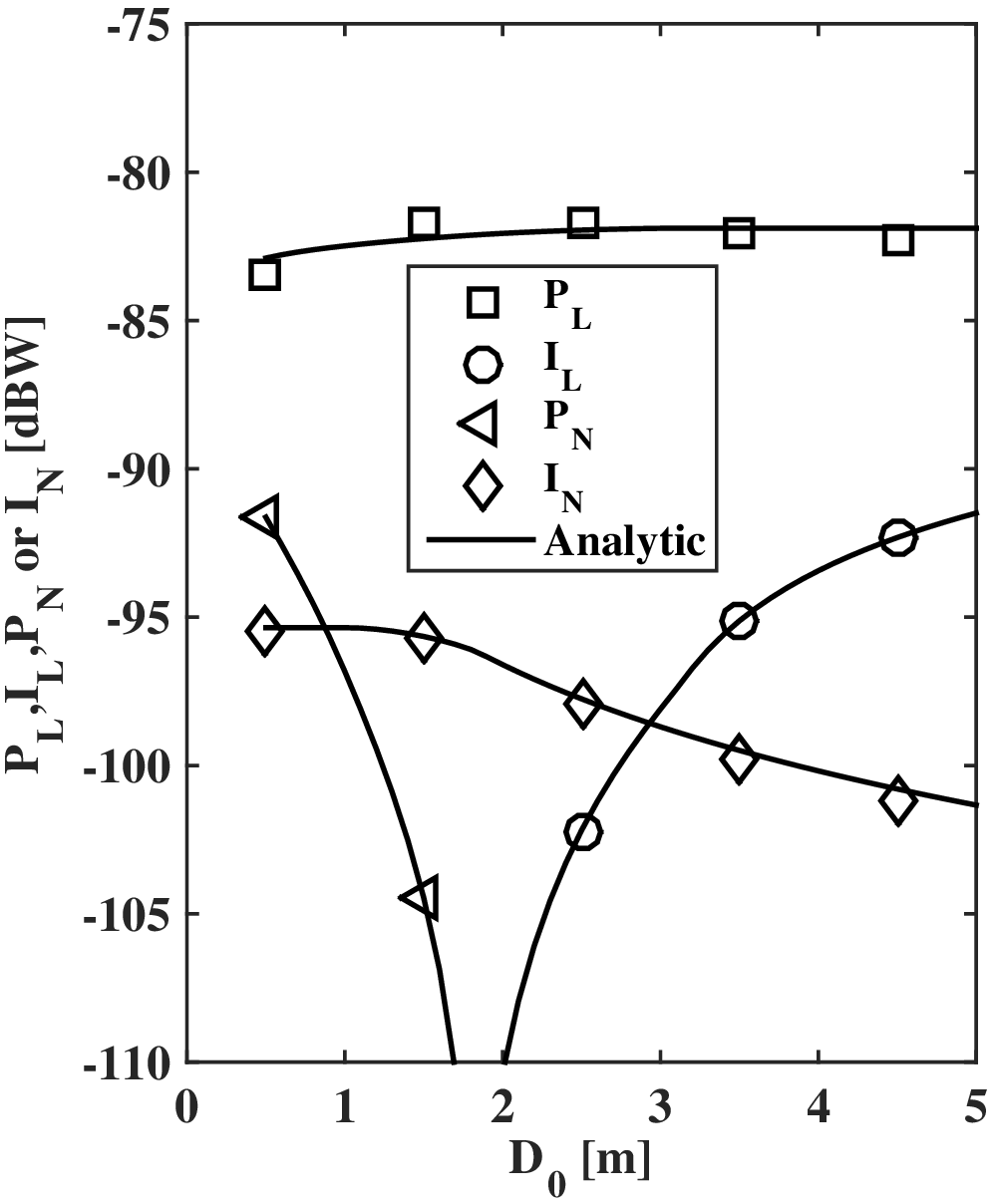}}
\subfigure[$\theta_{\mathrm{r}}=0.4$]{\includegraphics[width=1.5in]{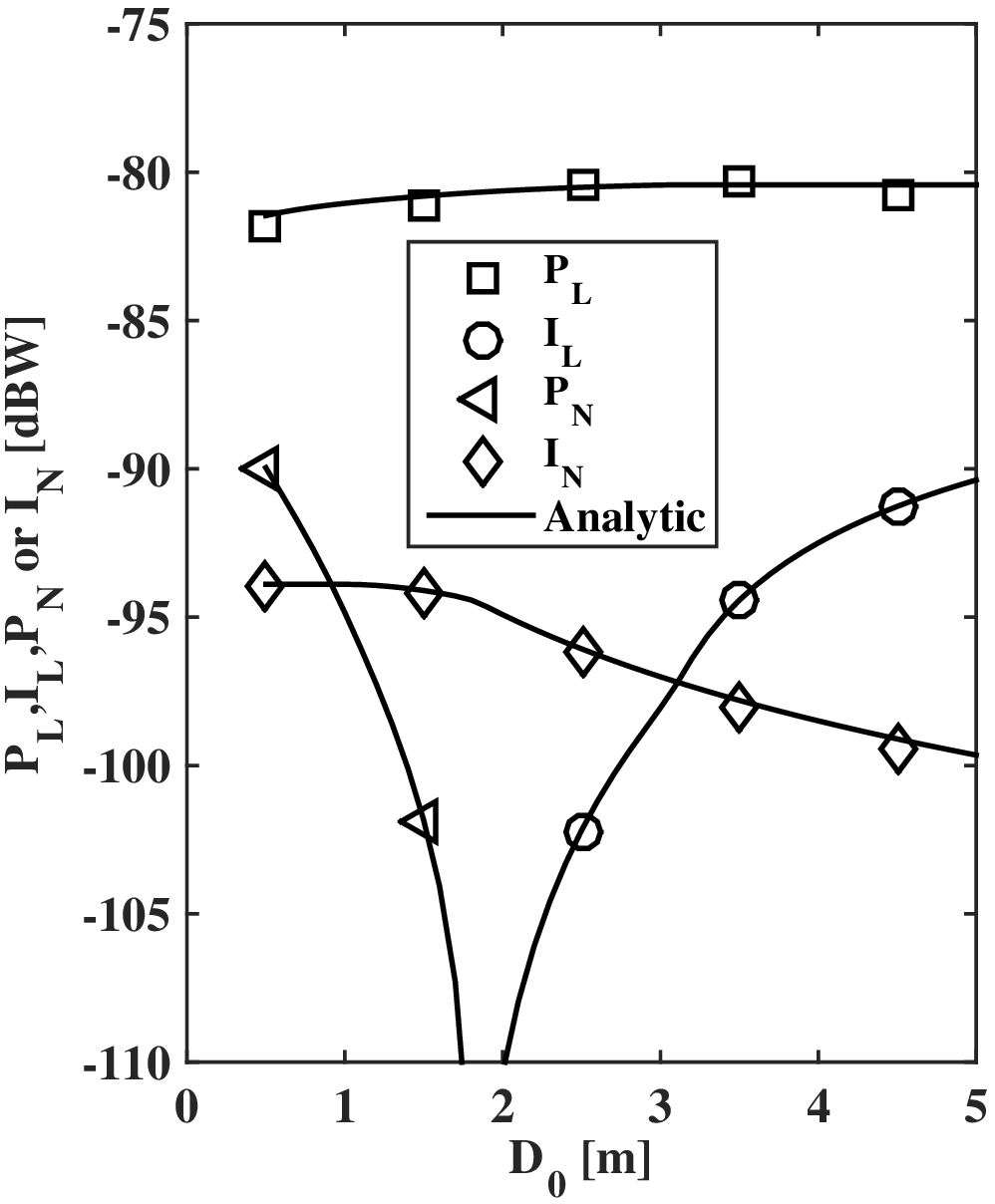}}
\subfigure[$\theta_{\mathrm{r}}=1$]{\includegraphics[width=1.5in]{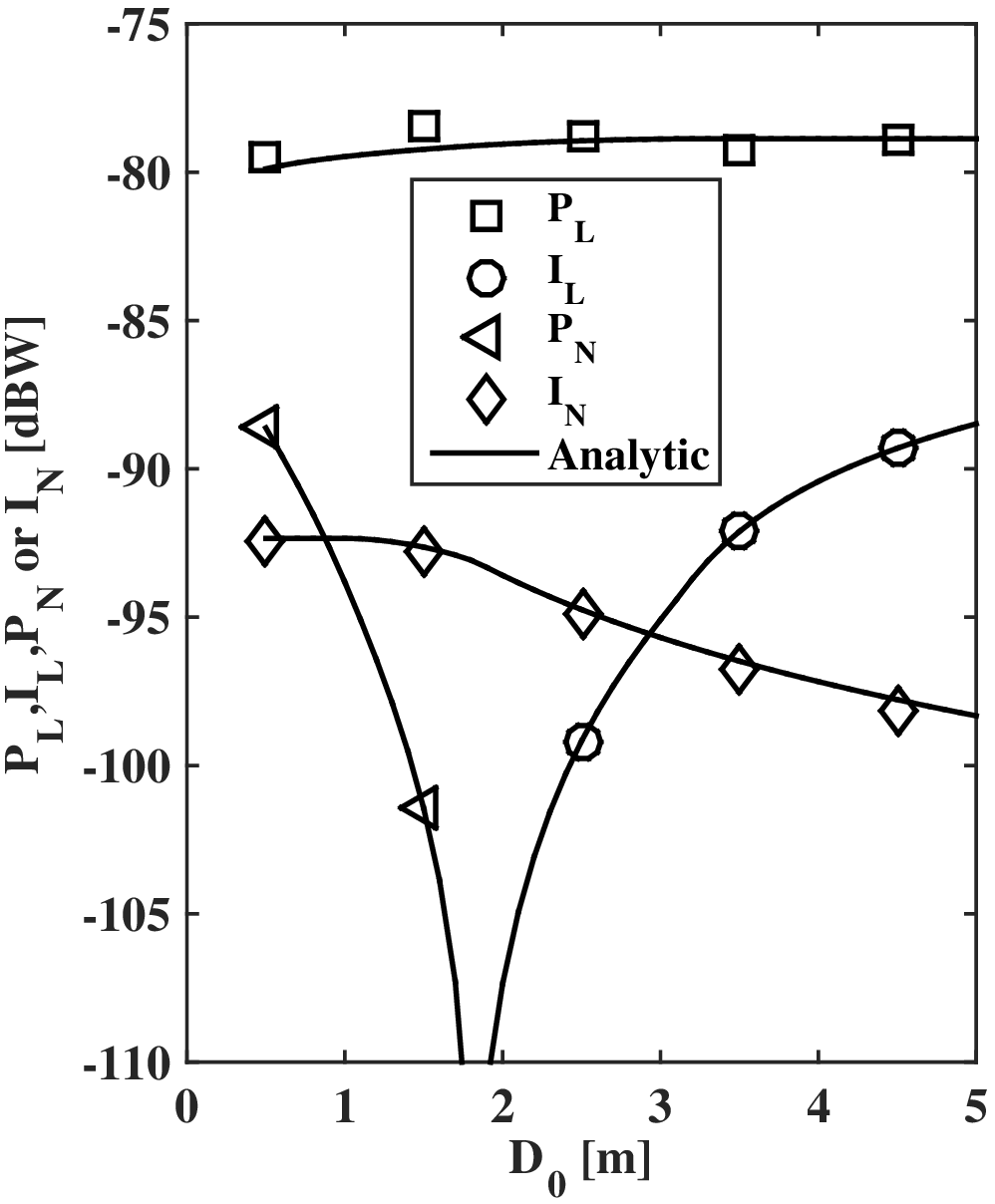}}
\caption{Comparison of analytic $P_{\mathrm{L}}$, $P_{\mathrm{N}}$, $I_{\mathrm{L}}$ and $I_{\mathrm{N}}$, computed by Theorems 2-5, with Monte-Carlo simulations. $f_c=28 \mathrm{GHz}$ $P_{\mathrm{T}}=-30$ $\mathrm{dBWm^{-2}}$, $P_{\mathrm{th}}=-100$ $\mathrm{dBWm^{-2}}$, and $\theta_{\mathrm{l}}=-1$ rad. Markers are generated from Monte Carlo simulations. Solid lines are analytic results. }\label{validationtoymodelmmwave}
\end{figure}

Let's now consider two generic wireless systems operating at the $1$ GHz and the $28$ GHz bands, respectively. The parameters $P_{\mathrm{L}}$, $I_{\mathrm{L}}$,  $P_{\mathrm{N}}$, and $I_{\mathrm{N}}$ are validated in Figs. \ref{validationtoymodel}-\ref{validationtoymodelmmwave}, respectively, where analytic $P_{\mathrm{L}}$, $I_{\mathrm{L}}$, $P_{\mathrm{N}}$, and $I_{\mathrm{N}}$ are provided in TABLEs \ref{TABLEPRL}-\ref{TABLEIN}. The analytic results match Monte Carlo simulations very well. With an increasing $D_0$, more transmitted power is considered as LOS power, and therefore   $P_{\mathrm{L}}$ and  $I_{\mathrm{L}}$ increase, while $P_{\mathrm{N}}$ and $I_{\mathrm{N}}$ decrease.

Following Remark \ref{finalremark}, the analytic $g_{\mathrm{I}}$ and $g_{\mathrm{P}}$ of a rectangular room of size $5\ \mathrm{m}\times10\ \mathrm{m}$ are validated in Figs. \ref{rect} and \ref{rectmmwave} for the 1 GHz and the 28 GHz bands,  respectively. Also, $g_{\mathrm{I}}$ and $g_{\mathrm{P}}$ of an L-shaped corner room are validated in Figs. \ref{cornor} and \ref{cornormmwave}. Floor plans of the considered rooms are given in Fig. \ref{examples}. Numerical results show that the analytic $g_{\mathrm{I}}$ and $g_{\mathrm{P}}$ agree well with the Monte Carlo simulations.

\begin{figure} [!t]
\begin{minipage}[t]{0.48\textwidth}
\subfigure[$g_{\mathrm{I}}$]{\includegraphics[width=1.5in]{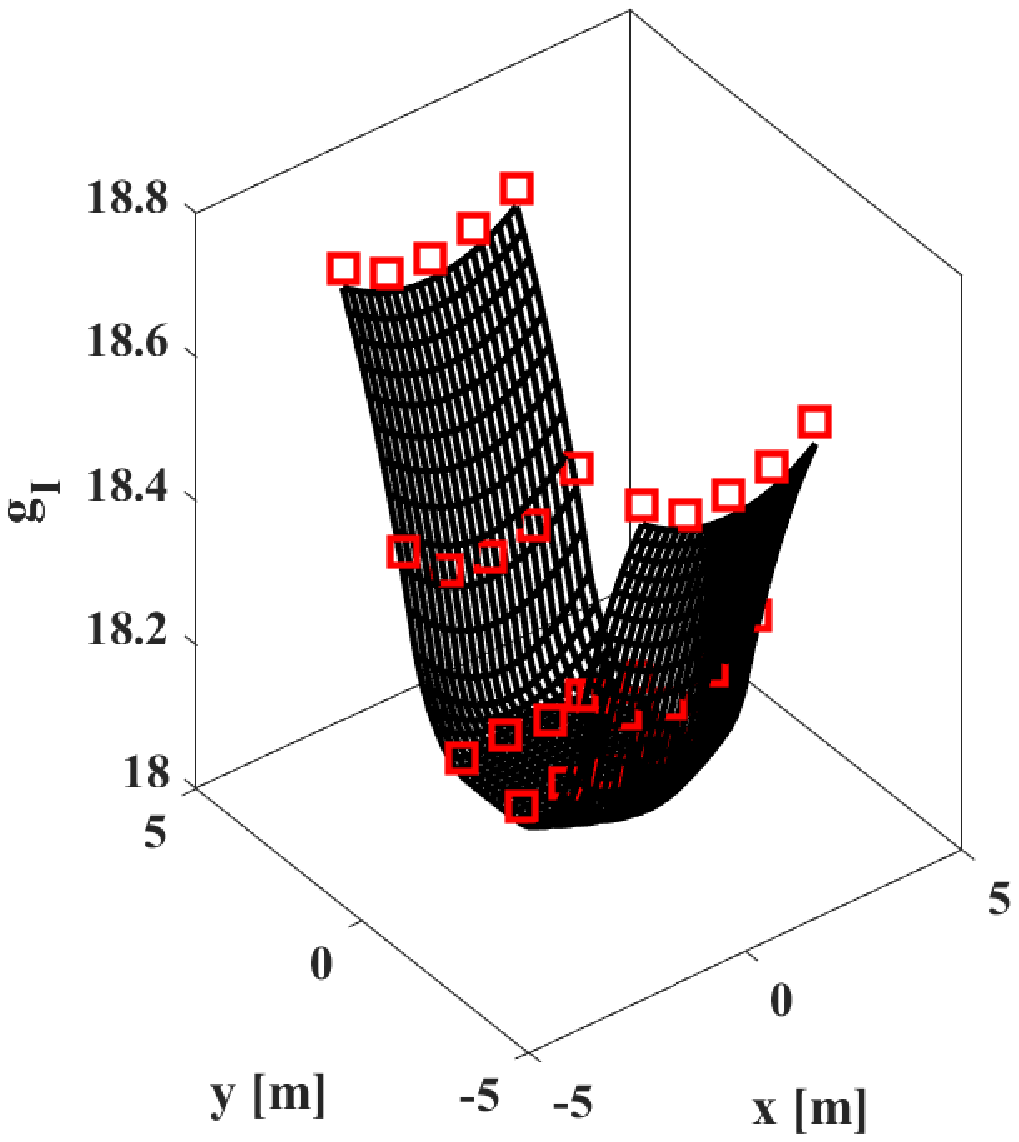}}
\subfigure[$g_{\mathrm{P}}$]{\includegraphics[width=1.5in]{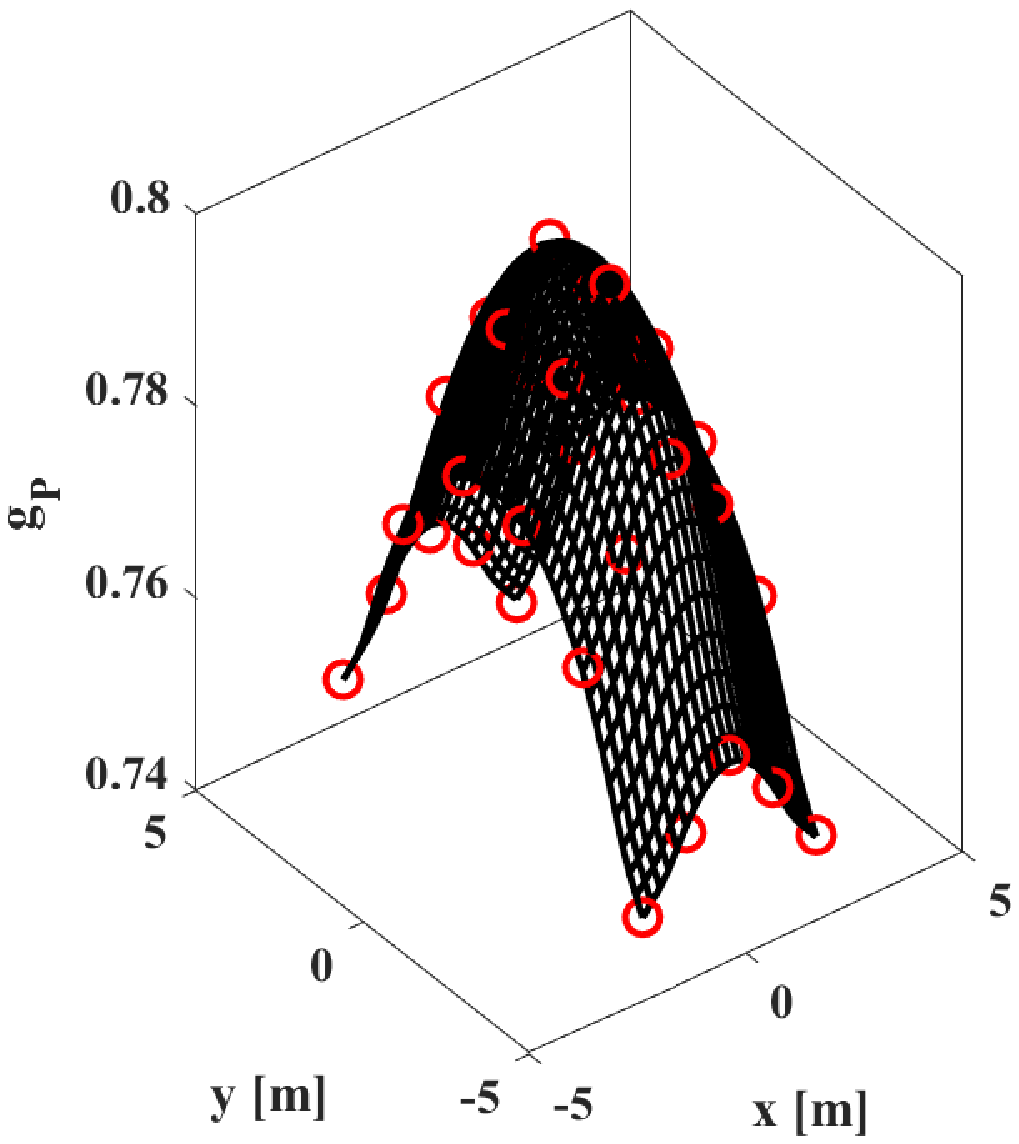}}
\caption{Comparison of analytic gains $g_{\mathrm{I}}$ and $g_{\mathrm{P}}$ with Monte-Carlo simulations at 1 GHz in a $5\mathrm{m}\times10\mathrm{m}$ rectangular room.  $P_{\mathrm{T}}=-30$ $\mathrm{dBWm^{-2}}$, $P_{\mathrm{th}}=90$ $\mathrm{dBWm^{-2}}$. Markers are generated from Monte Carlo simulations. Solid lines are computed by Remark \ref{finalremark}.
}\label{rect}
\end{minipage}
\hspace{0.1in}
\begin{minipage}[t]{0.48\textwidth}
\subfigure[$g_{\mathrm{I}}$]{\includegraphics[width=1.5in]{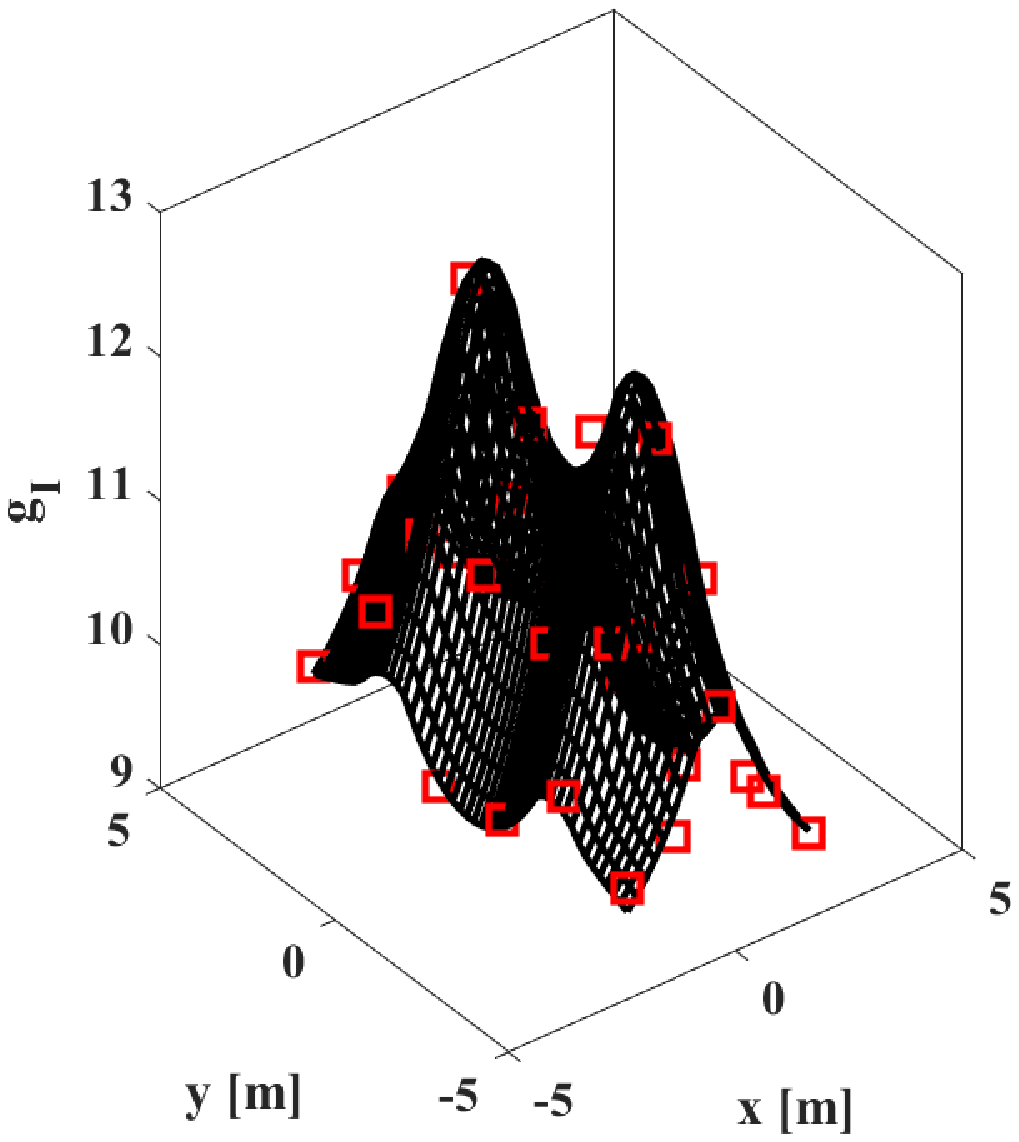}}
\subfigure[$g_{\mathrm{P}}$]{\includegraphics[width=1.5in]{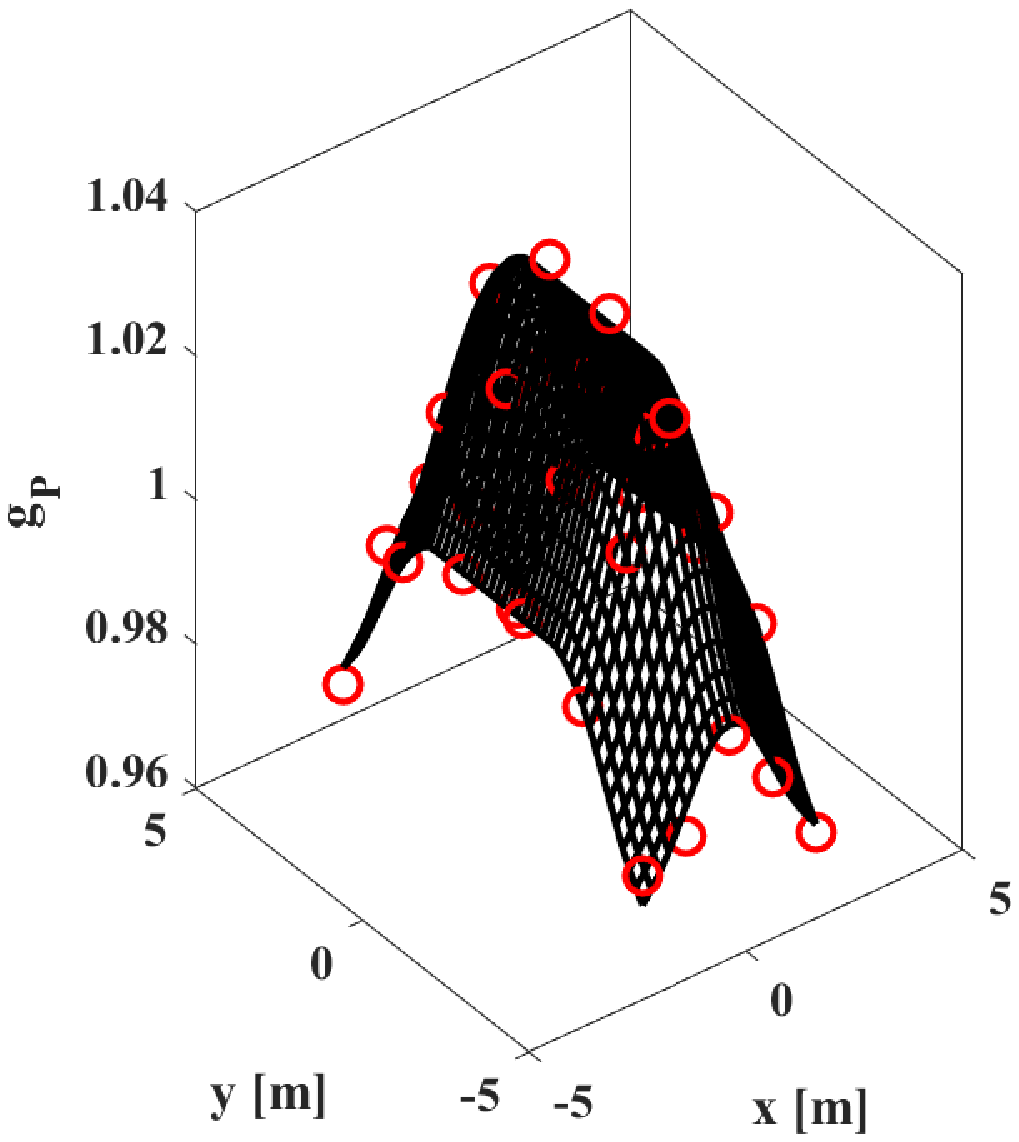}}
\caption{Comparison of analytic gains $g_{\mathrm{I}}$ and $g_{\mathrm{P}}$ with Monte-Carlo simulations at 28 GHz in a $5\mathrm{m}\times10\mathrm{m}$ rectangular room.   $P_{\mathrm{T}}=-30$ $\mathrm{dBWm^{-2}}$, $P_{\mathrm{th}}=-100$ $\mathrm{dBWm^{-2}}$. Markers are generated from Monte Carlo simulations. Solid lines are computed by Remark \ref{finalremark}.
}\label{rectmmwave}
\end{minipage}
\begin{minipage}[t]{0.48\textwidth}
\subfigure[$g_{\mathrm{I}}$]{\includegraphics[width=1.5in]{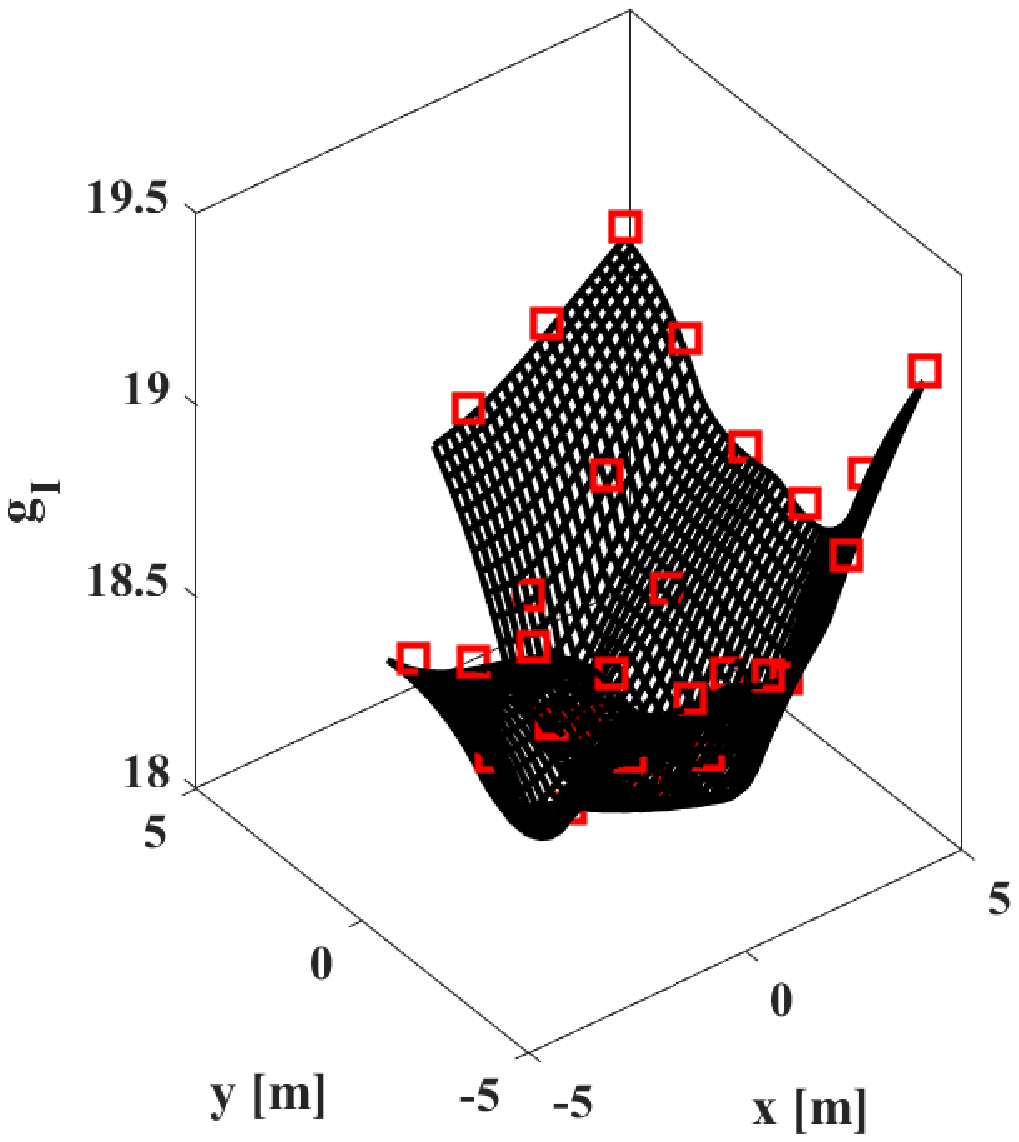}}
\subfigure[$g_{\mathrm{P}}$]{\includegraphics[width=1.5in]{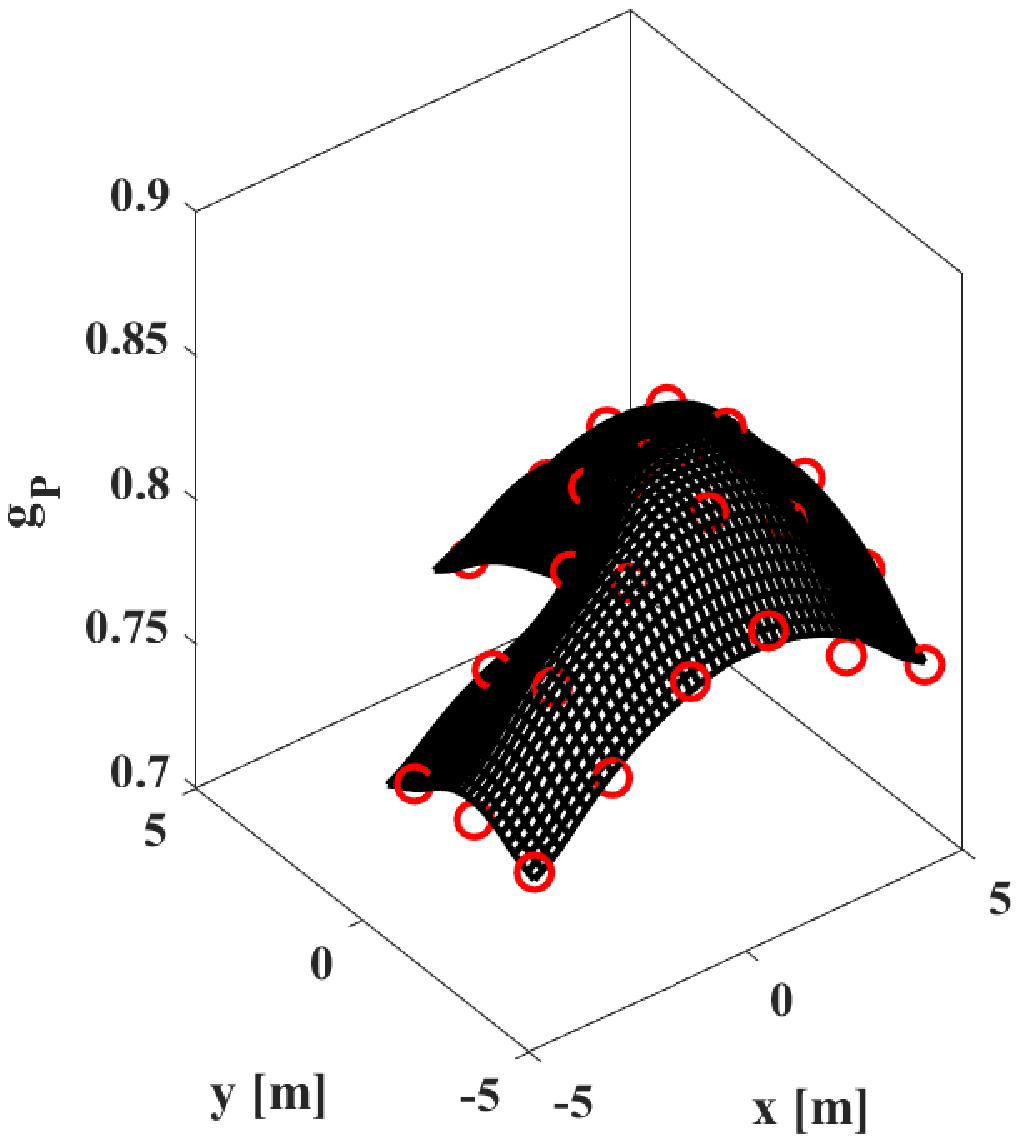}}
\caption{Comparison of analytic gains $g_{\mathrm{I}}$ and $g_{\mathrm{P}}$ with Monte-Carlo simulations at 1 GHz in the L-shaped corner in Fig. \ref{examples}.  $P_{\mathrm{T}}=-30$ $\mathrm{dBWm^{-2}}$, $P_{\mathrm{th}}=-90$ $\mathrm{dBWm^{-2}}$. Markers are generated from Monte Carlo simulations. Solid lines are computed by Remark \ref{finalremark}.}\label{cornor}
\end{minipage}
\hspace{0.1in}
\begin{minipage}[t]{0.48\textwidth}
\subfigure[$g_{\mathrm{I}}$]{\includegraphics[width=1.5in]{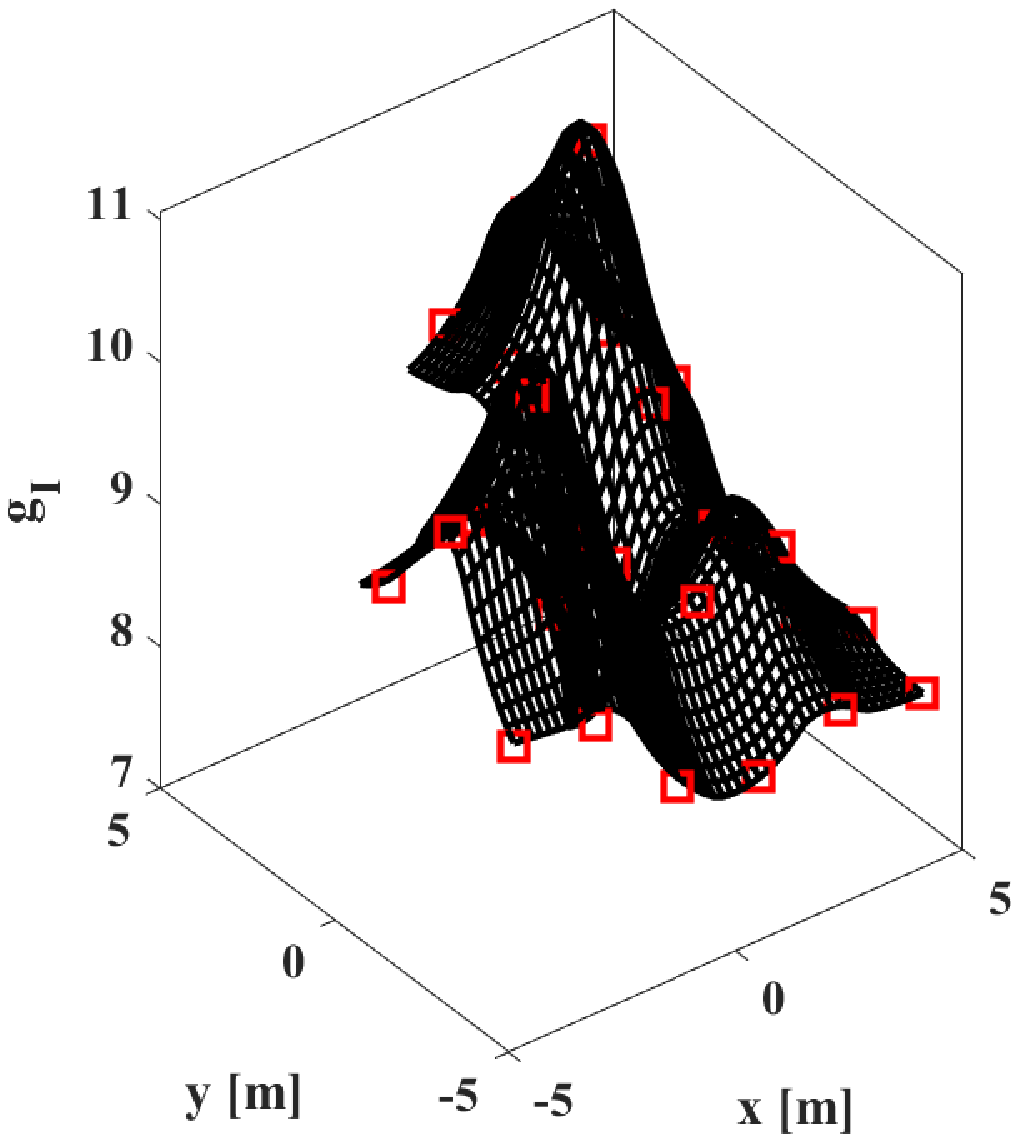}}
\subfigure[$g_{\mathrm{P}}$]{\includegraphics[width=1.5in]{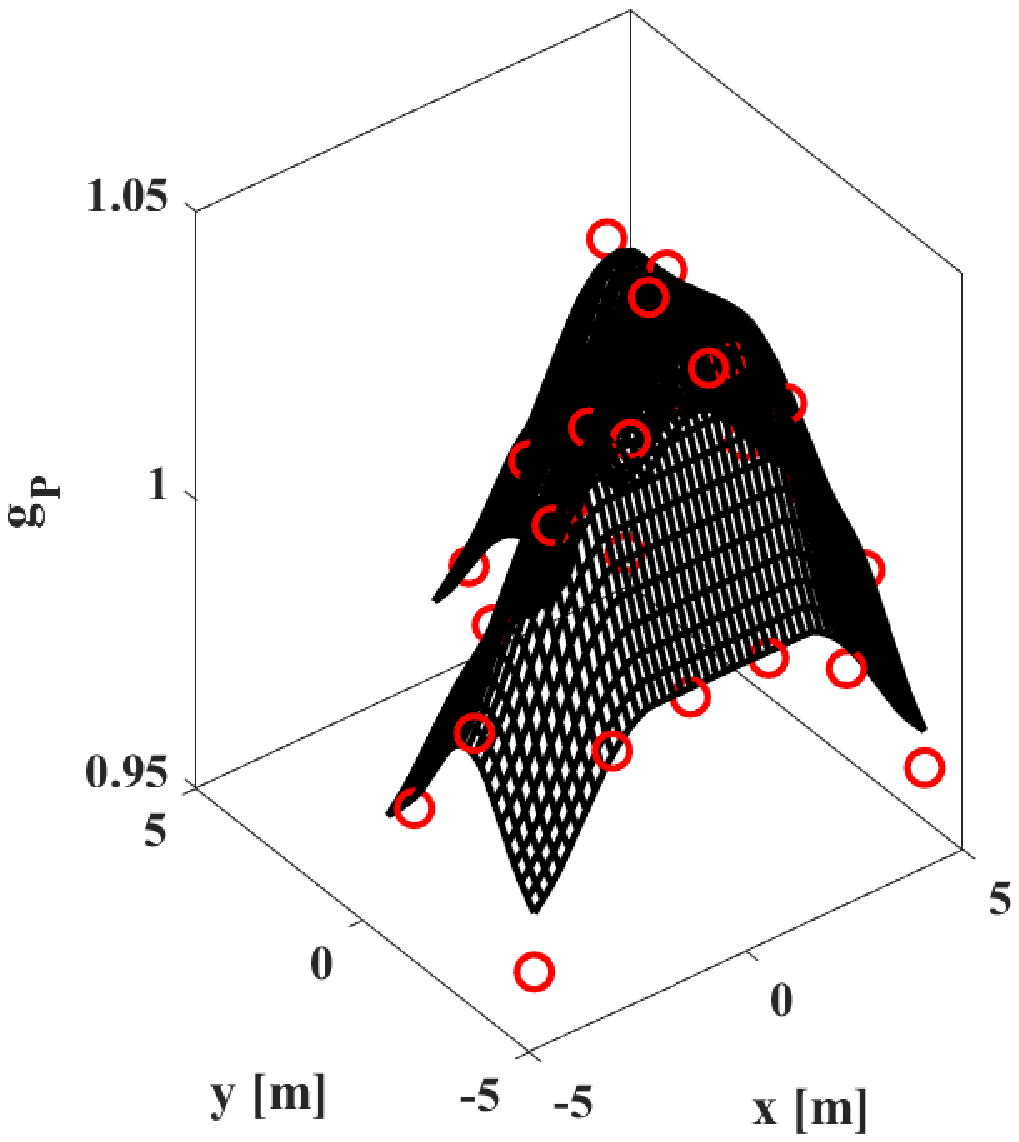}}
\caption{Comparison of analytic gains $g_{\mathrm{I}}$ and $g_{\mathrm{P}}$ with Monte-Carlo simulations at 28 GHz in the L-shaped corner in Fig. \ref{examples}.   $P_{\mathrm{T}}=-30$ $\mathrm{dBWm^{-2}}$, $P_{\mathrm{th}}=-100$ $\mathrm{dBWm^{-2}}$. Markers are generated from Monte Carlo simulations. Solid lines are computed by Remark \ref{finalremark}.
}\label{cornormmwave}
\end{minipage}
\end{figure}

In the following sections, we use only the analytic results to propose some examples of building wireless performance analysis since the computation of $g_{\mathrm{I}}$ and $g_{\mathrm{P}}$ by Monte Carlo simulation is computationally expensive.

\section{Simulations}
In this section, we give examples of the evaluation of building wireless performance of two reference floor plans: a single rectangular room layout and an office building layout with rectangular rooms.

\subsection{Performance evaluation for a single room}

The impact of the area of a rectangular room on $g_{\mathrm{I}}$ and $g_{\mathrm{P}}$ is analyzed. It is needed to emphasize that $g_{\mathrm{I}}$ and $g_{\mathrm{P}}$ have different values  at different positions in the room. To facilitate the comparison of rooms in terms of wireless performance, for each considered room, we compute the average
$g_{\mathrm{I}}$ and the average $g_{\mathrm{P}}$ for that room.

\begin{figure} [!t]
\begin{minipage}[t]{0.48\textwidth}
\subfigure[$g_{\mathrm{I}}$]{\includegraphics[width=1.5in]{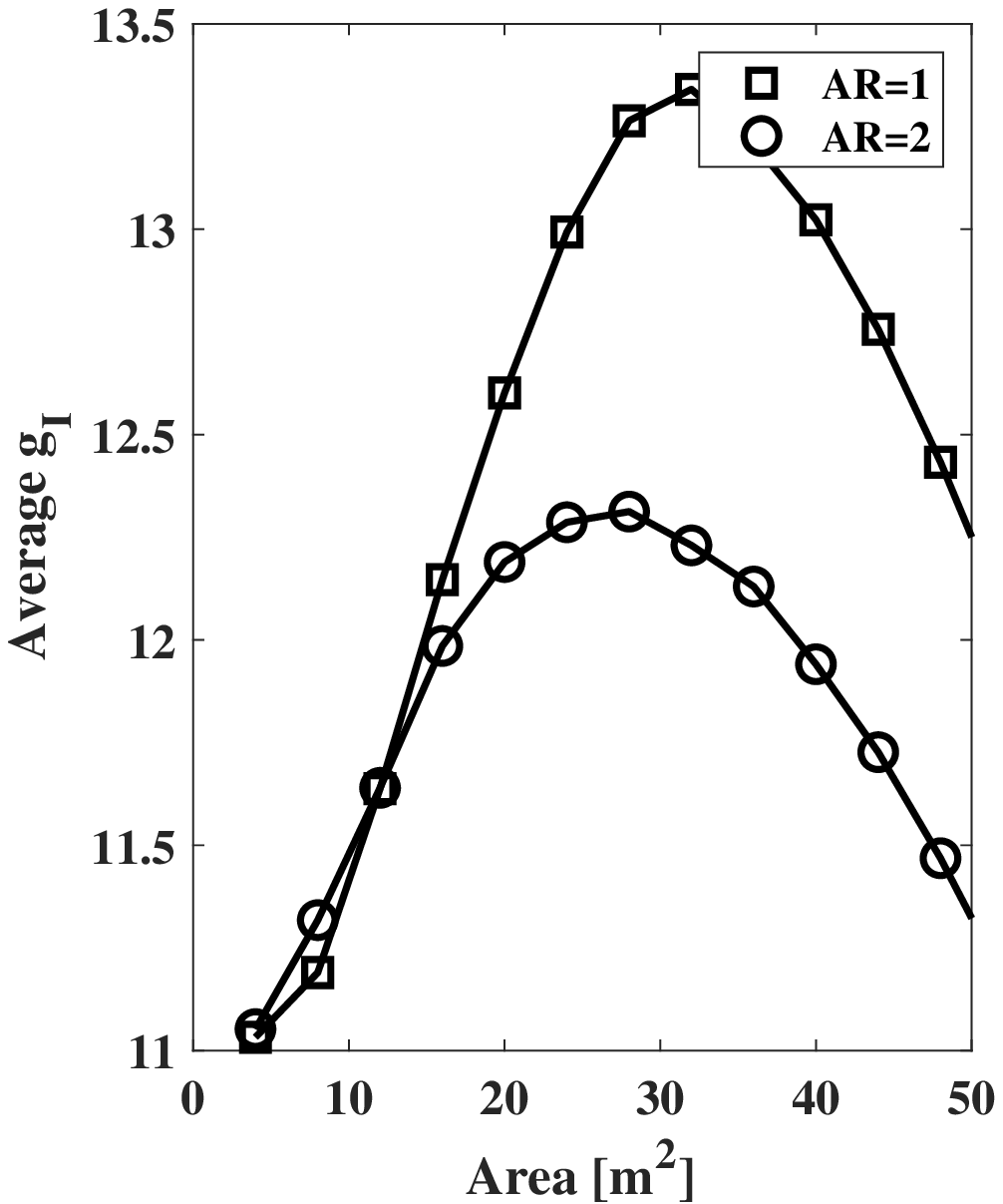}}
\subfigure[$g_{\mathrm{P}}$]{\includegraphics[width=1.5in]{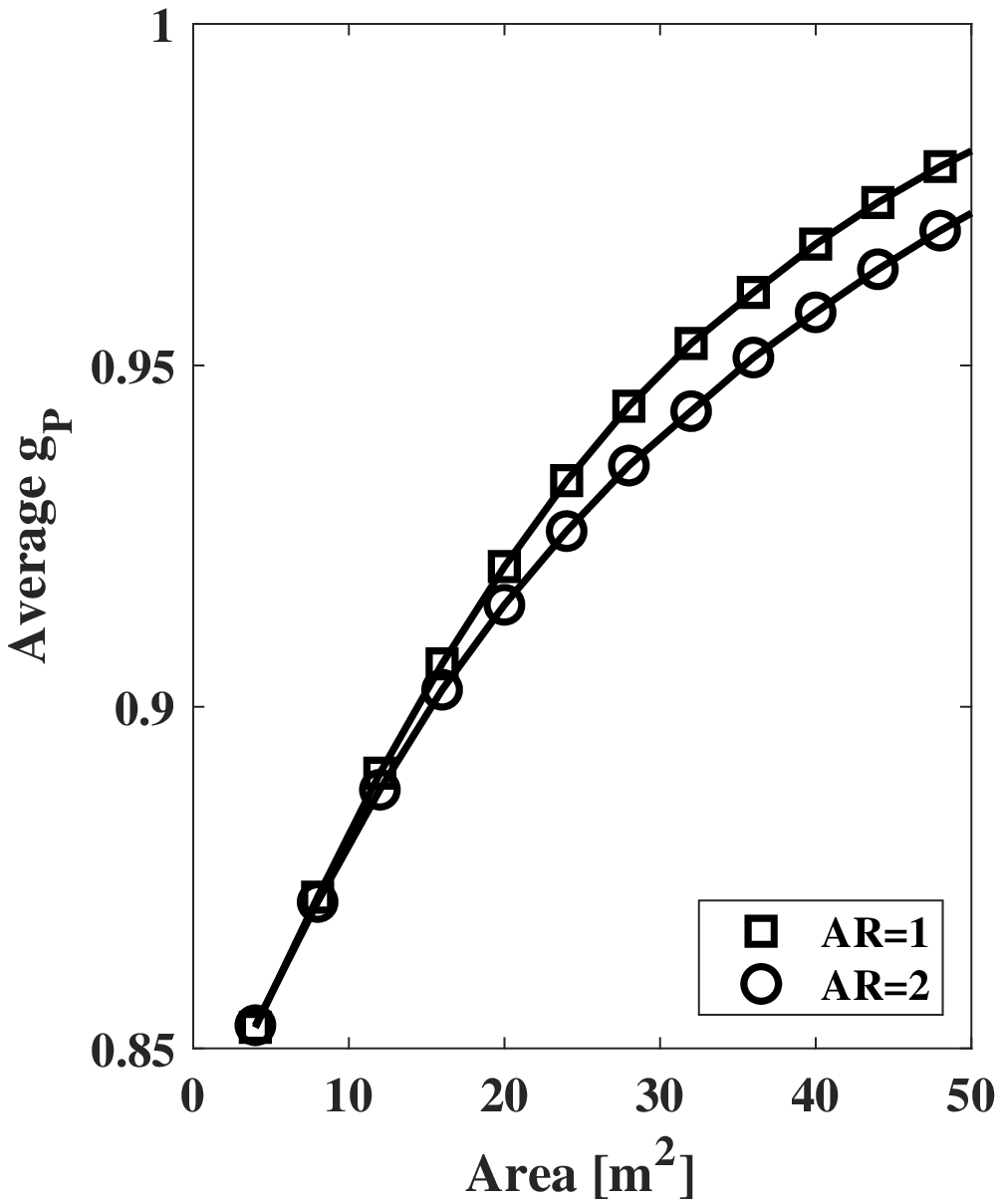}}
\caption{Impact of room size on $g_{\mathrm{I}}$ and $g_{\mathrm{P}}$ at 1 GHz in a rectangular room.  $P_{\mathrm{T}}=-30$ $\mathrm{dBWm^{-2}}$, $P_{\mathrm{th}}=-75$ $\mathrm{dBWm^{-2}}$. }
\label{agisize}
\end{minipage}
\hspace{0.1in}
\begin{minipage}[t]{0.48\textwidth}
\subfigure[$g_{\mathrm{I}}$]{\includegraphics[width=1.5in]{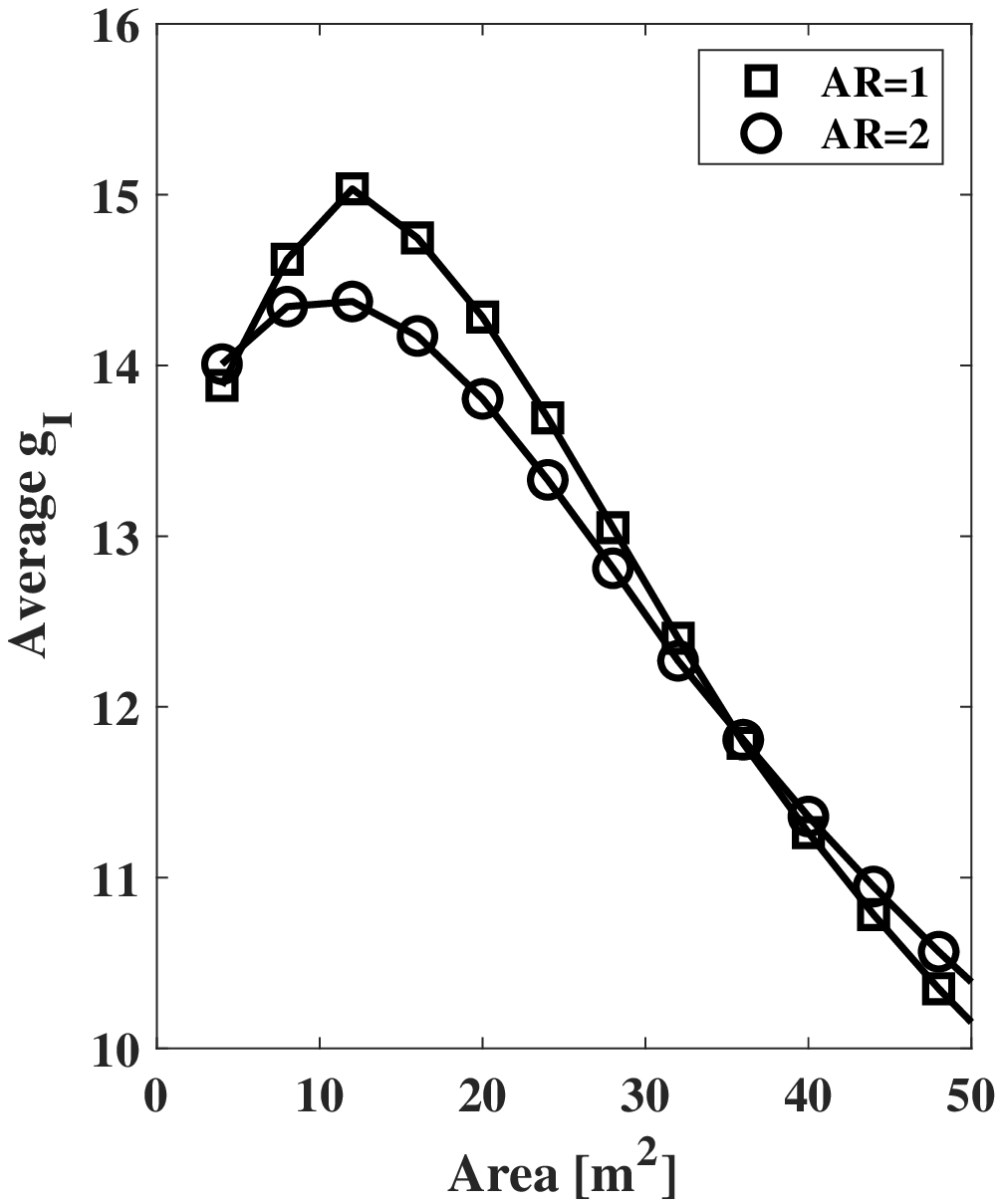}}
\subfigure[$g_{\mathrm{P}}$]{\includegraphics[width=1.5in]{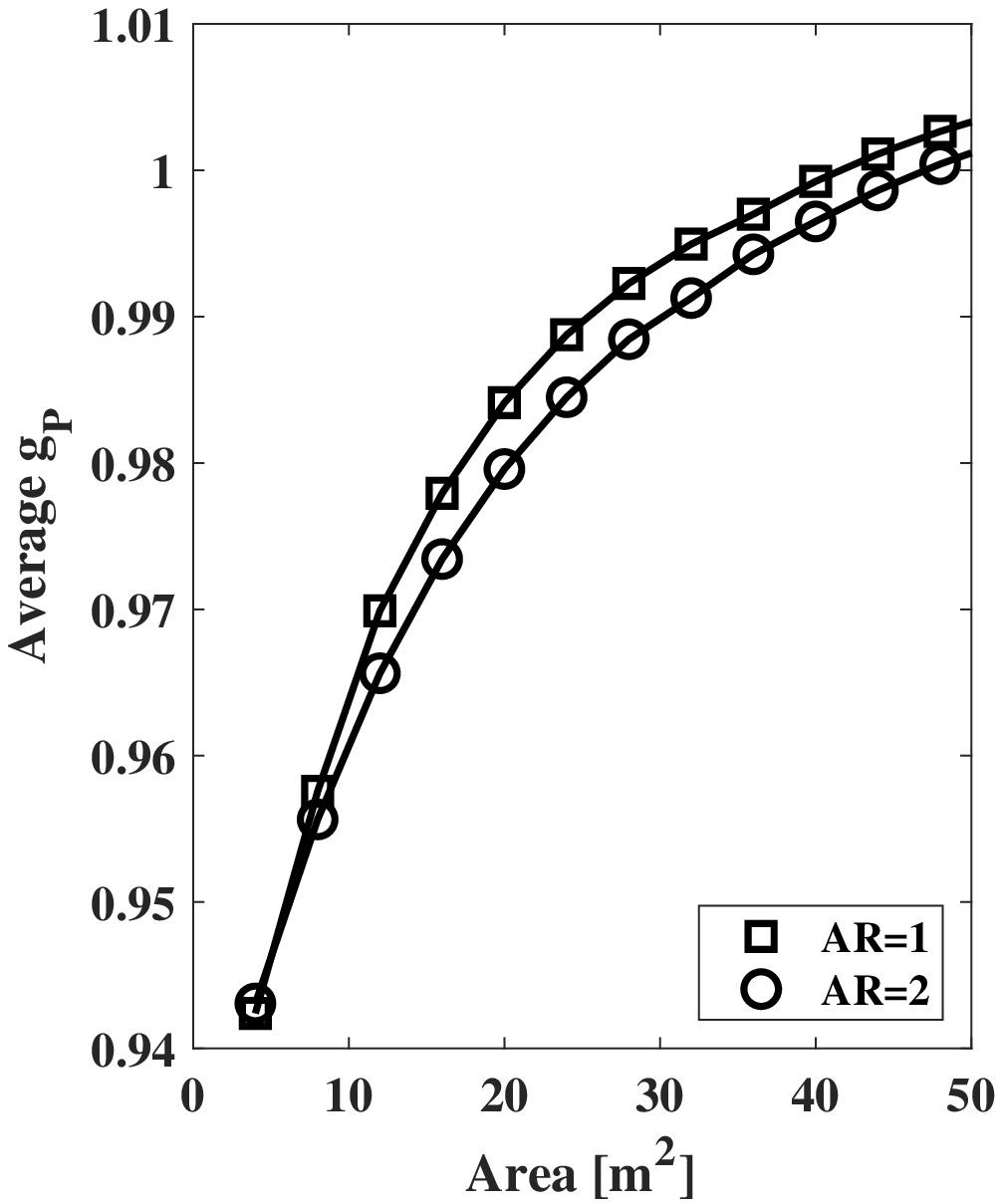}}
\caption{Impact of room size on $g_{\mathrm{I}}$ and $g_{\mathrm{P}}$ at 28 GHz in a rectangular room.   $P_{\mathrm{T}}=-30$ $\mathrm{dBWm^{-2}}$, $P_{\mathrm{th}}=-100$ $\mathrm{dBWm^{-2}}$.
}\label{agpsizem}
\end{minipage}
\centering
\subfigure[$g_{\mathrm{I}}$]{\includegraphics[width=1.6in]{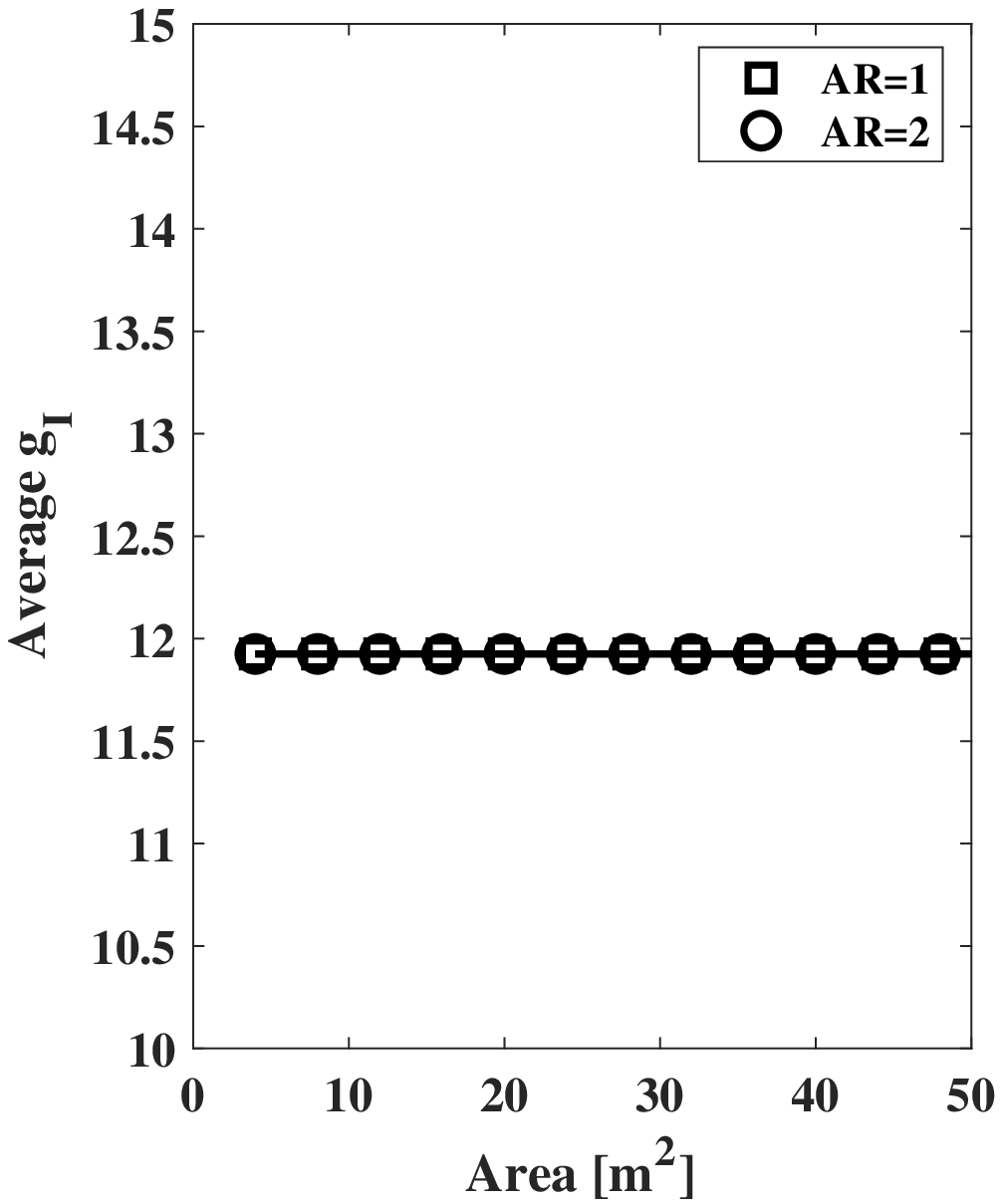}}
\subfigure[$g_{\mathrm{P}}$]{\includegraphics[width=1.6in]{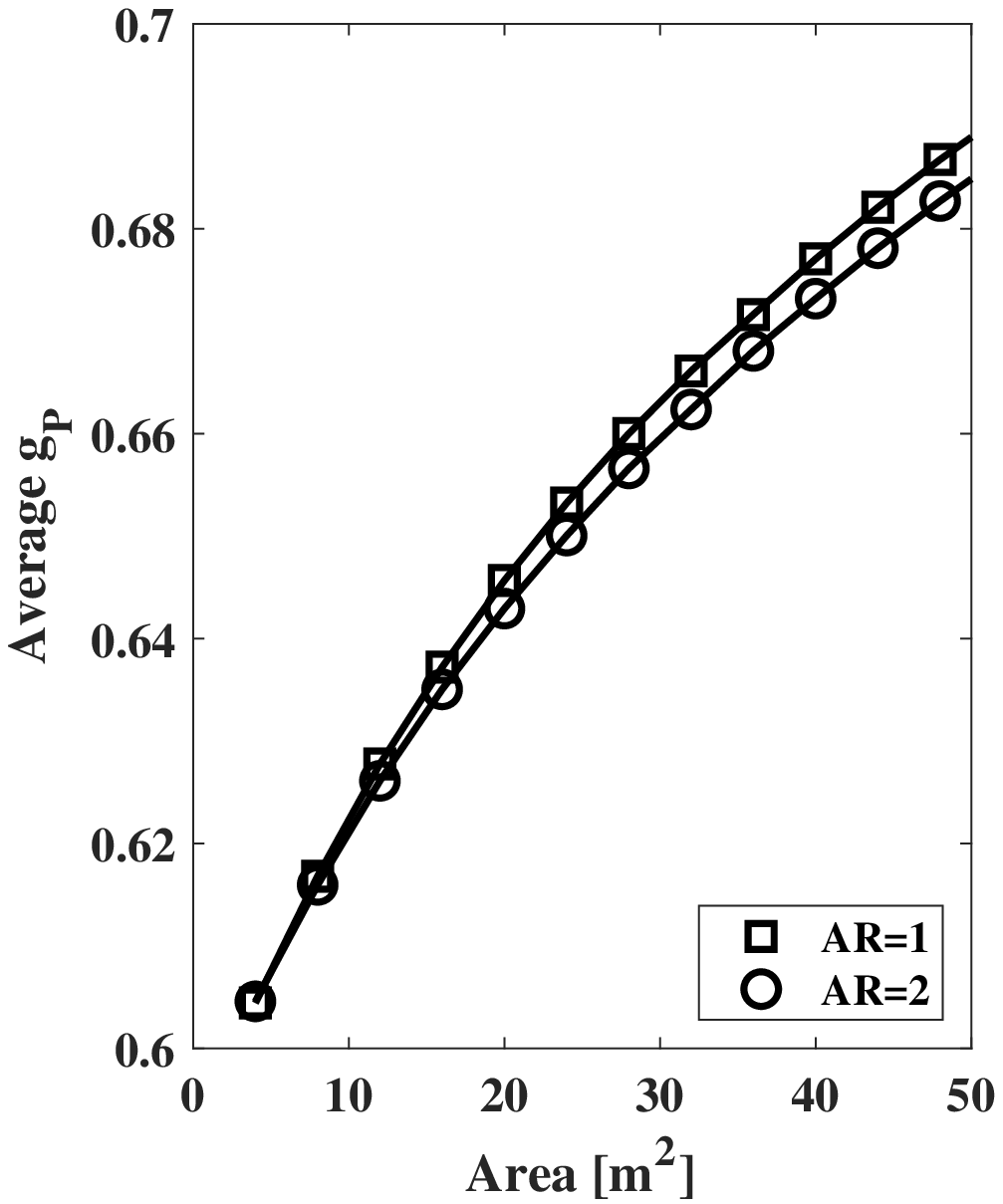}}
\caption{Impact of room size on $g_{\mathrm{I}}$ and $g_{\mathrm{P}}$ at 1 GHz in a rectangular room.  $P_{\mathrm{T}}=-30$ $\mathrm{dBWm^{-2}}$, $P_{\mathrm{th}}=-100$ $\mathrm{dBWm^{-2}}$. }
\label{larger}
\end{figure}

Figs. \ref{agisize}-\ref{agpsizem} show  the average $g_{\mathrm{I}}$ and the average $g_{\mathrm{P}}$  as a function of the indoor coverage area for two different AR, i.e., $AR\in\{1,2\}$  at 1 GHz and 28 GHz.
While covering a small room in the open space scenario, the power of the interference signals becomes larger in proportion to the intended signals due to the strong attenuation, and the limited receive sensitivity. Therefore, the significant increase of $I_{\mathrm{N}}$ takes place resulting in a decrease of $g_{\mathrm{I}}$. Whereas, in a larger room, the interference is more likely to be
LOS, then a larger $I_{\mathrm{L}}$ significantly reduces $g_{\mathrm{I}}$. Therefore, it is observed that $g_{\mathrm{I}}$ can be optimized subject to the size of the room by balancing $I_{\mathrm{N}}$ and $I_{\mathrm{L}}$ in an appropriate way. For different values of AR, the optimum size of a room and the maximized $g_{\mathrm{I}}$ are different. With an increasing AR, the maximum $g_{\mathrm{I}}$ decreases, and therefore a better wireless performance is achieved with a lower AR.

In Fig. \ref{agisize}, shown results correspond to $R_{\mathrm{O}}=4.2$ m, $R_{\mathrm{L}}=5.3$ m, and $R_{\mathrm{N}}=2.5$ m due to (\ref{RO})-(\ref{RN}).
In Fig. \ref{agpsizem}, we have $R_{\mathrm{O}}=2.7$ m, $R_{\mathrm{L}}=3.2$ m, and $R_{\mathrm{N}}=1.9$ m. Since the coverage area is in general less than the size of the room, the intended signal is less likely to be blocked. Thus, the value of the average $g_{\mathrm{P}}$ is close to one. Therefore, with a small coverage area, e.g., a relative small $\frac{P_\mathrm{T}}{P_\mathrm{th}}$, the impact of $g_{\mathrm{P}}$ is relatively slight. On the other hand, for a large $\frac{P_\mathrm{T}}{P_\mathrm{th}}$, e.g., $P_{\mathrm{T}}=-30\ \mathrm{dBWm^{-2}}$, $P_{\mathrm{th}}=-100\ \mathrm{dBWm^{-2}}$ at 1 GHz, the average $g_{\mathrm{I}}$ and the average $g_{\mathrm{P}}$ are shown in Fig. \ref{larger}. Following (\ref{RO})-(\ref{RN}), we have $R_{\mathrm{O}}=75$ m, $R_{\mathrm{L}}=148$ m, and $R_{\mathrm{N}}=15$ m. Even though for a room with an area of 50 m$^2$ and an AR$=2$, the maximum length of LOS link is $5\sqrt{2}$ m, which is less than $R_{\mathrm{N}}$. Thus, all the interference power is NLOS, i.e., $I_{\mathrm{L}}=0$. Also, the $I_{\mathrm{N}}$ becomes independent of the size and the AR of the room because the NLOS interference is not impacted by any wall. Therefore, if the maximum possible distance between two points of walls around the polygon room is less than $R_{\mathrm{N}}$, $I_{\mathrm{N}}$ does not change with the size and the AR of the room and $g_{\mathrm{I}}$ is a constant.

However, when the size of a room is less than $R_{\mathrm{N}}$, the blockage of the room attenuates the power of intended signals significantly. Reducing the size of the room, more intended signals are blocked and thus the $g_{\mathrm{P}}$ decreases. Therefore, under this condition, a decreasing size reduces the wireless performance of a room.

\begin{figure}[!t]
\centering
\subfigure[$g_{\mathrm{I}}$ at 1 GHz]{\includegraphics[width=1.5in]{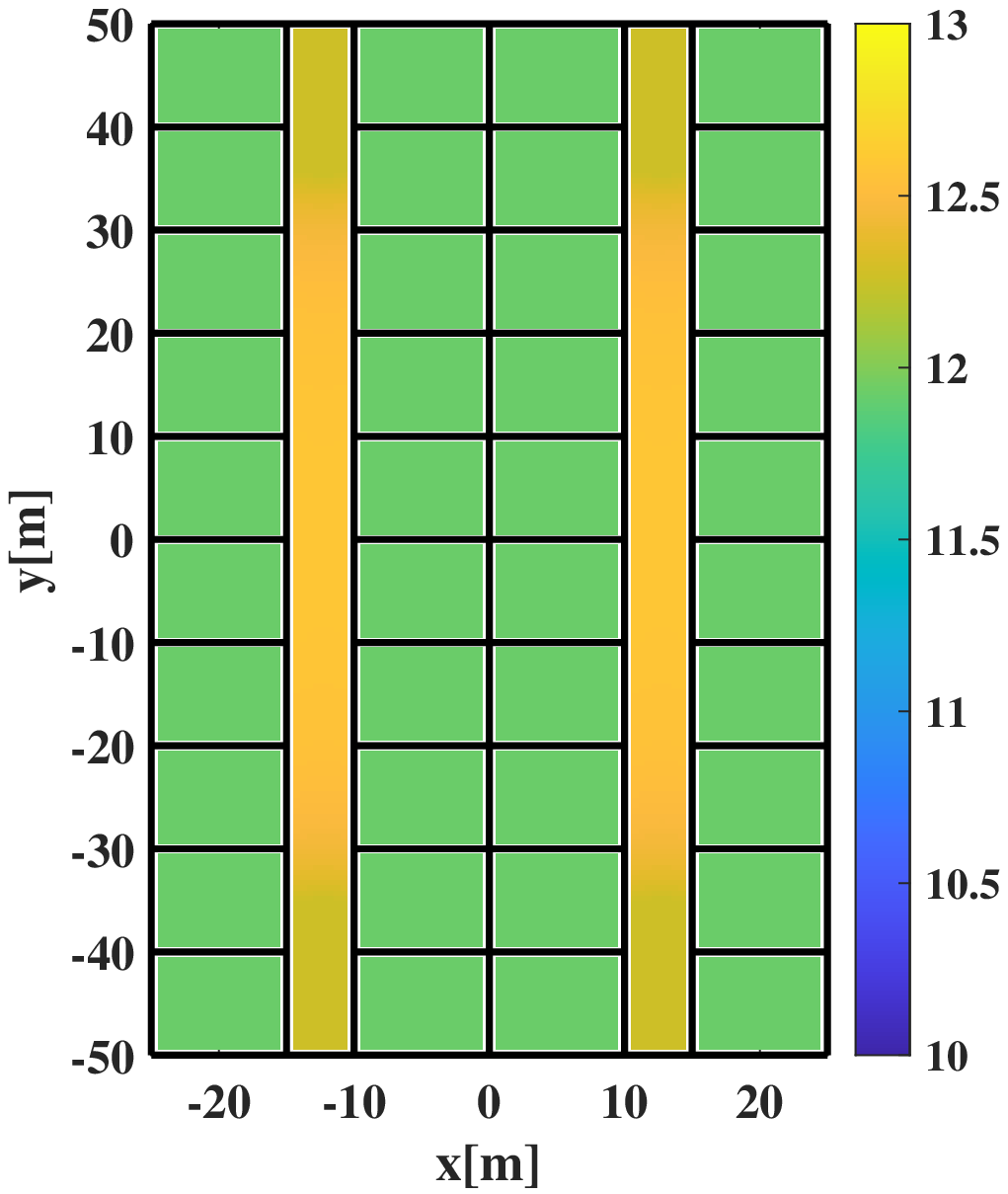}}
\subfigure[$g_{\mathrm{I}}$ at 28 GHz]{\includegraphics[width=1.5in]{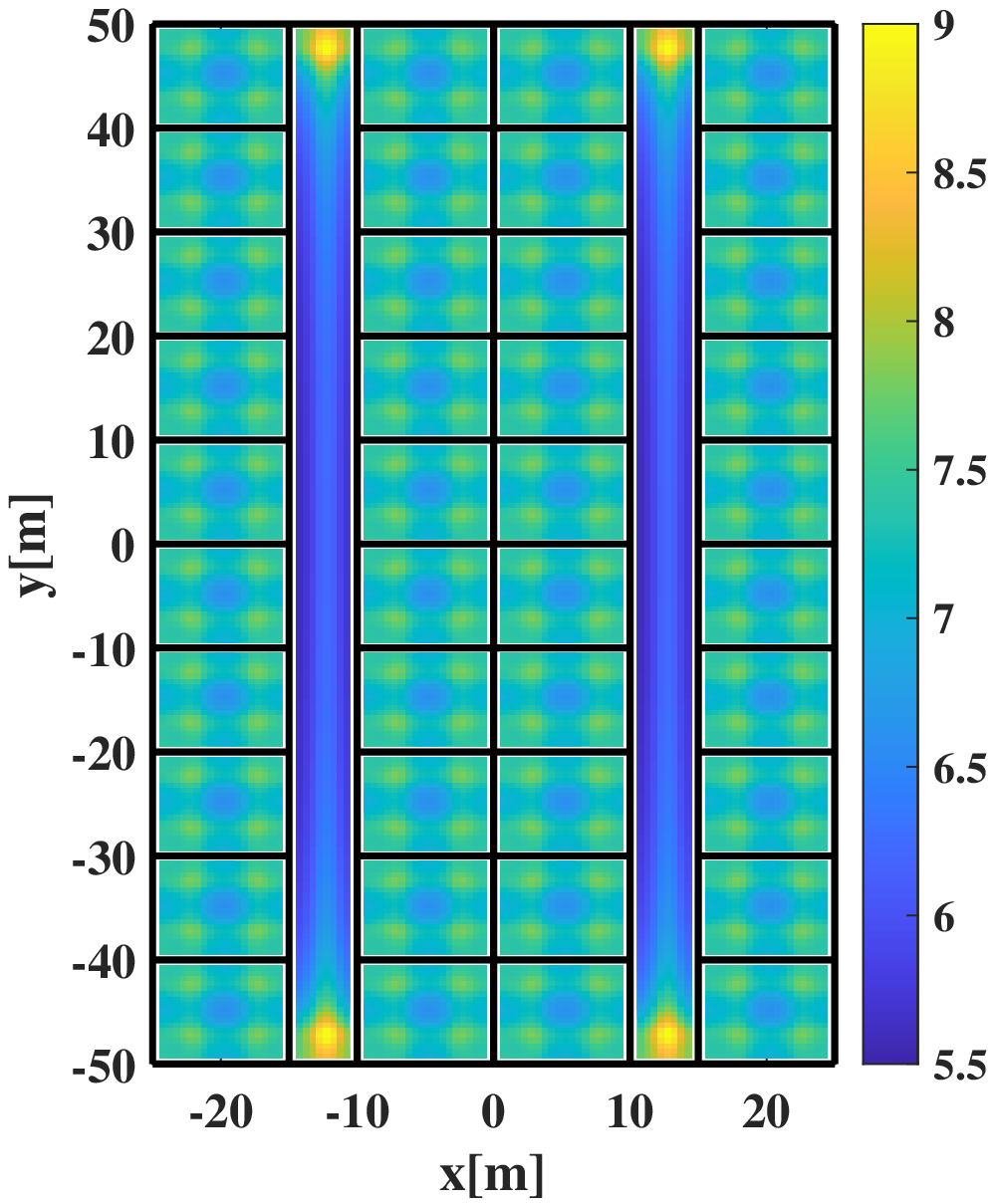}}
\subfigure[$g_{\mathrm{P}}$ at 1 GHz]{\includegraphics[width=1.5in]{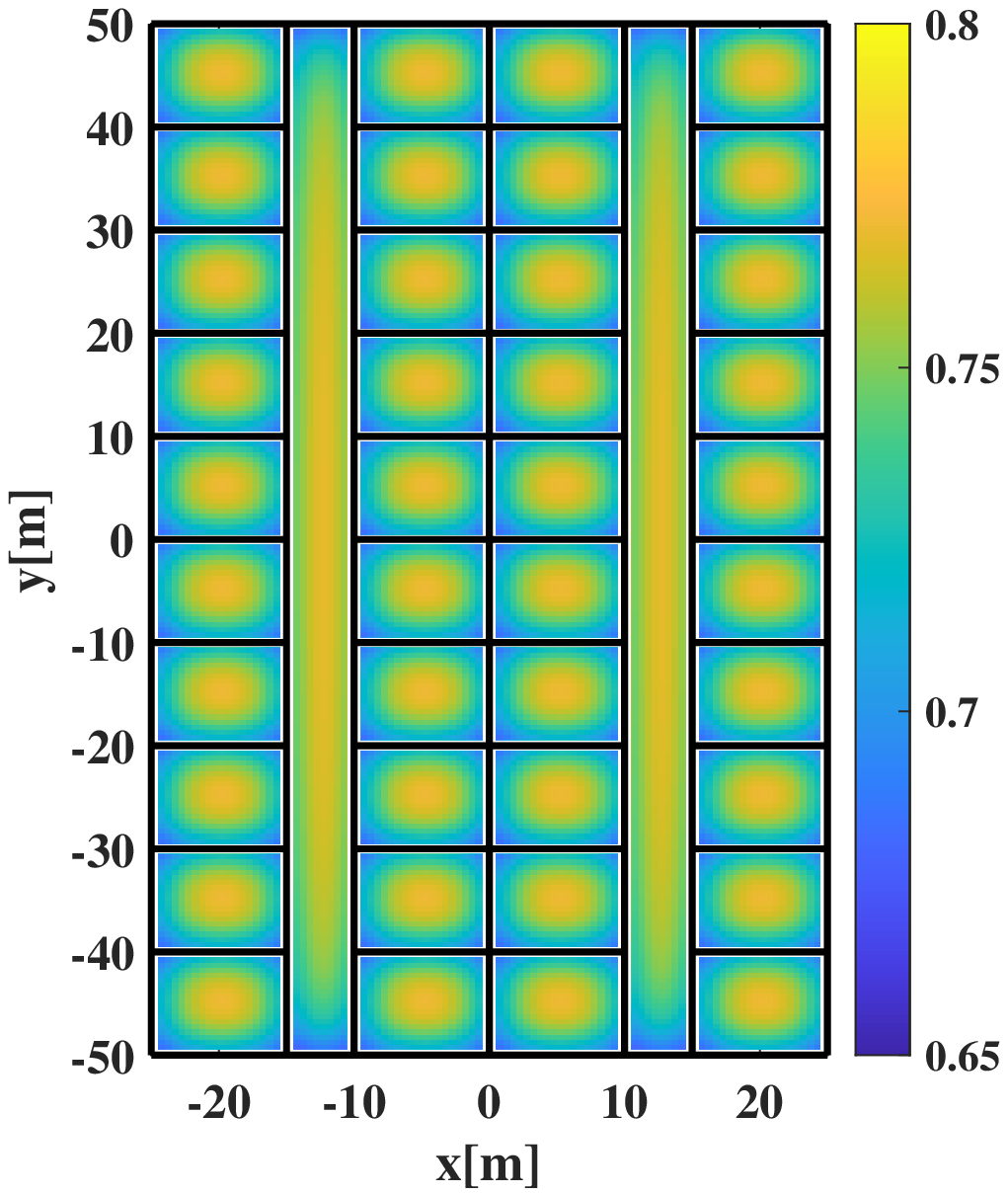}}
\subfigure[$g_{\mathrm{P}}$ at 28 GHz]{\includegraphics[width=1.5in]{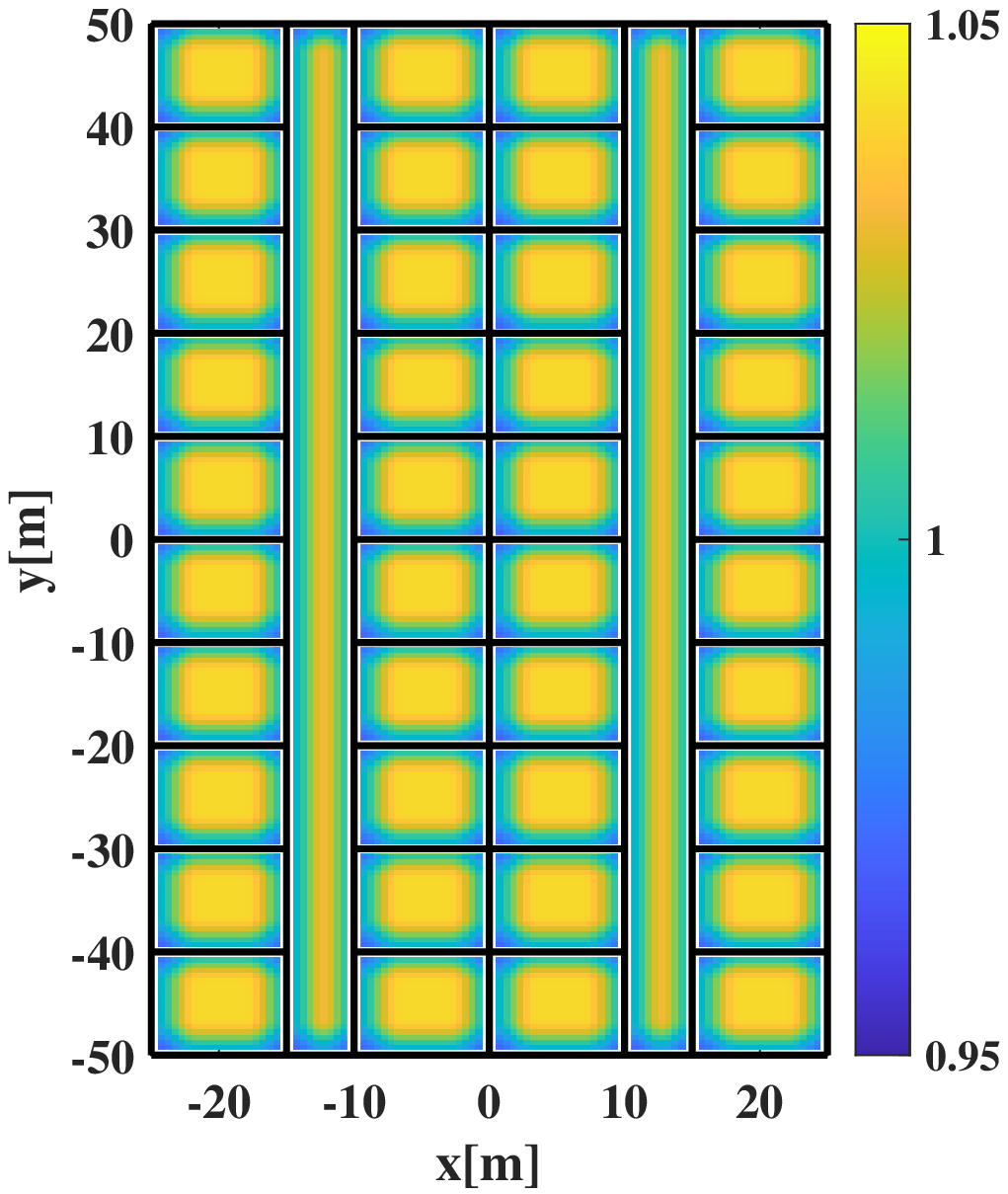}}
\caption{$g_{\mathrm{I}}$ and $g_{\mathrm{P}}$ in a typical office scenario. $P_{\mathrm{T}}=-30$ $\mathrm{dBWm^{-2}}$, $P_{\mathrm{th}}=-100$ $\mathrm{dBWm^{-2}}$. }
\label{A1analysis100}
\end{figure}
\subsection{An example of building evaluation}
In this subsection, the wireless performance evaluation for the A1 scenario of WINNER II channel model \cite[Fig. 2.1]{winnerII} is taken as an example. 
In the computations, we use $P_{\mathrm{T}}=-30$ $\mathrm{dBWm^{-2}}$, and $P_{\mathrm{th}}=-100$ $\mathrm{dBWm^{-2}}$. For each room or corridor, $g_{\mathrm{I}}$ and $g_{\mathrm{P}}$ are computed by evoking Remark \ref{finalremark}. The obtained $g_{\mathrm{I}}$ and $g_{\mathrm{P}}$ are illustrated in Fig. \ref{A1analysis100}. The average $g_{\mathrm{I}}=12.03$ and the average $g_{\mathrm{P}}= 0.74$, respectively. Whereas for  $f_{\mathrm{c}}=28$ GHz, i.e., at mm-Wave band, the  average $g_{\mathrm{I}}=7.08$ and the average $g_{\mathrm{P}}= 1.01$, respectively. For $P_{\mathrm{T}}=-30$ $\mathrm{dBWm^{-2}}$ and $P_{\mathrm{th}}=-100$ $\mathrm{dBWm^{-2}}$, the WINNER II A1 at 1 GHz outperforms that at 28 GHz in terms of $g_{\mathrm{I}}$ because it blocks almost all interference signals.  Whereas in terms of $g_{\mathrm{P}}$, the WINNER II A1 building has better wireless performance at 28 GHz. Therefore, for a building wireless performance evaluation in the future, the tradeoff between $g_{\mathrm{I}}$, $g_{\mathrm{P}}$, and $f_{\mathrm{c}}$ has to be taken into account carefully.


\section{Conclusions}


The intrinsic wireless performance characteristics of buildings have been formally defined as the interference gain and the power gain figures of merit. An analytical approach has been proposed to compute these figures of merit reflecting the wireless performance of buildings with polygonal-shaped rooms. The analytic approach has been validated through Monte Carlo simulations for typical indoor environments with well-established empirical path gain models showing a good agreement between them. The proposed analytic approach allows to quantify the impact of building properties, e.g., the aspect ratio and the area of rooms, on both the interference gain and the power gain. Numerical examples are provided based on different propagation scenarios at different frequencies, i.e., 1 GHz and 28 GHz. Based on the presented analysis it is shown that wireless building design need to consider the trade-off between the interference gain, the power gain and the frequency band of the wireless system deployed in a specific building. This finding has tremendous implications on the way new buildings shall be designed taking into consideration wireless performance already at the design phase. Indeed, it is expected that architects and civil engineers will now be able to carefully take the interference gain and the power gain into account before designing or retrofitting a building to provide adequate performance margins for future indoor wireless networks. The proposed approach bridges the building design and wireless communications industries. It has the potential to open up a wide range of innovative solutions offering wireless friendly buildings, leading to improved wireless services indoors where most of the communications traffic takes place.



\section*{Appendices}

\subsection{Proof of Theorem \ref{osPItheorem}}
\label{osPItheoremproof}

\subsubsection{Closed-form $I_{\mathrm{O}}$  (\ref{Iosfinal})}

$I_{\mathrm{O}}$ is defined by \cite{patent,tvtbwp}
\begin{eqnarray}
\label{IOdefeq}
I_{\mathrm{O}}=\int_0^{2\pi} \int_{R_{\mathrm{O}}}^{+\infty} P_{\mathrm{T}}G_{\mathrm{O}}(R) R\mathrm{d}R\mathrm{d}\theta.
\end{eqnarray}

Substituting (\ref{RO}) into (\ref{IOdefeq}), considering the case $\frac{P_{\mathrm{T}}}{P_{\mathrm{th}}}<\frac{(4\pi)^4\left({h_{\mathrm{T}}h_{\mathrm{R}}}\right)^2}{\lambda^4}$, we have
\begin{eqnarray}
\label{Ios2}
I_{\mathrm{O}}=\int_{0}^{2\pi}\int_{\frac{\lambda}{4\pi}\sqrt{\frac{P_{\mathrm{T}}} {P_{\mathrm{th}}}}}^{+\infty}{P_{\mathrm{T}}}{G_{\mathrm{O}}(R)}R\mathrm{d}R\mathrm{d}\theta.
\end{eqnarray}
Substituting (\ref{PL_os}) into (\ref{Ios2}), we obtain the closed-form $I_{\mathrm{O}}$ as
\begin{eqnarray}
\label{Iosfinal1}
I_{\mathrm{O}}=\frac{P_{\mathrm{T}}\lambda^2}{8\pi}\left\{
\frac{1}{2}
+  \ln\left(\frac{16\sqrt{P_{\mathrm{th}}}\pi^2 h_{\mathrm{T}}h_{\mathrm{R}}}{\sqrt{P_{\mathrm{T}}}\lambda^2}\right)\right\}.
\end{eqnarray}

For $\frac{P_{\mathrm{T}}}{P_{\mathrm{th}}}\geq\frac{(4\pi)^4\left({h_{\mathrm{T}}h_{\mathrm{R}}}\right)^2}{\lambda^4}$, we have
\begin{eqnarray}
\label{Ios3}
I_{\mathrm{O}}=\int_{0}^{2\pi}\int_{
\left(\frac{P_{\mathrm{T}}}{P_{\mathrm{th}}}\right)^{\frac{1}{4}}\sqrt{h_{\mathrm{T}}h_{\mathrm{R}}}
}^{+\infty}{P_{\mathrm{T}}}{G_{\mathrm{O}}(R)}R\mathrm{d}R\mathrm{d}\theta.
\end{eqnarray}
Substituting (\ref{PL_os}) into (\ref{Ios3}) and following some algebraic manipulations, we obtain
\begin{eqnarray}
\label{Iosfinal2}
I_{\mathrm{O}}=\pi\sqrt{P_{\mathrm{th}}P_{\mathrm{T}}} h_{\mathrm{T}}h_{\mathrm{R}}.
\end{eqnarray}

Combining (\ref{Iosfinal1}), and (\ref{Iosfinal2}), we obtain (\ref{Iosfinal}).

\subsubsection{Closed-form $P_{\mathrm{O}}$ (\ref{PRosfinal})}

$P_{\mathrm{O}}$ is defined by \cite{patent}
\begin{eqnarray}
\label{POdefeq}
P_{\mathrm{O}}=\int_0^{2\pi} \int_0^{R_{\mathrm{O}}} P_{\mathrm{T}}G_{\mathrm{O}}(R) R\mathrm{d}R\mathrm{d}\theta.
\end{eqnarray}

Substituting (\ref{RO}) into (\ref{POdefeq}), for $\frac{P_{\mathrm{T}}}{P_{\mathrm{th}}}<\frac{(4\pi)^4\left({h_{\mathrm{T}}h_{\mathrm{R}}}\right)^2}{\lambda^4}$, we have
\begin{eqnarray}
\label{pos2}
P_{\mathrm{O}}=\int_{0}^{2\pi}\int_{0}^{\frac{\lambda}{4\pi}
\sqrt{\frac{P_{\mathrm{T}}}{P_{\mathrm{th}}}}}{P_{\mathrm{T}}}{G_{\mathrm{O}}(R)}R\mathrm{d}R\mathrm{d}\theta.
\end{eqnarray}
Substituting (\ref{PL_os}) into (\ref{pos2}), $P_{\mathrm{O}}$ is then computed by
\begin{eqnarray}
\label{pos3}
P_{\mathrm{O}}=2\pi
\int_{0}^{\frac{\lambda}{4\pi}}P_{\mathrm{T}}R\mathrm{d}R
+
2\pi\int_{\frac{\lambda}{4\pi}}^{\frac{\lambda}{4\pi}
\sqrt{\frac{P_{\mathrm{T}}}{P_{\mathrm{th}}}}}P_{\mathrm{T}}\left(\frac{\lambda}{4\pi} \right)^2R^{-1}\mathrm{d}R
.
\end{eqnarray}
Following some algebraic manipulations, we obtain (\ref{posfinal1}).
\begin{eqnarray}
\label{posfinal1}
P_{\mathrm{O}}=\frac{P_{\mathrm{T}}\lambda^2}{16\pi}
\left[1+ \ln \left(\frac{P_{\mathrm{T}}}{P_{\mathrm{th}}}\right)\right].
\end{eqnarray}

For $\frac{P_{\mathrm{T}}}{P_{\mathrm{th}}}\geq\frac{(4\pi)^4\left({h_{\mathrm{T}}h_{\mathrm{R}}}\right)^2}{\lambda^4}$, we have
\begin{eqnarray}
\label{pos4}
P_{\mathrm{O}}=\int_{0}^{2\pi}\int_0^{
\left(\frac{P_{\mathrm{T}}}{P_{\mathrm{th}}}\right)^{\frac{1}{4}}\sqrt{h_{\mathrm{T}}h_{\mathrm{R}}}
}{P_{\mathrm{T}}}{G_{\mathrm{O}}(R)}R\mathrm{d}R\mathrm{d}\theta.
\end{eqnarray}
Substituting (\ref{PL_os}) into (\ref{pos4}), we obtain
\begin{eqnarray}
\label{pos5}
P_{\mathrm{O}}=2\pi
\int_{0}^{\frac{\lambda}{4\pi}}P_{\mathrm{T}}R\mathrm{d}R
+
2\pi\int_{\frac{\lambda}{4\pi}}^{\frac {4\pi h_{\mathrm{T}}h_{\mathrm{R}}}{\lambda}}P_{\mathrm{T}}\left(\frac{\lambda}{4\pi} \right)^2R^{-1}\mathrm{d}R
\\
\hspace{0.35in}+2\pi\int_{\frac {4\pi h_{\mathrm{T}}h_{\mathrm{R}}}{\lambda}}^{
\left(\frac{P_{\mathrm{T}}}{P_{\mathrm{th}}}\right)^{\frac{1}{4}}\sqrt{h_{\mathrm{T}}h_{\mathrm{R}}}
}{P_{\mathrm{T}}}{(h_{\mathrm{T}}h_{\mathrm{R}})^2 R^{-3}}\mathrm{d}R
.
\end{eqnarray}

Following some algebraic manipulations, we obtain
\begin{eqnarray}
\label{posfinal2}
P_{\mathrm{O}}=\frac{P_{\mathrm{T}}\lambda^2}{16\pi}+
\frac{P_{\mathrm{T}}\lambda^2}{8\pi}\ln\left(\frac{16\pi^2 h_{\mathrm{T}}h_{\mathrm{R}}}{\lambda^2}\right)
-\frac{\pi P_{\mathrm{th}}}{2}
+ \frac{\lambda^4{P_{\mathrm{T}}}}{2^9\pi^3(h_{\mathrm{T}}h_{\mathrm{R}})^{2}}.
\end{eqnarray}
Combining (\ref{posfinal1}), and (\ref{posfinal2}), we obtain (\ref{PRosfinal}).

\begin{figure}[!t]
\begin{minipage}[t]{0.48\textwidth}
  \centering
  \includegraphics [width=2.5in]{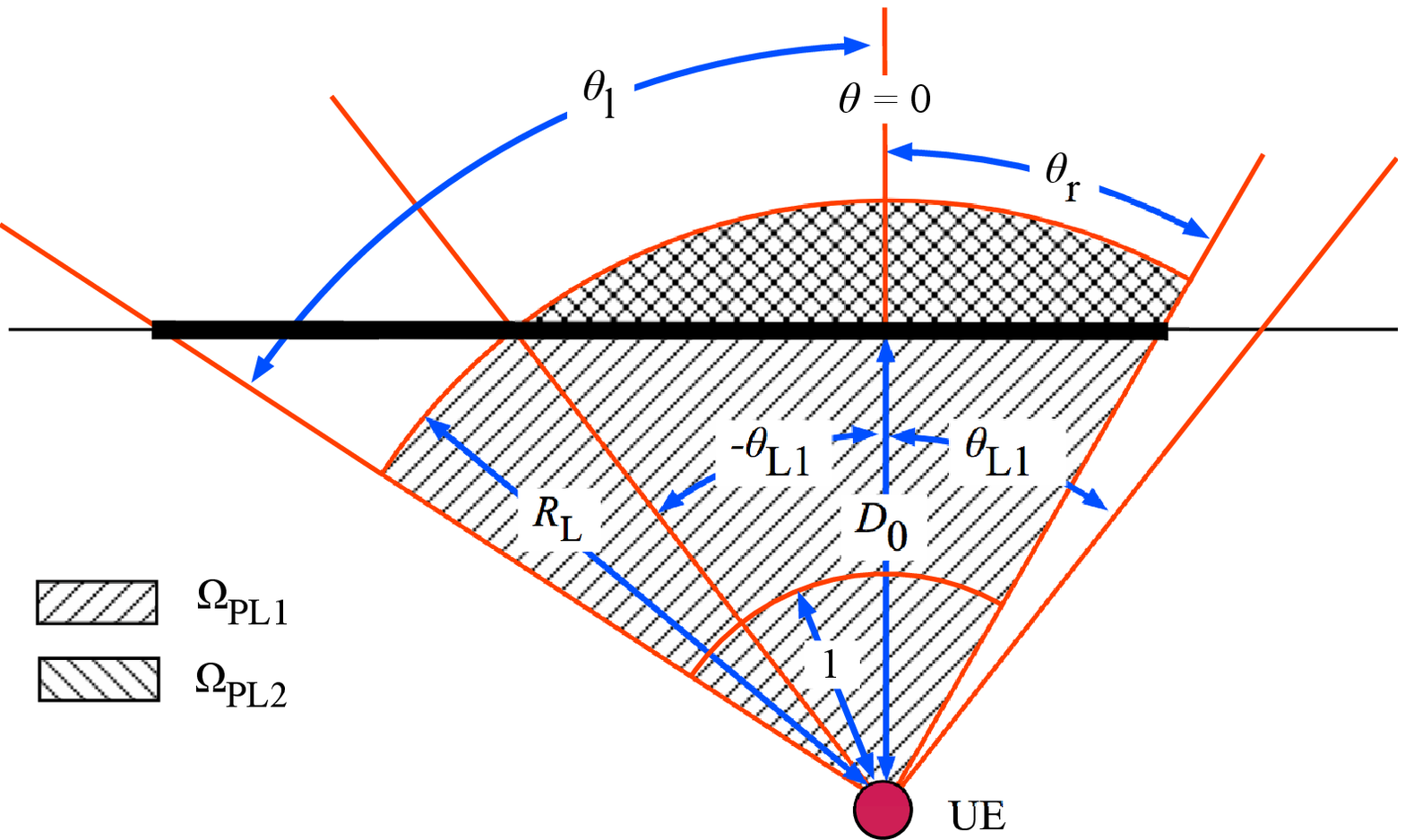}
\caption{Computation of $P_{\mathrm{L}}$ for $1\leq D_0< R_{\mathrm{L}}$.} \label{PLcase1}
\end{minipage}
\hspace{0.1in}
\begin{minipage}[t]{0.48\textwidth}
  \centering
  \includegraphics [width=2.5in]{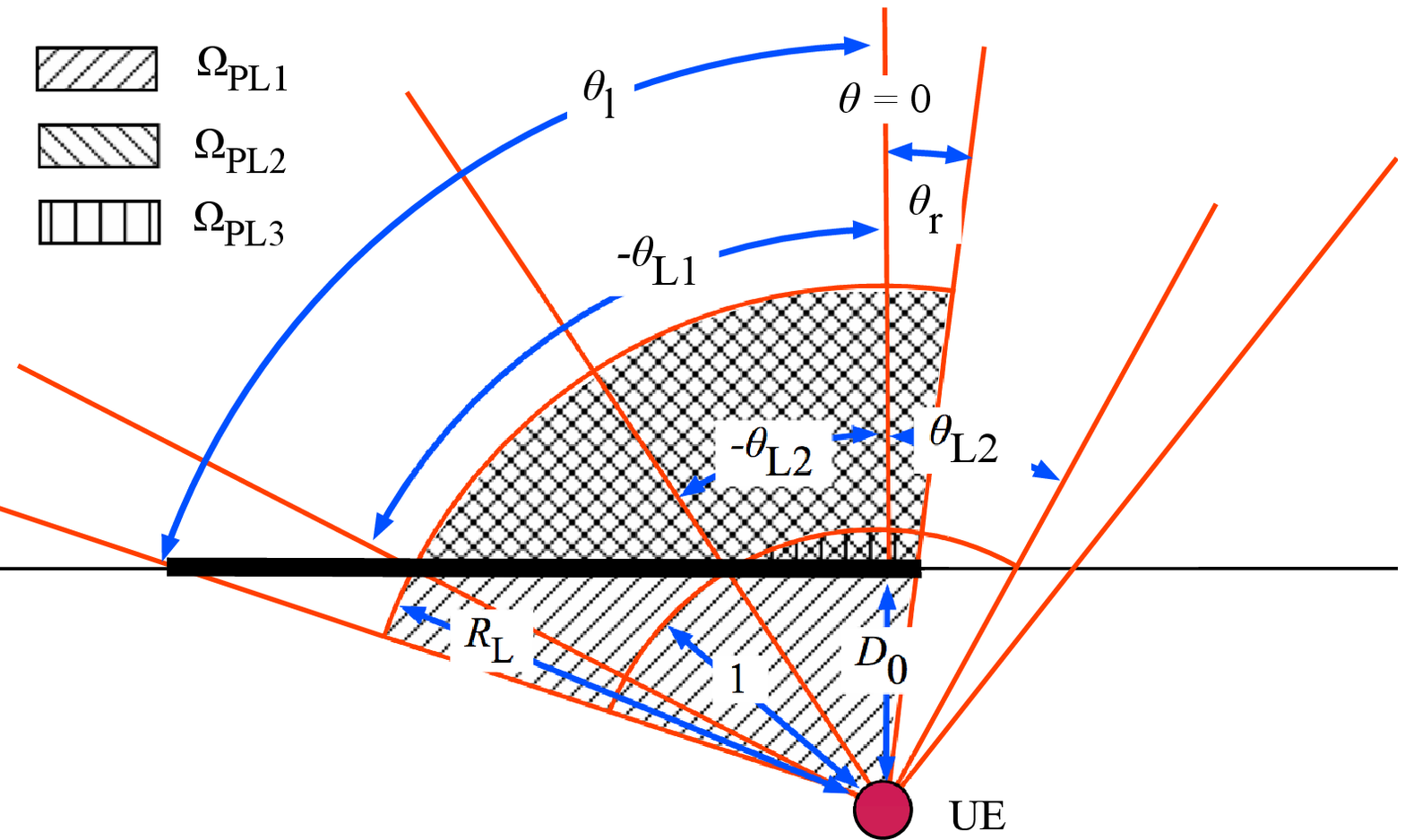}
\caption{Computation of $P_{\mathrm{L}}$ for $\frac{\lambda}{4\pi}\leq D_0< 1$.} \label{PLcase2}
\end{minipage}
\end{figure}

\subsection{Proof of Theorem \ref{PRLF}}
\label{PRLFa}
Here, we derive $P_{\mathrm{L},n_{\mathrm{TM}}}$ in a closed form. Substituting (\ref{omegaPL})  into (\ref{PLdefin}), we have
\begin{eqnarray}
P_{\mathrm{L},n_{\mathrm{TM}}}
=\int_{\theta_{\mathrm{l}}}^{\theta_{\mathrm{r}}}\int_{0}^{\min\left\{R_{\mathrm{L}},\frac{D_0}{\cos\theta}\right\}}
{P_{\mathrm{T}}}{G_{\mathrm{L}}(R)}R\mathrm{d}R\mathrm{d}\theta.
\end{eqnarray}

\subsubsection{$D_0\geq R_{\mathrm{L}}$}

For $D_0>R_{\mathrm{L}}$, we have $\min\left\{R_{\mathrm{L}},\frac{D_0}{\cos\theta}\right\}=R_{\mathrm{L}}$, and therefore
\begin{eqnarray}
P_{\mathrm{L},n_{\mathrm{TM}}}
=\int_{\theta_{\mathrm{l}}}^{\theta_{\mathrm{r}}}\int_{0}^{R_{\mathrm{L}}}
{P_{\mathrm{T}}}{G_{\mathrm{L}}(R)}R\mathrm{d}R\mathrm{d}\theta.
\end{eqnarray}
Following some algebraic manipulations, using (\ref{PL_os}), and the following Lemma \ref{lemmaz0}, we obtain
\begin{eqnarray}
\label{RLcase1final}
P_{\mathrm{L},n_{\mathrm{TM}}}=P_{\mathrm{L,1}},
\end{eqnarray}
where $P_{\mathrm{L,1}}$ is given in (\ref{PRLC1}).

\begin{lemma}
\label{lemmaz0}
The function $Z_0$ is defined as
\begin{eqnarray}
\label{z0defination}
Z_0(z_1,z_2,z_3,z_4,z_5)\triangleq \int_{z_1}^{z_2}\int_{z_3}^{z_4}  R^{-z_5}  \mathrm{d}R\mathrm{d}\theta,
\end{eqnarray}
where $-\frac{\pi}{2}<z_1<z_2<\frac{\pi}{2}$, $z_3>0$, $z_4>0$, and $z_5>-1$. The closed-form expression is given by 
\begin{eqnarray}
\label{z0final}
Z_0(z_1,z_2,z_3,z_4,z_5)\equiv\left\{
\begin{array}{ll}
(z_2-z_1) \ln\left(\frac{z_4}{z_3}\right),&z_5=1,\\
\frac{(z_2-z_1)(z_4^{1-z_5}-z_3^{1-z_5})}{1-z_5},&z_5\neq 1.
\end{array}
\right.
\end{eqnarray}

The proof of this Lemma is straightforward and therefore omitted here.
\end{lemma}

\subsubsection{$1\leq D_0< R_{\mathrm{L}}$}

For $1\leq D_0< R_{\mathrm{L}}$, an example of the toy model is shown in Fig. \ref{PLcase1}. Under this condition,
\begin{eqnarray}
\label{PRLcase2derive3}
P_{\mathrm{L},n_{\mathrm{TM}}}
=\int_{\Omega_{\mathrm{PL1}}}
{P_{\mathrm{T}}}{G_{\mathrm{L}}(R)}\mathrm{d}\Omega
-\int_{\Omega_{\mathrm{PL2}}}
{P_{\mathrm{T}}\left(\frac{\lambda}{4\pi} \right)^2}{R^{-n_{\mathrm{L}}}}\mathrm{d}\Omega,
\end{eqnarray}
where $\Omega_{\mathrm{PL1}}$ and $\Omega_{\mathrm{PL2}}$ are given by
\begin{eqnarray}
&&
\label{omegapl1}
\Omega_{\mathrm{PL1}}\triangleq \left\{\left(R,\theta\right)\left|
\left(R<R_{\mathrm{L}}\right)\land
\left(-\frac{\pi}{2}<\theta_{\mathrm{l}}<\theta<\theta_{\mathrm{r}}<\frac{\pi}{2}\right)
\right.\right\},
\\
&&
\label{omegapl2}
\Omega_{\mathrm{PL2}}\triangleq \left\{\left(R,\theta\right)\left|
\left(\frac{D_0}{\cos\theta}\leq R<R_{\mathrm{L}}\right)\land
\left(\max\{-\theta_{\mathrm{L1}},\theta_{\mathrm{l}}\}<\theta<\min\{\theta_{\mathrm{L1}},\theta_{\mathrm{r}}\}\right)
\right.\right\},
\end{eqnarray}
where $\theta_{\mathrm{L1}}$ is defined in (\ref{thata_L1}). Next, following some algebraic manipulations, we obtain
\begin{eqnarray}
\label{RL1derive}
\int_{\Omega_{\mathrm{PL1}}}
{P_{\mathrm{T}}}{G_{\mathrm{L}}(R)}\mathrm{d}\Omega=P_{\mathrm{L,1}},
\end{eqnarray}
and
\begin{eqnarray}
&&
\label{RL2derive}
\hspace{-0.1in}
\int_{\Omega_{\mathrm{PL2}}}
{P_{\mathrm{T}}}{G_{\mathrm{L}}(R)}\mathrm{d}\Omega
=
\left\{
\begin{array}{ll}
\int_{\max\{-\theta_{\mathrm{L1}},\theta_{\mathrm{l}}\}}^{\min\{\theta_{\mathrm{L1}},\theta_{\mathrm{r}}\}}\int_{ \frac{D_0}{\cos\theta}}^{R_{\mathrm{L}}}
\frac{P_{\mathrm{T}}\left(\frac{\lambda}{4\pi} \right)^2}{R^{n_{\mathrm{L}}-1}}\mathrm{d}R\mathrm{d}\theta, &(\theta_{\mathrm{l}}<\theta_{\mathrm{L1}})\land(\theta_{\mathrm{r}}>-\theta_{\mathrm{L1}}),
\\
0,&\textrm{else},
\end{array}
\right.
\nonumber\\&&
\hspace{1.3in}
=
\left\{
\begin{array}{ll}
-P_{\mathrm{L,2}}, &(\theta_{\mathrm{l}}<\theta_{\mathrm{L1}})\land(\theta_{\mathrm{r}}>-\theta_{\mathrm{L1}}),
\\
0,&\textrm{else},
\end{array}
\right.
\end{eqnarray}
where $P_{\mathrm{L,2}}$ is given in (\ref{PRLC2}), and $Z_1$ is  computed by the following Lemma \ref{lemmaz1}. 
\begin{lemma}
\label{lemmaz1}
The function $Z_1$ is defined as follows
\begin{eqnarray}
\label{z1defination}
Z_1(z_1,z_2,z_3,z_4,z_5)\triangleq \int_{z_1}^{z_2}\int_{z_3}^{\frac{z_4}{\cos(\theta)}}  R^{-z_5}  \mathrm{d}R\mathrm{d}\theta,
\end{eqnarray}
where $-\frac{\pi}{2}<z_1<z_2<\frac{\pi}{2}$, $z_3>0$, $z_4>0$, and $z_5\geq-1$. The closed-form expression is given by
\begin{eqnarray}
\label{z1final}
Z_1(z_1,z_2,z_3,z_4,z_5)\equiv\left\{
\begin{array}{ll}
\frac{z_4^2[\tan(z_2)-\tan(z_1)]}{2}+\frac{(z_1-z_2)z_3^2}{2}, &z_5=-1,\\ 
z_4\ln\left(\frac{\tan(z_2) + \sec(z_2) }{\tan(z_1) + \sec(z_1) }\right)+(z_1-z_2)z_3, &z_5=0,\\ 
(z_2-z_1)\ln \left(\frac{2z_4}{z_3}\right)+\frac{\mathrm{Im}\left[\mathrm{Li}_2(-e^{2jz_2})-\mathrm{Li}_2(-e^{2jz_1})\right]}{2}, &z_5=1,\\ 
\left\{
\begin{array}{l}
\frac{B\left(\frac{1}{2},\frac{z_5}{2}\right) \left[\mathrm{sgn}(z_2)-\mathrm{sgn}(z_1)\right] z_4^{1-z_5}}{2(1-z_5)}
\\+\frac{(z_1-z_2)z_3^{1-z_5}
}{1-z_5}  +
\frac{z_4^{1-z_5}}{(1-z_5)z_5}\times
\\
\left[
 \mathrm{sgn}(z_1)\cos^{z_5}(z_1)\ _2F_1\left(\frac{z_5}{2},\frac{1}{2}; \frac{z_5+2}{2}, \cos^2(z_1)\right)
\right.
 \\
 \left.-\mathrm{sgn}(z_2)\cos^{z_5}(z_2)\ _2F_1\left(\frac{z_5}{2},\frac{1}{2}; \frac{z_5+2}{2}, \cos^2(z_2)\right)
\right],
\end{array}
\right. &\mathrm{else}.
\end{array}
\right.
\end{eqnarray}
 
The Lemma is straightforward to verify and the proof of this Lemma is omitted here.
\end{lemma}

Substituting (\ref{RL1derive}) and (\ref{RL2derive}) into (\ref{PRLcase2derive3}), we obtain
\begin{eqnarray}
\label{RLcase2final}
P_{\mathrm{L},n_{\mathrm{TM}}}
=
\left\{
\begin{array}{ll}
P_{\mathrm{L,1}}+P_{\mathrm{L,2}}, &(\theta_{\mathrm{l}}<\theta_{\mathrm{L1}})\land(\theta_{\mathrm{r}}>-\theta_{\mathrm{L1}}),
\\
P_{\mathrm{L,1}},&\textrm{else}.
\end{array}
\right.
\end{eqnarray}
\subsubsection{$\frac{\lambda}{4\pi}\leq D_0< 1$}

An example of the toy model is shown
in Fig. \ref{PLcase2}, for $\frac{\lambda}{4\pi}\leq D_0< 1$. Under this condition, $P_{\mathrm{L},n_{\mathrm{TM}}}$ is given by 
\begin{eqnarray}
\label{PRLcase2derive4}
\begin{array}{l}
P_{\mathrm{L},n_{\mathrm{TM}}}
=\int_{\Omega_{\mathrm{PL1}}}
{P_{\mathrm{T}}}{G_{\mathrm{L}}(R)}\mathrm{d}\Omega
-\int_{\Omega_{\mathrm{PL2}}}
{P_{\mathrm{T}}\left(\frac{\lambda}{4\pi} \right)^2}{R^{-n_{\mathrm{L}}}}\mathrm{d}\Omega
\\\hspace{0.5in}+\int_{\Omega_{\mathrm{PL3}}}
{P_{\mathrm{T}}\left(\frac{\lambda}{4\pi} \right)^2}{R^{-n_{\mathrm{L}}}}\mathrm{d}\Omega
-\int_{\Omega_{\mathrm{PL3}}}
{P_{\mathrm{T}}\left(\frac{\lambda}{4\pi} \right)^2}R^{-2}\mathrm{d}\Omega
,
\end{array}
\end{eqnarray}
where the first term $\int_{\Omega_{\mathrm{PL1}}}
{P_{\mathrm{T}}}{G_{\mathrm{L}}(R)}\mathrm{d}\Omega$ and the second term $\int_{\Omega_{\mathrm{PL2}}}{P_{\mathrm{T}}\left(\frac{\lambda}{4\pi} \right)^2}{R^{-n_{\mathrm{L}}}}\mathrm{d}\Omega$ are computed following steps of computing (\ref{RL1derive}) and (\ref{RL2derive}), respectively.
For the third and the forth terms, $\Omega_{\mathrm{PL3}}$ is defined as 
\begin{eqnarray}
\label{omegapl3defeq}
\Omega_{\mathrm{PL3}}\triangleq \left\{\left(R,\theta\right)\left|
\left(\frac{D_0}{\cos\theta}\leq R<1\right)\land
\left(\max\{-\theta_{\mathrm{L2}},\theta_{\mathrm{l}}\}<\theta<\min\{\theta_{\mathrm{L2}},\theta_{\mathrm{r}}\}\right)
\right.\right\},
\end{eqnarray}
where $\theta_{\mathrm{L2}}$ is defined in (\ref{thata_L2}). Further substituting (\ref{omegapl3defeq}) into 
\begin{eqnarray}
\int_{\Omega_{\mathrm{PL3}}}
{P_{\mathrm{T}}\left(\frac{\lambda}{4\pi} \right)^2}{R^{-n_{\mathrm{L}}}}\mathrm{d}\Omega
-\int_{\Omega_{\mathrm{PL3}}}
{P_{\mathrm{T}}\left(\frac{\lambda}{4\pi} \right)^2R^{-2}}\mathrm{d}\Omega,
\end{eqnarray}
and following some straightforward
mathematical manipulations, we have
\begin{eqnarray}
\label{RL3derive}
\int_{\Omega_{\mathrm{PL3}}}
{P_{\mathrm{T}}\left(\frac{\lambda}{4\pi} \right)^2}{R^{-n_{\mathrm{L}}}}\mathrm{d}\Omega
-\int_{\Omega_{\mathrm{PL3}}}
{P_{\mathrm{T}}\left(\frac{\lambda}{4\pi} \right)^2}R^{-2}\mathrm{d}\Omega
=\left\{
\begin{array}{ll}
P_{\mathrm{L,3}}, &(\theta_{\mathrm{l}}<\theta_{\mathrm{L2}})\land(\theta_{\mathrm{r}}>-\theta_{\mathrm{L2}}),
\\
0,&\textrm{else},
\end{array}
\right.
\end{eqnarray}
where $P_{\mathrm{L,3}}$ is given in (\ref{PRLC3}). Therefore, substituting (\ref{RL1derive}), (\ref{RL2derive}) and (\ref{RL3derive}) into (\ref{PRLcase2derive4}), we have 
\begin{eqnarray}
\label{RLcase3final}
P_{\mathrm{L},n_{\mathrm{TM}}}
=
\left\{
\begin{array}{ll}
P_{\mathrm{L,1}}+P_{\mathrm{L,2}}+P_{\mathrm{L,3}}, &(\theta_{\mathrm{l}}<\theta_{\mathrm{L2}})\land(\theta_{\mathrm{r}}>-\theta_{\mathrm{L2}}),
\\
P_{\mathrm{L,1}}+P_{\mathrm{L,2}}, &(\theta_{\mathrm{l}}<\theta_{\mathrm{L1}})\land(\theta_{\mathrm{r}}>-\theta_{\mathrm{L1}})
\land\left[(\theta_{\mathrm{l}}>\theta_{\mathrm{L2}})\lor(\theta_{\mathrm{r}}<-\theta_{\mathrm{L2}})\right],
\\
P_{\mathrm{L,1}},&\textrm{else}.
\end{array}
\right.
\end{eqnarray}
Combining (\ref{RLcase1final}), (\ref{RLcase2final}) and (\ref{RLcase3final}), we obtain TABLE \ref{TABLEPRL} and therefore Theorem \ref{PRLF} is proved.

\subsection{Proof of Theorem \ref{ILF}}
\label{ILFa}
Here, we derive $I_{\mathrm{L},n_{\mathrm{TM}}}$ in closed form.
Since $R_{\mathrm{L}}>1$ is assumed, for $R>R_{\mathrm{L}}$, we have $G_{\mathrm{L}}(R)=\left(\frac{\lambda}{4\pi}\right)^{2}R^{-n_{\mathrm{L}}}$.
Substituting (\ref{omegaIL})  into (\ref{ILdefin}), we have
\begin{eqnarray}
\label{ILderive0}
I_{\mathrm{L},n_{\mathrm{TM}}}
=\int_{(\theta_{\mathrm{l}}<\theta<\theta_{\mathrm{r}})\land( R_{\mathrm{L}}<\frac{D_0}{\cos\theta})}\int_{R_{\mathrm{L}}}^{\frac{D_0}{\cos\theta}}
{P_{\mathrm{T}}}\left(\frac{\lambda}{4\pi}\right)^{2}R^{1-n_{\mathrm{L}}}\mathrm{d}R\mathrm{d}\theta.
\end{eqnarray}

\subsubsection{$R_{\mathrm{L}}\leq{D_0}$}

For $R_{\mathrm{L}}<{D_0}$, $(R_{\mathrm{L}}<\frac{D_0}{\cos\theta})=\top$, and
\begin{eqnarray}
\label{ILcase1final}
I_{\mathrm{L},n_{\mathrm{TM}}}
=\int_{\theta_{\mathrm{l}}}^{\theta_{\mathrm{r}}}\int_{R_{\mathrm{L}}}^{\frac{D_0}{\cos\theta}}
{P_{\mathrm{T}}}{G_{\mathrm{L}}(R)}R\mathrm{d}R\mathrm{d}\theta=I_{\mathrm{L},1},
\end{eqnarray}
where $I_{\mathrm{L},1}$ is computed by (\ref{ILC1}).

\subsubsection{$R_{\mathrm{L}}>{D_0}$}

For $R_{\mathrm{L}}>{D_0}$,
we can show (\ref{ILderive1})  by using straightforward mathematical manipulations.
\begin{eqnarray}
\label{ILderive1}
&&\left(\theta_{\mathrm{l}}<\theta<\theta_{\mathrm{r}}\right)\land\left( R_{\mathrm{L}}<\frac{D_0}{\cos\theta}\right)
=\left(\theta_{\mathrm{l}}<\theta<\theta_{\mathrm{r}}\right)\land\left( |\theta|>|\theta_{\mathrm{L1}}|\right)
\nonumber\\
&&\hspace{2.1in}
=\left[\left(\theta_{\mathrm{l}}<-\theta_{\mathrm{L1}}<\theta_{\mathrm{r}}\right)\land\left(\theta_{\mathrm{l}}<\theta<-\theta_{\mathrm{L1}}\right)\right]
\nonumber\\
&&\hspace{2.1in}
\lor
\left[\left(\theta_{\mathrm{l}}<\theta_{\mathrm{L1}}<\theta_{\mathrm{r}}\right)\land\left(\theta_{\mathrm{L1}}<\theta<\theta_{\mathrm{r}}\right)\right]
\nonumber\\
&&\hspace{2.1in}
\lor
\left\{\left[\left(\theta_{\mathrm{r}}<-\theta_{\mathrm{L1}}\right)\lor\left(\theta_{\mathrm{L1}}<\theta_{\mathrm{l}}\right)\right]\land\left(\theta_{\mathrm{l}}<\theta<\theta_{\mathrm{r}}\right)\right\}.
\end{eqnarray}
Substituting (\ref{ILderive1}) into (\ref{ILderive0}), we obtain  the closed-form expression of $I_{\mathrm{L},n_{\mathrm{TM}}}$ for $R_{\mathrm{L}}>{D_0}$ as 
\begin{eqnarray}
\label{ILcase2final}
I_{\mathrm{L},n_{\mathrm{TM}}}
=
\left\{
\begin{array}{l l}
I_{\mathrm{L},1},&\left(\theta_{\mathrm{r}}<-\theta_{\mathrm{L1}}\right)\lor\left(\theta_{\mathrm{L1}}<\theta_{\mathrm{l}}\right),\\
I_{\mathrm{L},2},&\left(\theta_{\mathrm{l}}<-\theta_{\mathrm{L1}}<\theta_{\mathrm{r}}\right)\lor\left(\theta_{\mathrm{l}}<\theta_{\mathrm{L1}}<\theta_{\mathrm{r}}\right),\\
I_{\mathrm{L},3},&\left(\theta_{\mathrm{l}}<-\theta_{\mathrm{L1}}<\theta_{\mathrm{r}}\right)\lor\left(\theta_{\mathrm{l}}<\theta_{\mathrm{L1}}<\theta_{\mathrm{r}}\right),\\
I_{\mathrm{L},2}+I_{\mathrm{L},3},&\left(\theta_{\mathrm{l}}<-\theta_{\mathrm{L1}}<\theta_{\mathrm{r}}\right)\lor\left(\theta_{\mathrm{l}}<\theta_{\mathrm{L1}}<\theta_{\mathrm{r}}\right).
\end{array}
\right.
\end{eqnarray}
where $I_{\mathrm{L},2}$ and $I_{\mathrm{L},3}$ are respectively computed by (\ref{ILC2}) and (\ref{ILC3}). Combining (\ref{ILcase1final}) and (\ref{ILcase2final}), we obtain TABLE \ref{TABLEIL}, and therefore Theorem \ref{ILF} is proved.

\subsection{Proof of Theorem \ref{PRNF}}
\label{PRNFa}
We now derive $P_{\mathrm{N},n_{\mathrm{TM}}}$ in a closed form. Substituting (\ref{omegaPN})  into (\ref{PNdefin}), we have
\begin{eqnarray}
P_{\mathrm{N},n_{\mathrm{TM}}}
=
\int_{(\theta_{\mathrm{l}}<\theta<\theta_{\mathrm{r}})\land(R_{\mathrm{N}}>\frac{D_0}{\cos\theta})}\int_{\frac{D_0}{\cos\theta}}^{R_{\mathrm{N}}}
{P_{\mathrm{T}}}{G_{\mathrm{N}}(R)}R\mathrm{d}R\mathrm{d}\theta.
\end{eqnarray}

\subsubsection{$D_0\geq R_{\mathrm{N}}$}

For $D_0>R_{\mathrm{N}}$, $(R_{\mathrm{N}}>\frac{D_0}{\cos\theta})=\bot$, and therefore
\begin{eqnarray}
\label{PNcase1final}
P_{\mathrm{N},n_{\mathrm{TM}}}
=0.
\end{eqnarray}
\subsubsection{$1\leq D_0< R_{\mathrm{N}}$}

For $1\leq D_0< R_{\mathrm{N}}$
\begin{eqnarray}
\label{PNcase2d1}
P_{\mathrm{N},n_{\mathrm{TM}}}
=
\int_{\Omega_{\mathrm{PN1}}}
{P_{\mathrm{T}}\left(\frac{\lambda}{4\pi} \right)^2}{R^{-n_{\mathrm{N}}}}\mathrm{d}\Omega,
\end{eqnarray}
where $\theta_{\mathrm{L1}}$ is defined in (\ref{thata_L1}), and $\Omega_{\mathrm{PN1}}$ is defined as
\begin{eqnarray}
\label{omegapn1defeq}
\Omega_{\mathrm{PN1}}\triangleq \left\{\left(R,\theta\right)\left|
\left(\frac{D_0}{\cos\theta}\leq R<R_{\mathrm{N}}\right)\land
\left(\max\{-\theta_{\mathrm{N1}},\theta_{\mathrm{l}}\}<\theta<\min\{\theta_{\mathrm{N1}},\theta_{\mathrm{r}}\}\right)
\right.\right\}.
\end{eqnarray}

Following some algebraic manipulations, we obtain
\begin{eqnarray}
\label{PNcase2final}
P_{\mathrm{N},n_{\mathrm{TM}}}
=\int_{\Omega_{\mathrm{PN1}}}
{P_{\mathrm{T}}\left(\frac{\lambda}{4\pi} \right)^2}{R^{-n_{\mathrm{N}}}}\mathrm{d}\Omega
=
\left\{
\begin{array}{ll}
P_{\mathrm{N,1}}, &(\theta_{\mathrm{l}}<\theta_{\mathrm{N1}})\land(\theta_{\mathrm{r}}>-\theta_{\mathrm{N1}}),
\\
0,&\textrm{else},
\end{array}
\right.
\end{eqnarray}
where $Z_1$ is defined in (\ref{z1defination}) and computed by (\ref{z1final}). $P_{\mathrm{N,1}}$ is computed by (\ref{PRNC1}).
\subsubsection{$\frac{\lambda}{4\pi}\leq D_0< 1$}

\begin{figure}[!t]
\begin{minipage}[t]{0.48\textwidth}
  \centering
  \includegraphics [width=2.5in]{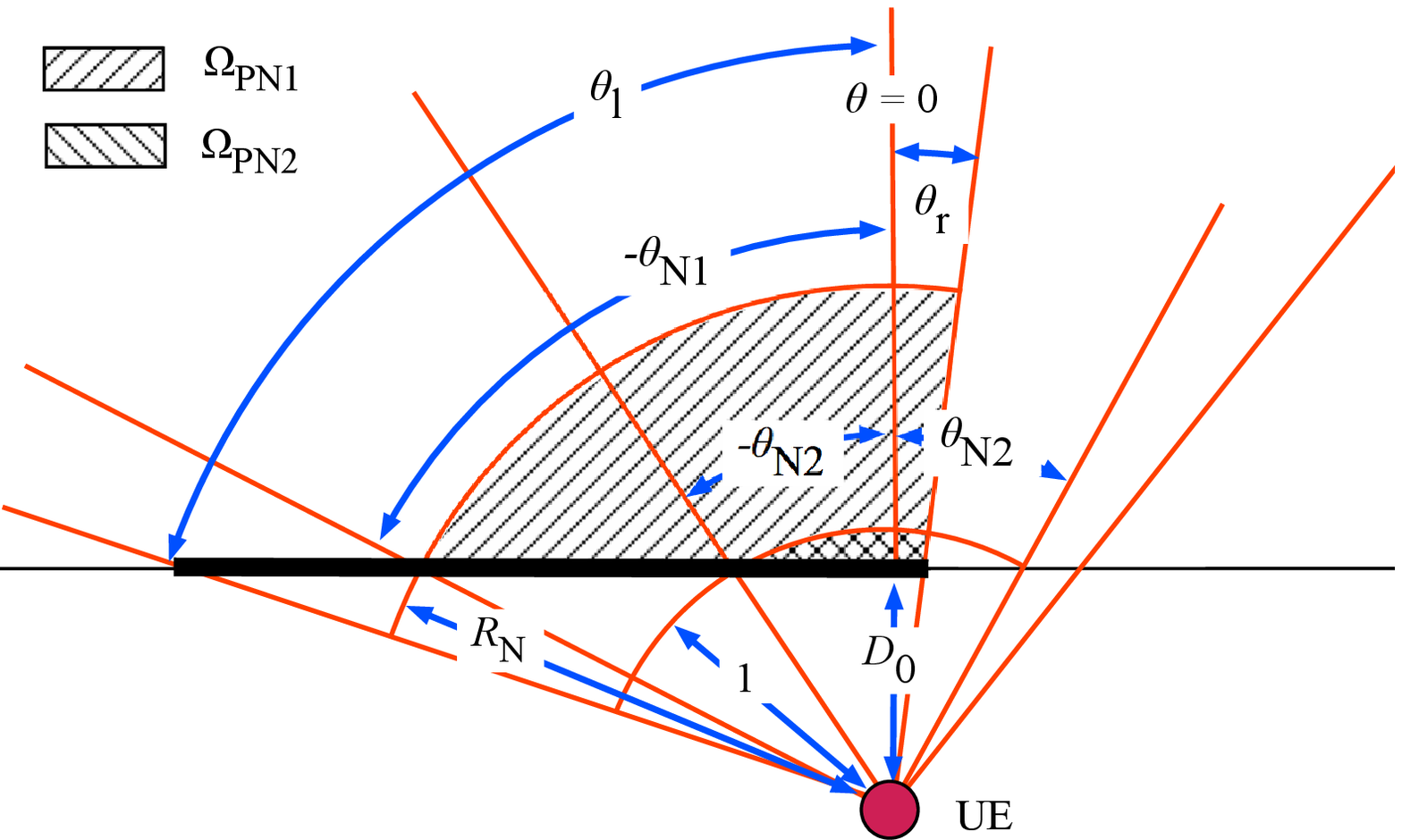}
\caption{Computation of $P_{\mathrm{N}}$ for $\frac{\lambda}{4\pi}\leq D_0< 1$.} \label{PNcase2}
\end{minipage}
\begin{minipage}[t]{0.48\textwidth}
  \centering
  \includegraphics [width=2.5in]{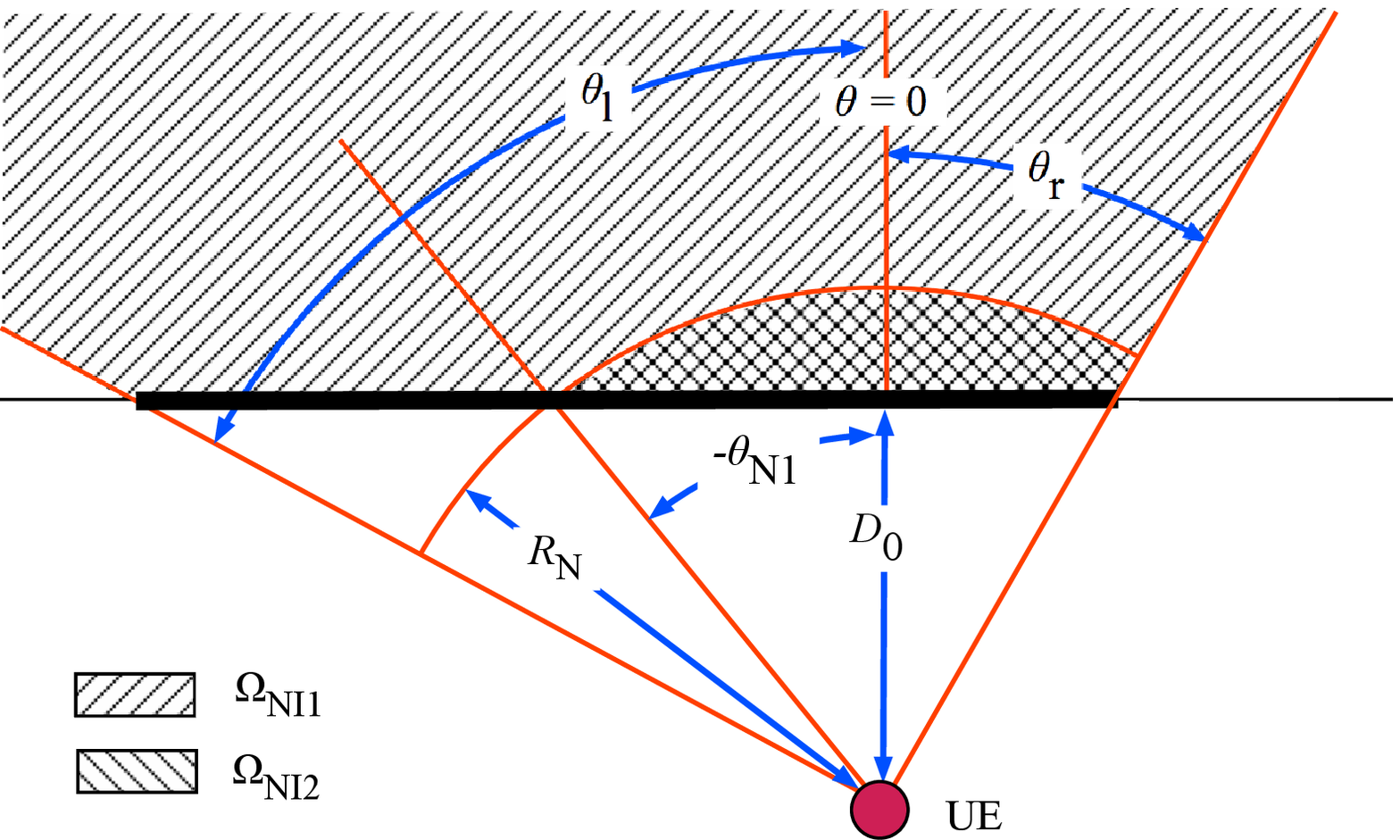}
\caption{Computation of $I_{\mathrm{N}}$ for $1\leq D_0< R_{\mathrm{N}}$.} \label{INcase1}
\end{minipage}
\end{figure}

For $\frac{\lambda}{4\pi}\leq D_0< 1$, an example of the toy model is shown in Fig. \ref{PNcase2}. Under this condition,
\begin{eqnarray}
\label{Ncase2d3}
P_{\mathrm{N},n_{\mathrm{TM}}}
=\int_{\Omega_{\mathrm{PN1}}}
{P_{\mathrm{T}}\left(\frac{\lambda}{4\pi} \right)^2}{R^{-n_{\mathrm{N}}}}\mathrm{d}\Omega
-\int_{\Omega_{\mathrm{PN2}}}
{P_{\mathrm{T}}\left(\frac{\lambda}{4\pi} \right)^2}{R^{-n_{\mathrm{N}}}}\mathrm{d}\Omega
+\int_{\Omega_{\mathrm{PN2}}}
{P_{\mathrm{T}}\left(\frac{\lambda}{4\pi} \right)^2}R^{-2}\mathrm{d}\Omega
,
\end{eqnarray}
where
 $\theta_{\mathrm{N2}}$ is defined in (\ref{thata_N2}),  $\int_{\Omega_{\mathrm{PN2}}}{P_{\mathrm{T}}\left(\frac{\lambda}{4\pi} \right)^2}{R^{-n_{\mathrm{N}}}}\mathrm{d}\Omega$ is computed by (\ref{PNcase2final}), and $\Omega_{\mathrm{PN2}}$ is defined as
 \begin{eqnarray}
\label{omegapn2defeq}
\Omega_{\mathrm{PN2}}\triangleq \left\{\left(R,\theta\right)\left|
\left(\frac{D_0}{\cos\theta}\leq R<1\right)\land
\left(\max\{-\theta_{\mathrm{N2}},\theta_{\mathrm{l}}\}<\theta<\min\{\theta_{\mathrm{N2}},\theta_{\mathrm{r}}\}\right)
\right.\right\}.
 \end{eqnarray}
 
Following some straightforward computations by substituting (\ref{omegapn2defeq}) into the second and the third terms of (\ref{Ncase2d3}), we have
\begin{eqnarray}
\label{PNcase3d4}
\int_{\Omega_{\mathrm{PN2}}}
{P_{\mathrm{T}}\left(\frac{\lambda}{4\pi} \right)^2}{R^{-2}}\mathrm{d}\Omega
-\int_{\Omega_{\mathrm{PN2}}}
{P_{\mathrm{T}}\left(\frac{\lambda}{4\pi} \right)^2}{R^{-n_{\mathrm{N}}}}\mathrm{d}\Omega
=
\left\{
\begin{array}{ll}
P_{\mathrm{N,2}}, &(\theta_{\mathrm{l}}<\theta_{\mathrm{L2}})\land(\theta_{\mathrm{r}}>-\theta_{\mathrm{L2}}),
\\
0,&\textrm{else},
\end{array}
\right.
\end{eqnarray}
where $P_{\mathrm{N,2}}$ is computed by (\ref{PRNC2}). Therefore, for $\frac{\lambda}{4\pi}\leq D_0< 1$, $P_{\mathrm{N},n_{\mathrm{TM}}}$ is given by
\begin{eqnarray}
\label{PNcase3final}
P_{\mathrm{N},n_{\mathrm{TM}}}
=
\left\{
\begin{array}{ll}
P_{\mathrm{N,1}}+P_{\mathrm{N,2}}, &(\theta_{\mathrm{l}}<\theta_{\mathrm{N2}})\land(\theta_{\mathrm{r}}>-\theta_{\mathrm{N2}}),
\\
P_{\mathrm{N,1}}, &(\theta_{\mathrm{l}}<\theta_{\mathrm{N1}})\land(\theta_{\mathrm{r}}>-\theta_{\mathrm{N1}})
\land\left[(\theta_{\mathrm{l}}>\theta_{\mathrm{N2}})\lor(\theta_{\mathrm{r}}<-\theta_{\mathrm{N2}})\right],
\\
0,&\textrm{else}.
\end{array}
\right.
\end{eqnarray}

Combining (\ref{PNcase1final}), (\ref{PNcase2final}) and (\ref{PNcase3final}), we obtain TABLE \ref{TABLEPRN} and therefore Theorem \ref{PRNF} is proved.

\subsection{Proof of Theorem \ref{INF}}
\label{INFa}

Here, we derive $I_{\mathrm{N},n_{\mathrm{TM}}}$ in a closed form.
Substituting (\ref{omegaIN})  into (\ref{INdefin}), we have
\begin{eqnarray}
I_{\mathrm{N},n_{\mathrm{TM}}}
=\int_{\theta_{\mathrm{l}}}^{\theta_{\mathrm{r}}}\int_{\max\left\{R_{\mathrm{L}},\frac{D_0}{\cos\theta}\right\}}^{+\infty}
{P_{\mathrm{T}}}{G_{\mathrm{N}}(R)}R\mathrm{d}R\mathrm{d}\theta.
\end{eqnarray}

\subsubsection{$R_{\mathrm{N}}<{D_0}$}

For $R_{\mathrm{N}}<{D_0}$, we have $\left(\max\left\{R_{\mathrm{L}},\frac{D_0}{\cos\theta}\right\}=\frac{D_0}{\cos\theta}\right)=\top$. Therefore, we obtain
\begin{eqnarray}
\label{INcase1final}
I_{\mathrm{N},n_{\mathrm{TM}}}
=\int_{\theta_{\mathrm{l}}}^{\theta_{\mathrm{r}}}\int_{\frac{D_0}{\cos\theta}}^{+\infty}
{P_{\mathrm{T}}}\left(\frac{\lambda}{4\pi}\right)^2R^{1-n_{\mathrm{N}}}\mathrm{d}R\mathrm{d}\theta=I_{\mathrm{N},1},
\end{eqnarray}
where $I_{\mathrm{N},1}$ is computed by (\ref{INC1}).
\subsubsection{$R_{\mathrm{N}}>{D_0}$}

For $R_{\mathrm{N}}>{D_0}$, an example of the toy model is shown in Fig. \ref{INcase1}. Under this condition,
\begin{eqnarray}
\label{INcase1derive1}
I_{\mathrm{N},n_{\mathrm{TM}}}
=\int_{\Omega_{\mathrm{IN1}}}
{P_{\mathrm{T}}\left(\frac{\lambda}{4\pi} \right)^2}{R^{-n_{\mathrm{N}}}}\mathrm{d}\Omega
-\int_{\Omega_{\mathrm{IN2}}}
{P_{\mathrm{T}}\left(\frac{\lambda}{4\pi} \right)^2}{R^{-n_{\mathrm{N}}}}\mathrm{d}\Omega,
\end{eqnarray}
where $\Omega_{\mathrm{IN1}}$ and $\Omega_{\mathrm{IN2}}$ are defined as
\begin{eqnarray}
&&
\label{omegain1defeq}
\Omega_{\mathrm{IN1}}\triangleq \left\{\left(R,\theta\right)\left|
\left(R\geq \frac{D_0}{\cos\theta}\right)\land
\left(-\frac{\pi}{2}<\theta_{\mathrm{l}}<\theta<\theta_{\mathrm{r}}<\frac{\pi}{2}\right)
\right.\right\},
\\
&&
\label{omegain2defeq}
\Omega_{\mathrm{IN2}}\triangleq \left\{\left(R,\theta\right)\left|
\left(\frac{D_0}{\cos\theta}\leq R<R_{\mathrm{N}}\right)\land
\left(\max\left\{-\theta_{\mathrm{N1}},\theta_{\mathrm{l}}\}<\theta<\min\{\theta_{\mathrm{N1}},\theta_{\mathrm{r}}\right\}\right)
\right.\right\}.
\end{eqnarray}
Following some algebraic manipulations, we have
\begin{eqnarray}
\label{INcase2final}
I_{\mathrm{N},n_{\mathrm{TM}}}=
\left\{
\begin{array}{ll}
I_{\mathrm{N},1}+I_{\mathrm{N},2}, &(\theta_{\mathrm{l}}<\theta_{\mathrm{L2}})\land(\theta_{\mathrm{r}}>-\theta_{\mathrm{L2}}),
\\
I_{\mathrm{N},1},&(\theta_{\mathrm{l}}>\theta_{\mathrm{L2}})\lor(\theta_{\mathrm{r}}<-\theta_{\mathrm{L2}}),
\end{array}
\right.
\end{eqnarray}
where $I_{\mathrm{N},2}$ is computed by (\ref{INC2}). Combining (\ref{INcase1final}) and (\ref{INcase2final}), we obtain TABLE \ref{TABLEIN}, and therefore Theorem \ref{INF} is proved.


\begin{thebibliography}{1}


\bibitem{indoors80}
Cisco, ``Cisco vision: 5G-thriving indoors,'' [Online], \url{https://www.cisco.com/c/dam/en/us/solutions/collateral/service-provider/ultra-services-platform/5g-ran-indoor.pdf}.




\bibitem{6Gdefine}
M. Giordani, et al., ``Toward 6G networks: Use cases and technologies''  \textit{IEEE  Commun. Mag.}, vol. 58, no. 3, pp. 55-61, 2020.




\bibitem{sc2}
X. Ge, et al., ``5G ultra-dense cellular networks,'' \textit{IEEE Wireless Commun.}, vol. 23, no. 1, pp. 72-79, 2016.

\bibitem{sc3}
L. Wang, ``Wireless powered dense cellular networks: how many small cells do we need?''
\textit{IEEE J. Sel. Areas Commun.},
vol. 35, no. 9, pp. 2010-2024, 2017.

\bibitem{sc4}
J. Chen, et al., ``Coverage and handoff analysis of 5G fractal small cell networks,'' \textit{IEEE Trans. Wireless Commun.}, vol 18, no. 2, pp. 1263-1276, 2019. 
\bibitem{sc5}
X. Ge, et al., ``Small cell networks with fractal coverage characteristics,'' \textit{IEEE Trans. Commun.}, vol. 66, no. 11, pp. 5457-5469, 2018. 


\bibitem{jz1}
J. Zhang, et al., ``Femtocells-technologies and deployment,'' \textit{Wiley},  2010.
\bibitem{jz2}
D. Lopez, et al., ``OFDMA femtocells: a roadmap on interference avoidance,'' \textit{IEEE Commun. Mag.}, vol. 47, no 9, pp. 41-48, 2009.
\bibitem{jz3}
A. H. Jafari, et al., ``Performance analysis of dense small cell networks with practical antenna heights under Rician fading,''
\textit{IEEE Access}, vol. 6, pp. 9960-9974, 2017.

\bibitem{xg0}
X. Ge, et al., ``Joint optimization of computation and communication power in multi-user massive MIMO systems,'' \textit{IEEE Trans Wireless Commun.}, Vol. 17, No. 6, pp. 4051-4063, 2018.

\bibitem{massive1}

J. Hoydis, et al., ``Massive MIMO in the UL/DL of cellular networks: How many antennas do we need?'' \textit{IEEE J. Sel. Area Commun.}, vol. 31, no. 2, pp. 160-171, 2013.

\bibitem{massive2}
E. G. Larsson, et al., ``Massive MIMO for next generation wireless systems,''  \textit{IEEE Commun. Mag.}  vol. 52, no. 2, pp. 186-195, 2014.

\bibitem{massive3}

L. Lu, et al., ``An overview of massive MIMO: Benefits and challenges,'' \textit{IEEE J. Sel. Topics Sig. Proc.}, vol. 8, no. 5, pp. 742-758, 2014.

\bibitem{smartcity}
T. Nam and T. A. Pardo, ``Conceptualizing smart city with dimensions of technology, people, and institutions,'' in Proc. \textit{Annual International Conference on Digital Government Research}, pp. 282-291, 2011.

\bibitem{patent}
Ranplan Wireless Network Design,``Method for evaluating the building wireless performance,'' \textit{Patent}, 2019.

\bibitem{tvtbwp}
J. Zhang, et al., ``Wireless energy efficiency evaluation for buildings under design based on analysis of interference gain,'' \textit{IEEE Trans. Veh. Tech.}, vol. 69, no. 6, pp. 6310-6324, 2020.

\bibitem{tallbuilding}
 D. Yagaanbuant, et al., ``Some design issues of high-rise buildings,'' in Proc. \textit{IFOST}, 2007.
\bibitem{buildingenergy}
L. P\'erez-Lombard, et al., ``A review on buildings energy consumption information,'' \textit{Energy and buildings},  vol. 40, no. 3, pp. 394-398, 2008.

\bibitem{cibse}
CIBSE, ``TM54: evaluating operational energy performance of buildings at the design stage,'' \textit{CIBSE}, 2013.



\bibitem{r1} 
D. Feng, et al., ``A survey of energy-efficient wireless communications,'' \textit{IEEE Commun. S. \& Tutor.}, vol. 15, no. 1, pp. 167-178, 2013.

\bibitem{r2}
C. L. I, et al., ``Energy-efficient 5G for a greener future,'' \textit{Nat. Electron.}, vol. 3, pp. 182-184, 2020.

\bibitem{r3}
A. Abrol and R. K. Jha, ``Power optimization in 5G Networks: a step towards green communication,'' \textit{IEEE Access}, vol. 4, pp. 1355-1374, 2016.
\bibitem{LOS1}
H. Zheng, et al., ``Exact line-of-sight probability for channel modelling in typical indoor environments,'' \textit{IEEE Ant. Wireless Prop. Lett.}, vol. 17, no. 7, pp. 1359-1362, 2018.
\bibitem{LOS2}
W. Yang, et al., ``Line-of-sight probability for channel modeling in 3D indoor environments,'' \textit{IEEE Ant. Wireless Prop. Lett.}, vol. 19, no. 7, pp. 1182-1186, 2020.
\bibitem{LOS3}
W. Yang, et al., ``Machine learning based indoor line-of-sight probability prediction,'' in Proc. \textit{IEEE ISAP}, 2019.
\bibitem{TB}
T. Bai, et al., ``Coverage and rate analysis for millimeter-wave cellular networks,'' \textit{IEEE Trans. Wireless Commun.}, vol. 4, no. 2, pp. 1100-1114, 2015.

\bibitem{wallgeneration}
M. K. M$\rm\ddot{u}$ller, et al., ``Analyzing wireless indoor communications by blockage models,'' \textit{IEEE Access}, vol. 5, pp. 2172-2186,  2017.



\bibitem{raytracing1}
K. Haneda, et al., ``Spatial coexistence of millimeter-wave distributed indoor channels,'' in Proc. \textit{IEEE VTC-Spring}, 2015.
\bibitem{raytracing2}
L. Azpilicueta, et al., ``A hybrid ray launching-diffusion equation approach for propagation prediction in complex indoor environments,'' \textit{IEEE Ant. Wireless Prop. Lett.}, vol. 16, pp. 214-217, 2017.
\bibitem{raytracing3}
P. H. Tseng, et al.,``Ray-tracing-assisted fingerprinting based on channel impulse response measurement for indoor positioning,'' \textit{IEEE Trans. Ins. Mea.}, vol. 66, no. 5, pp. 1032-1045, 2017.
\bibitem{raytracing4}
F. Casino, et al., ``Optimized wireless channel characterization in large complex environments by hybrid ray launching-collaborative filtering approach,'' \textit{IEEE Ant. Wireless Prop. Lett.}, vol. 16, pp. 780-783, 2017.
\bibitem{raytracing5}
Z. Lai, et al., ``Modelling the mmWave channel based on intelligent ray launching model,'' in Proc. \textit{Eucap}, 2015.
\bibitem{raytracing6}
W. Yang, et al., ``Indoor measurement based verification of ray launching algorithm at the Ka-band,'' in Proc. \textit{URSI GASS}, 2020.
\bibitem{ZHL}
Z. Lai, et al., ``Intelligent ray launching algorithm for indoor scenarios,'' \textit{Radioengineering}, vol. 20, no. 2, 2011.

\bibitem{eurostars}
EUROSTARS Project 11088 BuildWise: The development of a software tool to design buildings with tailored wireless performance (Ranked the 16th among 331 applications), [Online]. https://www.eurostars-eureka.eu/project/id/11088, 2017.



\bibitem{twolobe}
X. Zhang and J. Andrews. ``Downlink cellular network analysis with multi-slope path loss models,'' \textit{IEEE Trans. Commun.}, vol. 63, no. 5, pp. 1881-1894, 2015.

\bibitem{tworay}

C. Sommer and F. Dressler, ``Using the right two-ray model? a measurement based evaluation of PHY models in VANETs,''
in Proc. \textit{ACM MobiCom}, 2011.

\bibitem{3GPP}
3GPP, ``Study on channel model for frequencies from 0.5 to 100 GHz,'' \textit{3GPP TR 38.901 version 14.0.0 Release 14}, 2017.

\bibitem{dmmimo}
U. Madhow, et al.,``Distributed massive MIMO: algorithms, architectures and concept systems,'' in Proc. \textit{IEEE ITA}, 2014.


\bibitem{comp1}
S. Fu, et al.,``Distributed transmission scheduling and power allocation in CoMP,'' \textit{IEEE Sys. J.}, vol. 12, no. 4, pp. 3096-3107, 2018.
\bibitem{comp2}
E. Pateromichelakis et al., ``On the evolution of multi-cell scheduling in 3GPP LTE/LTE-A,'' \textit {IEEE Commun. Surveys Tuts.}, vol. 15, no. 2, pp. 701-717, 2013.


%
%
%
%

\bibitem{5Gdefine}

J. G.  Andrews, et al., ``What will 5G be?''  \textit{IEEE J. Sel. Area Commun.}, vol. 32, no. 6, pp. 1065-1082, 2014.

\bibitem{winnerII}
D. S. Baum, et al., ``IST-WINNER D5.4, final report on link and system level channel models,''  \url{https://www.researchgate.net/publication/229031750_IST-2003-507581_WINNER_D5_4_v_14_Final_Report_on_Link_Level_and_System_Level_Channel_Models}, 2005.

%
\end{thebibliography}
\end{document}